\RequirePackage{fix-cm}
\documentclass[smallextended]{svjour3}       
\smartqed  
\usepackage{graphicx}
\usepackage{amsmath}

\usepackage{amsthm}
\usepackage{algorithm}
\usepackage{algpseudocode}
\usepackage{subcaption}
\usepackage{amssymb}
\usepackage{subcaption}
\usepackage{grffile}
\usepackage{cite}
\usepackage[table,xcdraw]{xcolor}
\usepackage{sidecap}
\usepackage{xcolor}

\sidecaptionvpos{figure}{c}

\makeatletter
\newcommand\subparagraph{%
	\@startsection{subparagraph}{5}
	{\parindent}
	{3.25ex \@plus 1ex \@minus .2ex}
	{-1em}
	{\normalfont\normalsize\bfseries}}
\makeatother
\usepackage{titlesec}
\let\subparagraph\relax

\usepackage{titlesec}

\titleformat{\paragraph}
{\normalfont\normalsize\bfseries}{\theparagraph}{1em}{}
\titlespacing*{\paragraph}
{0pt}{3.25ex plus 1ex minus .2ex}{1.5ex plus .2ex}

\graphicspath{{./}}

\begin{document}

\title{Understanding and forecasting lifecycle events in information cascades
}


\author{Soumajyoti Sarkar         \and
        Ruocheng Guo \and
        Paulo Shakarian 
}

\institute{Soumajyoti Sarkar \at
		Arizona State University \\
	\email{ssarka18@asu.edu}           
	\and
Ruocheng Guo \at
Arizona State University \\
\email{rguosni@asu.edu} 
\and
Paulo Shakarian\at
Arizona State University \\
\email{shak@asu.edu} 
}


\date{Received: date / Accepted: date}

\maketitle

\begin{abstract}
Most social network sites allow users to reshare a piece of information posted by a user. As time progresses, the cascade of reshares grows, eventually saturating after a certain time period. While previous studies have focused heavily on one aspect of the cascade phenomenon, specifically predicting when the cascade would go viral, in this paper, we take a more holistic approach by analyzing the occurrence of two events within the cascade lifecycle - the period of maximum growth in terms of surge in reshares and the period where the cascade starts declining in adoption. We address the challenges in identifying these periods and then proceed to make a comparative analysis of these periods from the perspective of  network topology. We study the effect of several node-centric structural measures on the reshare responses using Granger causality which helps us quantify the significance of the network measures and understand the extent to which the network topology impacts the growth dynamics. This evaluation is performed on a dataset of 7407 cascades extracted from the Weibo social network.  Using our causality framework, we found that an entropy measure based on nodal degree causally affects the occurrence of these events in 93.95\% of cascades. Surprisingly, this outperformed clustering coefficient and PageRank which we hypothesized would be more indicative of the growth dynamics based on earlier studies. We also extend the Granger-causality Vector Autoregression (VAR) model to forecast the times at which the events occur in the cascade lifecycle.
 
\end{abstract}

\section{Introduction}
Sharing information in online social networks has become a widespread phenomenon where multimedia information can be in the form of text, photos or links to other information. When such a piece of information is shared among multiple people over a prolonged period of time, we obtain cascades of reshares for that information. There has been a growing interest in information cascades as they have wide range of applications in viral marketing \cite{kkt_viral, paulo_viral} and cascade prediction \cite{leskovic_bib}. The increasing availability of data identifying diffusion traces that lead to such cascades has allowed researchers to obtain empirical evidences of mechanisms through which information diffuses in social networks. One attribute of the cascade that has received widespread attention in the recent past has been the cascade size. The authors in \cite{leskovic_bib} address the question of whether the cascade size can be predicted at all. Following their positive results, there have been several attempts to predict the future size of the cascade and its extension to whether it would cross a threshold within a certain time \cite{guo_cascade}. 

While these papers present several interesting results on the temporal dynamics of the cascade progress that impacts the future size, they either predict the final cascade size or when the cascade would reach a certain size using social network characteristics or diffusion modeling mechanisms. However, most of these studies tend to overlook the time periods in which a cascade grows and attains a size which explains when the cascade would be experiencing fast adoption and when it would be nearing its saturation leading to inhibition of further progress. One of the closest work related to this problem of understanding the growth dynamics in different phases of the cascade lifecycle has been done in \cite{cui_behavior} where the authors model the cascading mechanism using behavioral dynamics thereby predicting the trajectory of the cascade growth in terms of the size attained over time. In this paper, we try to bridge this gap in understanding the predictive features of cascade size and the growth dynamics in various time periods in the cascade lifecycle by studying the cascade topology at various times.

Innovation diffusion of products has often been compared to information diffusion in social networks as in both cases, the market penetration of new products or services are hugely driven by social influences and trust between the users. As studied in \cite{product_innov}, analyzing the growth patterns and the turning points in the product lifecycle is always crucial for market evolution. Drawing motivation from such ideas, we identify these turning points in the cascade lifecycle and then understand the dynamics of growth in the phases encompassing these turning points. Similar to the product lifecycle, the lifecycle of a cascade progresses through different phases of growth and while previous studies have focused on the period where there is a fast rate of adoption shown by bursty growth \cite{burst_time}, there have been little to no work focusing on the period when the cascade starts saturating leading to a complete stop in reshares. This raises an interesting question: how do the social network attributes differ in the time period where there is a sudden rise in reshares compared to the period where the cascade nears complete inhibition and are there any early visible patterns in the network structure that can help to predict the future course of the cascade lifecycle ? 

To address the above, we describe the concept of ``events'' in the lifecycle of information cascades and then we describe the set of events that we study in this paper. Traditionally, the Product Life Cycle (PLC) concept \cite{plc} used in economics assumes 4 time phases to describe the product life span: introduction, growth, equilibrium and decline. The \textit{introduction} phase refers to the take-off starting period when the rate of adoption is slow because of being new to the market. The \textit{growth} phase refers to the phase when the adoption rate gains maximum momentum in the entire course of the lifecycle and is followed by the \textit{maturity} phase when the adoption rate starts saturating eventually leading to the \textit{decline} phase after which the adoption rate fails to resurge. A cascade lifecycle can be represented by a sequence of similar time phases which we term as \textit{lifecycle events} as they signify the occurrence of change points in the trajectory of the cascade. In this context, the main contributions of this paper are three fold:

\begin{itemize}
	\item We create a framework to identify two \textit{events} - the period of maximum growth or the ``steep'' interval and the start of inhibition period in the course of the cascade lifecycle using the concept of Hawkes self exciting point process model \cite{hawkes_zha, seismic}. 
	
	\item We study how the structural properties of the cascade network changes as it grows over time and we analyze the time intervals leading to the two events to explain the properties in the network structure that augments or creates a bottleneck in the information diffusion process. To this end, we use node-centric social network measures to explain the phenomena of growth and decay of the resharing process by focusing on individuals as the factor for information diffusion. Our framework of Granger causality to quantitatively evaluate the features \cite{granger_causality} shows that degree entropy as a measure of a node's neighbors' degree is a strong causal feature and node clustering coefficient is the weakest for measuring the future reshare response times indicated by the fact that while degree entropy Granger causes the response times in 93.95\% of cascades, clustering coefficient shows a similar causal effect in only 89.4\% of cascades.
	
	\item We use the node-centric measures individually in addition to extending the Vector Autoregression model (VAR) used in the Granger causality framework to forecast the occurrence of the steep and inhibition times in the lifecycle using the node-centric measures. We find that while degree entropy performs the best as an individual feature having a mean absolute forecasting error of 33.65 minutes for the steep time and 81.18 minutes for the inhibition time, clustering coefficient performs the worst having a mean absolute error of 224.04 and 400.03 minutes for the steep and inhibition times respectively, for a similar regression model. This also suggests that forecasting the inhibition time is a more difficult problem to tackle than forecasting the steep time in the cascade lifecycle.
\end{itemize}

The rest of the paper is organized as follows: we first introduce the framework for the identification of the event intervals followed by the statistical framework for evaluation of node measures using Granger causality in Section~\ref{sec:tech_prelim}. Then we describe our dataset in Section~\ref{sec:dataset} followed by the description of the social network measures used in this paper in Section~\ref{sec:net_feat}. Finally we provide the experimental results in Section~\ref{sec:quant_res} followed by the Conclusion. The key component in our approach is studying the cascade network structure at different time intervals within the lifecycle and to the best of our knowledge, this is a first comprehensive study on comparing and evaluating such \textit{events} in the cascade lifecycle.

\section{Related Work}
As mentioned before, there has been an increasing line of work surrounding the dynamics of cascade growth \cite{cui_behavior, leskovic_bib}, where the authors try to model the trajectory of the cascade growth and in the event also predict the future course of the cascades. Modeling the underlying diffusion process to characterize events in the lifecycle of the cascades is a widely studied problem \cite{burst_time, fall_patterns}. Two of the most widely used approaches to study such diffusion processes are: (1) using the social network graph topology to understand the position and importance of nodes and then using them for measuring or predicting the diffusion spread as done in \cite{diff_cent, deg_entropy, ugander, kitsak} and (2) using the temporal process of the formation of cascades to build parametric influence models and then use optimization methods to learn the parameters and use them for prediction of events such as done in \cite{cui_behavior, seismic}. There are numerous other methods also which have been studied in social network diffusion \cite{paulo_book} surrounding such cascades. Social network diffusion has been an important component in predicting cascade growth. To this end, \cite{hawkes_zha, zha_diffusion} used Hawkes model to measure time-varying social influence and to model viral network diffusion. Applications of Hawkes process dates back to studies describing self-exciting processes of earthquakes \cite{hawkes_pp}. Our approach to identify the period of maximum growth and start of the inhibition region in a cascade life based on Hawkes process is performed along the line of work introduced in \cite{seismic} where the authors use Hawkes point process model to predict the final number of reshares of a post. 

In this paper, we focus on using graph based measures in assessing the structural properties of the social network that explains why the cascade starts inhibition after a certain time period as opposed to its steep period where it experiences huge growth. Using social network features to identify superspreaders in information diffusion has been comprehensively studied in \cite{sei_spreaders, kitsak, paulo} where the authors compare various features like PageRank, degree centrality, core number in a $k$-core decomposition of network, to identify influential users.

Studying the position of nodes in the network structure has been a popular way of understanding the diffusion spread. Related work in \cite{kitsak} study such node centralities using the $k$-core measure where the authors show that the position of a node in the core structure of the network is more revealing of the diffusion spread than just the neighbor degree. But in such studies, the authors use the friendship or human contact networks to model the diffusion spread and simulate them over SIR or SIS models. Therefore most of these networks are static snapshots as opposed to our approach where we intend to study temporal networks which change rapidly within a short span. Temporal centralities studied in \cite{temp_kim} are defined based on dynamic networks where the network edges change over time. However the major drawback in defining such temporal centralities is the time granularity on which such centralities are defined and as such with large network size, the computation of such centralities would be expensive. To avoid such huge overheads, in our work we define the centralities on static networks but considering evolving time-ordered networks as static networks over a time range. Diffusion centrality introduced in \cite{DiffKang} measures the centrality of nodes with respect to different propagation properties where the authors postulate that the importance of a vertex may differ with respect to different topics in the same network. Since we ignore the content or topics that are propagated in this network, we use other widely used centralities on static networks evolving over time instead of using such centralities. Similarly the authors in \cite{DiffusionCent} study the effect of the top ranked nodes with respect to various centrality measures in spreading the infection in different phases of the diffusion cycle namely the sub-critical, critical and super-critical regimes. In this paper we use a data-driven study of diffusion mechanism avoiding any assumptions of a diffusion model, and we use evolving time networks to measure the importance of the nodes with respect to network centralities in two important phases of the cascade lifecycle described in the following sections. 

\begin{table}[!t]
	\centering
	\renewcommand{\arraystretch}{1}
	\caption{Table of Symbols}
	\begin{tabular}{|p{2cm}|p{10cm}|}
		\hline 
		{\bf Symbol} & {\bf Description}\\ 
		\hline\hline
		C           & Information cascade \\
		\hline
		$T_C $ & Total span of cascade C in minutes (time difference between the first and the last reposting)
		\\
		\hline
		$S_C $ & Total size of cascade $C$ equal to the total number of reshares for $C$.
		\\
		\hline
		$S_C^t $ & Size of cascade $C$ in the time range $[0, t]$.
		\\
		\hline
		$\tau_C $ & Sequence of reshare times within a cascade ordered by time. 
		\\
		\hline
		$\tau'_C $ & Subsequences within $\tau$ for the cascade $C$.
		\\
		\hline
		$\mathcal{Q}$ & Number of subsequences in a cascade.
		\\
		\hline
		$V^{\tau'}_C $ & Nodes which participated in the cascade $C $  in the subsequence $\tau'$. \\
		\hline
		$E^{\tau'}_C$ & The social interactions between pairs of individuals denoted by $e$=$(i,j)$ in the subsequence $\tau'$. \\    
		\hline 
	\end{tabular}
	\label{tab:table0}
\end{table}

\section{Framework for the empirical study of lifecycle events}
\label{sec:tech_prelim}
Following conventions established in previous work~\cite{guo_cascade, burst_time, leskovic_bib}, we will use the symbol $C$ to denote an arbitrary information cascade (i.e. a microblog that  spreads in the social network). Formally, a cascade is represented by a sequence of tuples $(u, v, t) $ such that the microblog was reshared by $v$ from $u$ at time $t$. We denote the sequence of reshare times for $C$ as $\tau_C$ = $\langle 0, \dots, t, \ldots T_C \rangle$ ordered by time, where $T_C$ denotes the time difference between the first and last posting. Here $t \in \tau_C$ denotes the reshare time offset by the starting time for that cascade and $S_C$=$|\tau_C|$ denotes the number of reshares. We will drop $C$ from all notations when they are applicable for all cascades. We will often use the notation $\tau'$ to denote a subsequence of $\tau$.  We slightly abuse the terms \textit{interval} and \textit{subsequences} in this paper - an \textit{interval} is considered here as a generic term for a range of time points not subject to any constraints whereas we define \textit{subsequences} formally later in Section~\ref{sec:net_analysis} as a contiguous subset of $\tau$ and are subject to a set of constraints. We use subsequences for cascade topology analysis and for forecasting purposes.

Rules for identifying time subsequences mapping the \textit{events} described before are generally not well defined in the context of information cascades. In the context of $C$, product adoption in a PLC refers to the resharing process by users and we refer to the subsequences mapping the \textit{growth} and \textit{decline} phases in PLC as the \textit{steep} interval and \textit{inhibition} interval respectively. Our entire work in this paper is centered around the identification and analysis of the \textit{growth} and \textit{decline} phases which we consider as our \textit{events} of interest and the corresponding subsequences mapping these set of events as \textit{event subsequences}. The problem we study in this paper is identification and forecast of such subsequences during which behavior change occurs in the context of cascade adoption using social network measures. In the literature of statistics, such problems fall in the area of change-point detection. 

\begin{figure}[!t]
	\centering
	\hfill
	\minipage{0.5\textwidth}%
	\includegraphics[width=6.5cm, height=3.5cm]{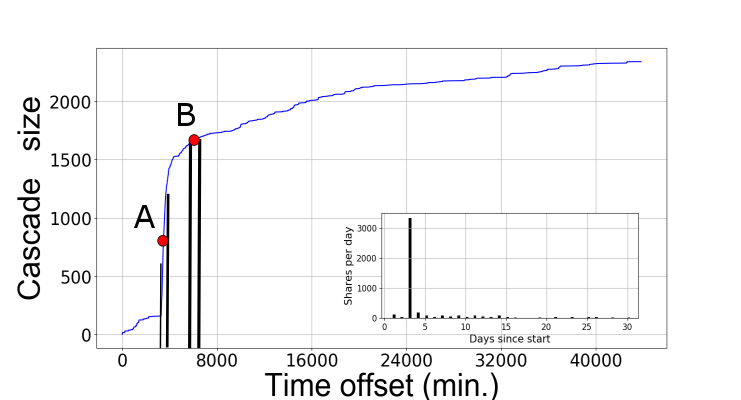}
	\hspace*{3.25cm}\subcaption{}
	\endminipage 
	\hfill
	\minipage{0.5\textwidth}
	\includegraphics[width=6.5cm, height=3.5cm]{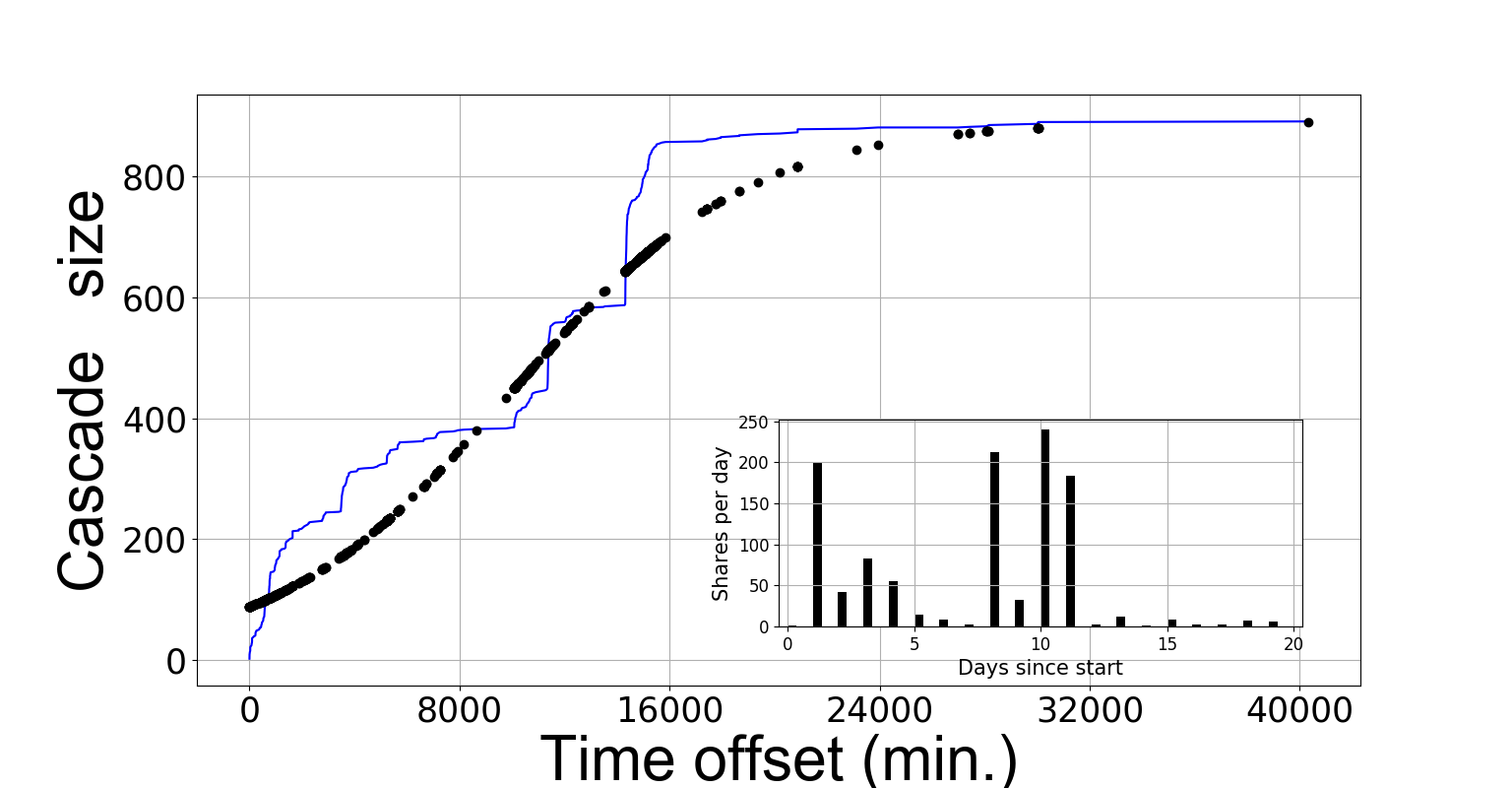}
	\hspace*{3.25cm}\subcaption{}
	\endminipage 
	\hfill
	
	\minipage{0.5\textwidth}%
	\includegraphics[width=6.5cm, height=3.5cm]{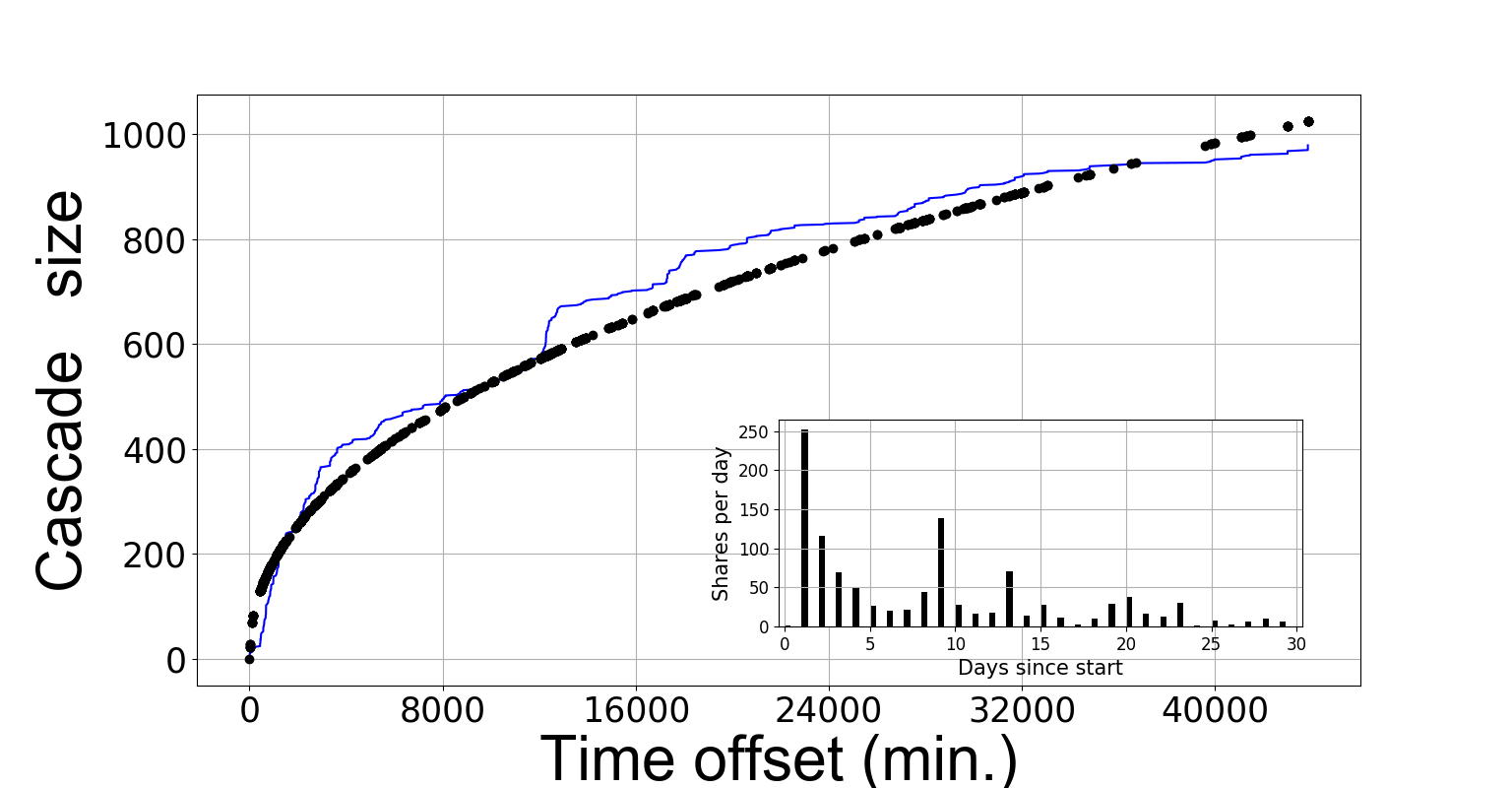}
	\hspace*{3.25cm}\subcaption{}
	\endminipage 
	\hfill
	\minipage{0.5\textwidth}
	\includegraphics[width=6.5cm, height=3.5cm]{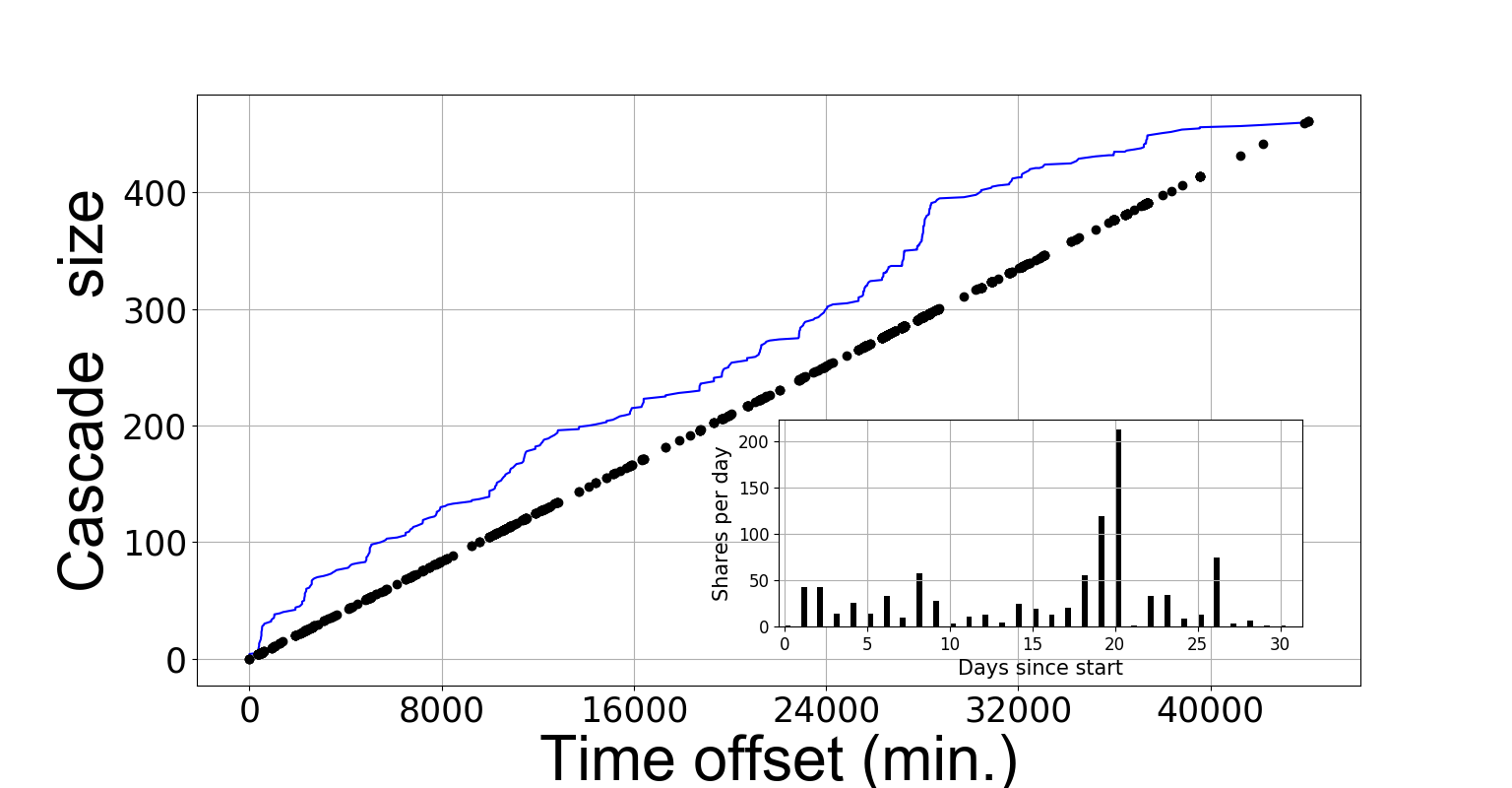}
	\hspace*{3.25cm}\subcaption{}
	\endminipage\hfill
	\caption{(a) \textit{Growth curve} depicting the progress of a Type I cascade. Points A and B marked in red denote $t_{steep}$ and $t_{inib}$ respectively and the subsequences bounding them denote $\tau'_{steep}$ and $\tau'_{inhib}$ respectively. (b) Type II curve fitted to logistic function (c) Type III curve fitted to a concave increasing function (d) Type III curve fitted to a straight line curve. The blue curve denotes the cascade curve while the black dotted curve denotes the estimated curve fit. The subplots inside depicts the number of shares per day over its lifecycle for that example cascade.}
	\label{fig:types_cascades}
\end{figure}

However, in information cascades we do not generally see such smooth transitions in all cascades explained by empirical observations where we find three types of Growth curves shown in the plots in Figure~\ref{fig:types_cascades}, each of which depict cumulative cascade size over time $t$. Apart from Type I cascades as illustrated in Figure~\ref{fig:types_cascades}(a), which depicts an ideal logistic S-shaped growth pattern, most transitions in the cascade lifecycle are not smooth. One such example is given by Type II cascades illustrated by Figure~\ref{fig:types_cascades}(b) which are characterized by multiple temporal patterns of growth, a problem which has been previously studied in the context of time series where convex and concave patterns are used to fit the phases within the lifecycle \cite{youtube, pattern_time}. Since our focus in this paper is in understanding the \textit{steep} and \textit{inhibition} subsequences from a network analysis perspective, we avoid such rigorous pattern fitting mechanisms and instead use point processes to model cascade generation and identify the subsequences mapping the two aforementioned events.

 \subsection{Steep and Inhibition intervals}\label{sec:steep_inhib}
As mentioned above, in Figure~\ref{fig:types_cascades}(a) we show an example plot of \textit{Growth curve} and show three major types we identified empirically:
 
 \begin{enumerate}
 	\item \textit{Type I cascades}: Cascades which follow the logistic function as shown in Figure~\ref{fig:types_cascades}(a).
 	\item \textit{Type II cascades}: These cascades exhibit a step-like pattern of growth shown in Figure~\ref{fig:types_cascades}(b).
 	\item \textit{Type III cascades}: These cascades do not follow the logistic function as shown in Figures~\ref{fig:types_cascades}(c) and (d) and one of the many reasons for these cascade curves is that they probably do not complete their lifecycle within the period of 1 month that we have considered for each cascade.
 \end{enumerate}
 
 The motivation behind defining these three types of \textit{Growth Curves} lies in the way we define \textit{steep} and \textit{inhibition} intervals.
 Intuitively, the \textit{steep interval} is the interval where maximum,``spikey'' diffusion activity occurs characterized by a sharp increase in the adoption over the previous intervals that is also maintained in subsequent intervals.  Likewise, the \textit{inhibition interval} represents a significant, ``spikey'' decline in adoption relative to previous intervals - which is followed by subsequent intervals where adoption continues to decay.
 Furthermore, the inhibition interval is a period whereafter the cascade fails to regain any surge in adoption rate. We aim to tag only a single period as the steep interval and a single period as an inhibition interval in the cascade, therefore we define these three types and only consider Type I.
 As shown in Figures~\ref{fig:types_cascades}(b),(c) and (d), other types of cascade have multiple regions with slackness in growth, making it difficult to tag only one inhibition region.
 In addition, Type II and Type III only make up minority of the whole set of cascades and capture most anomalies due to time scaling issues.
 Therefore, we only consider Type I cascades and assume a logsitic fit to the growth
 curve.
 
 The major challenge in our study is to identify the event subsequences that would allow us to explain the growth dynamics in the vicinity of those subsequences. To overcome the issue caused by the absence of ground truth, we use retrospective analysis on a few selected cascades to infer the parameters of a model based on point processes and maximum likelihood estimation and then calculate the approximate steep and inhibition times of all the cascades in our corpus. We briefly describe the procedure we apply on the selected cascades to develop the model and infer the parameters. Given scaling parameter $\alpha$, we divide $\tau_C$ into sequence of time intervals of uniform interval size $K_C= \ \alpha \log(T_C)$ (refer Table~\ref{tab:table0} for symbols). Amongst these intervals, two are of interest in this work: the \textit{steep interval} and the \textit{inhibition interval}, an example of which is shown in Figure~\ref{fig:types_cascades}(a). 



 \begin{figure*}[]
 	\caption{Flowchart for the steps used in the identification of the \textit{event} intervals - namely the \textit{steep} and the \textit{inhibition} intervals.}
	\centering
	\hfill
	\begin{minipage}{1.0\textwidth}
	\includegraphics[width=12.1cm, height=6cm]{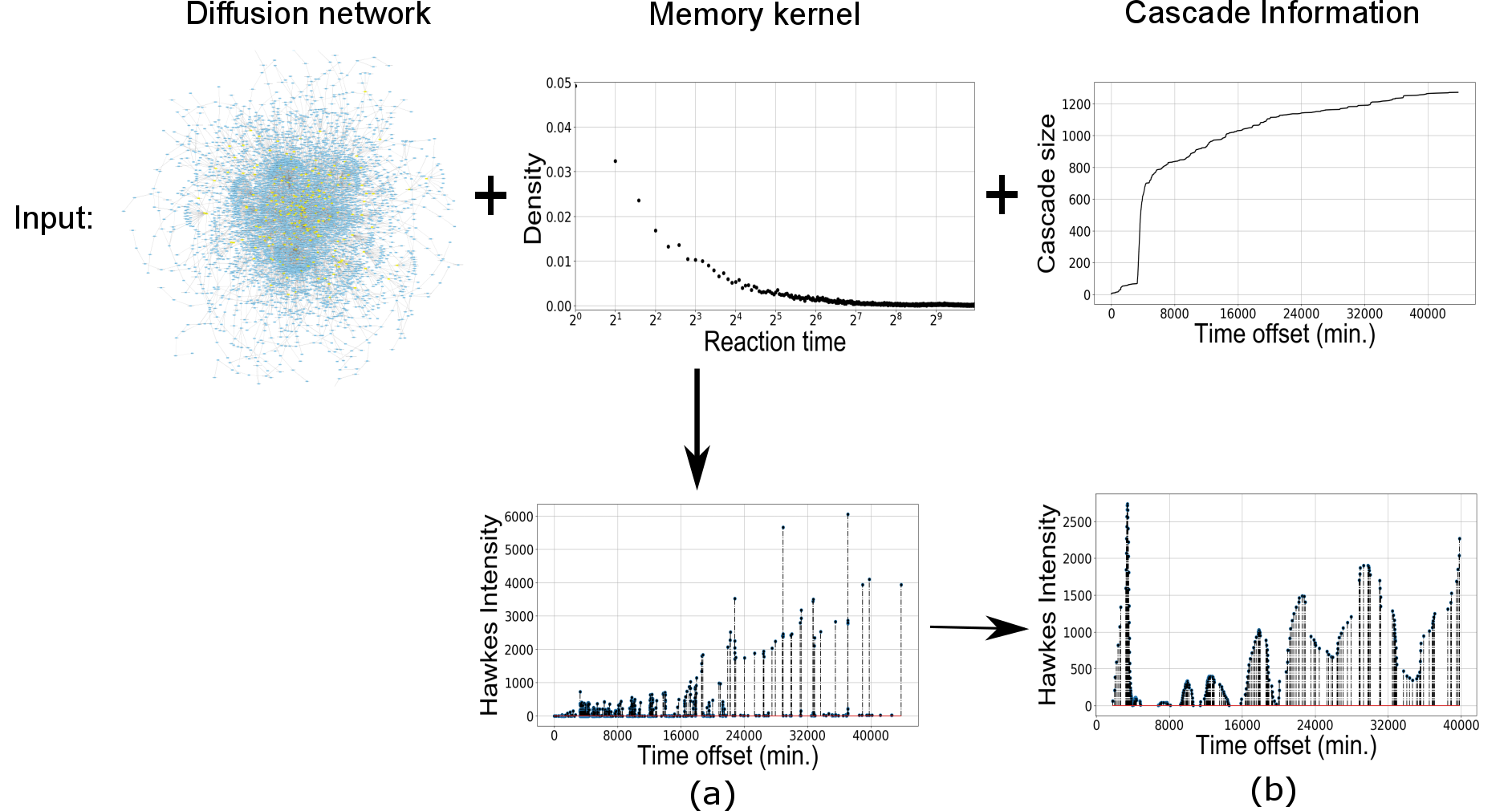}
	\subcaption{\textbf{Step 1}: We obtain the Hawkes intensity curve (fig. (a)) from the input as shown using the concept of point processes. We then compute the Hawkes interval curve $HI$ (fig. (b)) by splitting the Hawkes intensity curve (fig. (b)) into intervals of size $K_C$ and summing the intensities inside each interval.}
	\end{minipage}
	\\
	
	\begin{minipage}{1.0\textwidth}
		\includegraphics[width=12.1cm, height=3cm]{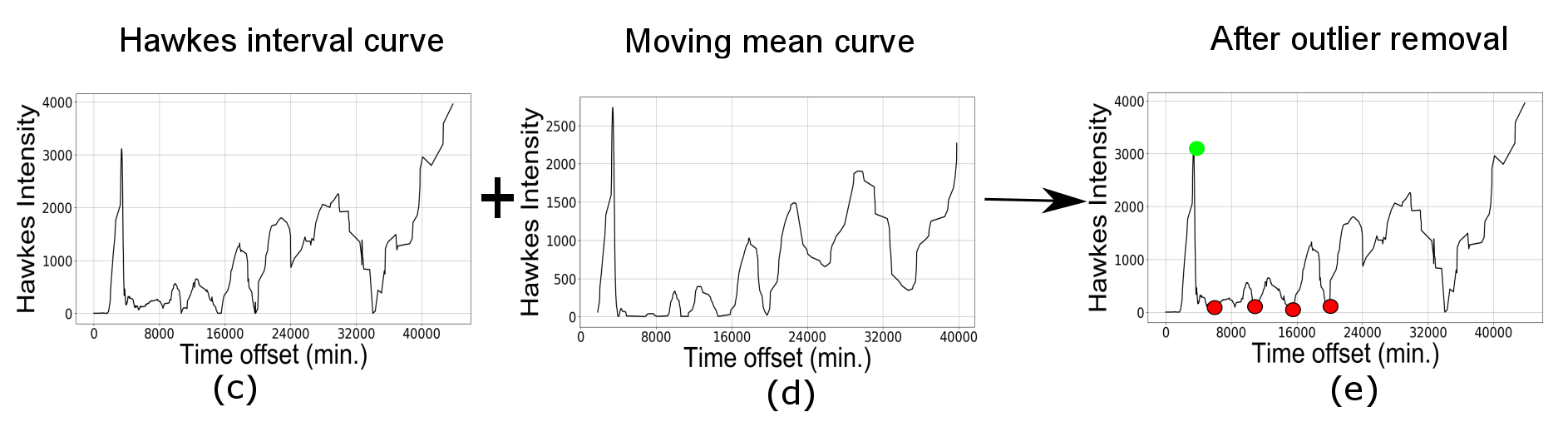}
		\subcaption{\textbf{Step 2}: We set  $t_{steep} $ (marked in green point in fig. (e)) by selecting the first local maxima in $HI$ (fig. (c)) and obtain a set of possible inhibition interval points (marked in red points) by selecting all the local minima in $HI$. Using the moving mean curve (fig. (d)) we filter out outliers. At the end of this step, we obtain $t_{steep} $ and a set $I_C$ of $t_{inhib} $ points shown in fig. (e).}
	\end{minipage}
	\\
	
	\begin{minipage}{1.0\textwidth}
	\includegraphics[width=12.1cm, height=3cm]{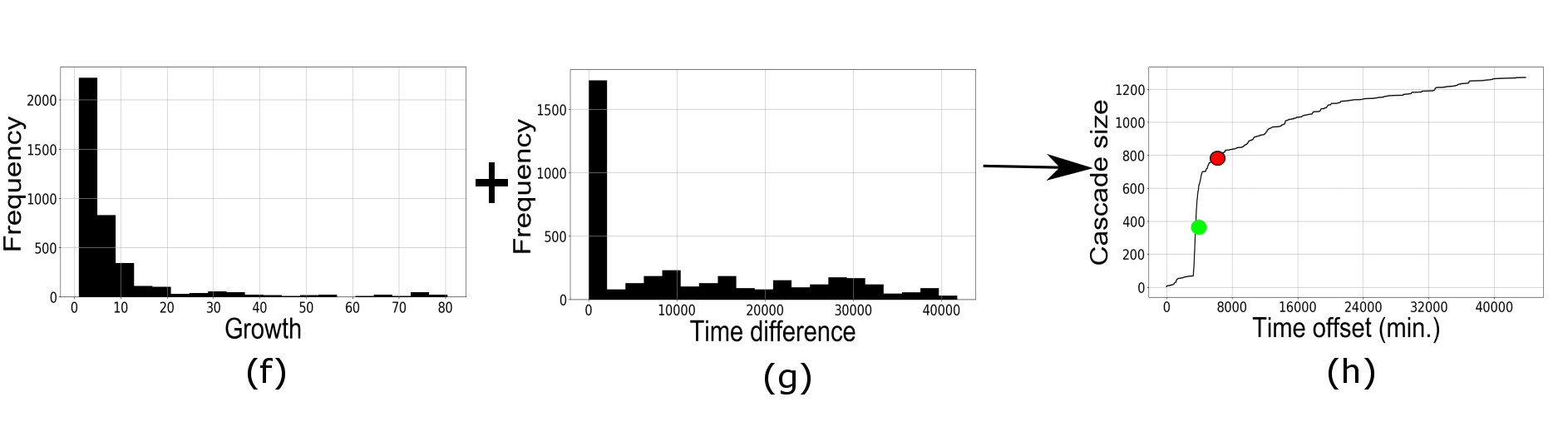}
	\subcaption{\textbf{Step 3}: We consider selected cascades in our corpus $C_s$ and obtain the \textit{Growth ratio} histogram (fig. (f)), where $Growth[t_{inhib}] = \frac{S^{t_{inhib}}}{S^{t_{steep}}} $ and the \textit{Time difference} histogram (fig. (g)), where $TG[t_{inhib}]$ = $t_{inhib} - t_{steep} $, $\forall$ $t_{inhib} \in I_C $, $\forall C \in C_s$. We use maximum likelihood estimation on the probability distribution of these two attributes to obtain the estimated thresholds of \textit{Growth} and \textit{Time difference}. We set the first time point $t$ in the lifecycle of cascade $C$, whose $Growth[t]$ and $TG[t]$ cross the estimated thresholds, as the final $t_{inhib}$ for $C$ shown in fig. (h).}
	\end{minipage}
	\hfill
	
	\label{fig:steps_hawkes}
\end{figure*}

 To formalize these ideas, the simplest notion would have been to find the slope or the first derivative of the cumulative cascade size with respect $t$ - but we found this method to have some significant drawbacks:
 
 \begin{enumerate}
 	\item It is difficult to define a threshold for the slope values at the start of the inhibition phase as the rate of adoption in that region for each cascade varies significantly.
 	\item There will be multiple regions in the same cascade with nearly equal slopes - though most intervals do not fall into our described category of \textit{inhibition interval}. 
 \end{enumerate}
 
 Put together, the first order derivative approach does not incorporate sufficient information about the time taken by users to adopt the cascade and only takes into account the cumulative size of cascade at each reshare time point which is insufficient to identify the event intervals.  We identify these intervals in a three-step process, which we provide technical details for in the Appendix A1. This process, illustrated in Figure~\ref{fig:steps_hawkes} is described intuitively below:
 \begin{enumerate}
 	\item Based on recent findings that relate point processes to network diffusion modeling(i.e. \cite{seismic}\cite{hawkes_zha}, we calculate the Hawkes intensity at each reshare time point $t$ as a function of the number of past interactions of the participating users for the current reshare, and the distribution of times taken by the users to adopt the cascade $C$. We convert this curve into \textit{Hawkes interval curve} shown in Figure~\ref{fig:steps_hawkes}(c) by  summing the intensities of time points in each interval.
 	\item\label{step2} We then identify intervals with local maxima (which are candidates for the steep interval) and local minima (which are candidates for the inhibition interval).
 	\item Based on ideas from \cite{hawkes_zha}, we then use a maximum-likelihood approach to filter the points in step~\ref{step2} shown in Step 3 of Figure~\ref{fig:steps_hawkes} and obtain the parameters that we use to infer the \textit{steep} and \textit{inhibition} times of the new cascades.
 \end{enumerate}
 
 \noindent Once we infer the parameters, we follow the above three steps for identifying the \textit{event times} (which we represent by the mean of the respective event intervals) of the rest of the cascades in the corpus except that we avoid the maximum likelihood step and directly use the inferred parameters  to compute the event times using a threshold technique. At the end of this procedure, we identify the time points $t_{steep}$ and $t_{inhib}$, identifying the approximate $growth$ and $decline$ phases respectively. We refer to either of these time points as $t_e$ when we generalize the operations for an event, where $e$ denotes the event of interest namely, the \textit{growth} and the \textit{decline} phases. Also $t_{steep}$ and $t_{inhib}$ refer to two time points with indices in the range spanned by $\tau$ but we use them to strictly point to two specific times in $\tau$. 
 
 \begin{figure}[t!]
 	\centering
 	\includegraphics[width=12.5cm, height=3cm]{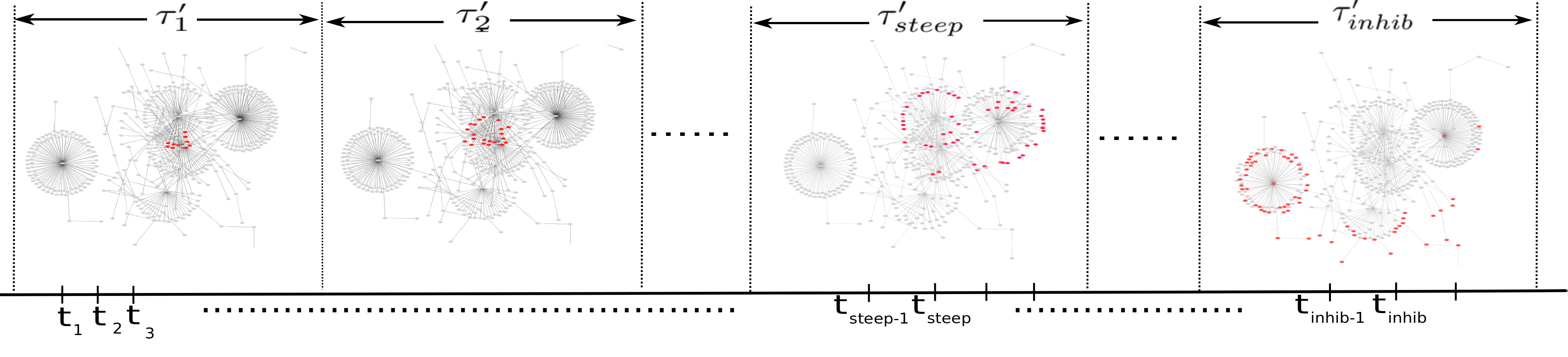}
 	\caption{Example showing how the network is partitioned over subsequences for network analysis. Observe that $\tau'_{steep}$ and $\tau'_{inhib}$contain $t_{steep}$ and $t_{inhib}$ respectively. Each $\tau'$ contains equal number of nodes for analysis and as the cascade progresses more nodes get activated marked in red. Although each $\tau'$ looks uniform in the time span, we note that each $\tau'$ may differ in the time range depending on how much time it took for $|V^{\tau'}|$ nodes to form the network.}
 	\label{fig:net_evolve}
 \end{figure}
 
 \subsection{Network analysis model}
 \label{sec:net_analysis}
 We represent a social network as a directed graph $G=(V,E)$ where $V$ is a population of individuals and the edge $(i,j)\in E$ refers to individual $i$ having the ability to influence individual $j$. These influence relationships are known a-priori and as a practical matter, we determine these relationships from previously observed microblog relationships that occur prior to the cascades analyzed in this paper. In our work, we denote this social network information by an undirected network $G_D$=$(V_D$, $E_D)$ where $V_D$ denotes the individuals involved in the historical diffusion process and an edge $e \in E_D$ denotes that information has been shared between a pair of individuals ignoring the direction of propagation. This is similar to our previous work~\cite{guo_cascade} and will be described in detail in Section \ref{sec:dataset}.  
 
 We denote the network produced by the participants of a cascade $C$ by $G^{\tau}_C$ = $(V^{\tau}_C, E^{\tau}_C)$ where $V^{\tau}_C$ denotes all the individuals who participated in the diffusion spread of $C$ in its entire lifecycle spanned by $\tau$ and an edge in $e=(u, v) \in$ $E^{\tau}_C$ denotes that either $v$ reshared $C$ from $u$ or the interaction happened in the past denoted by the presence of $e$ in $E_D$, that is to say we add the influence of the propagation links from our historical social network in the diffusion mechanism of the cascades. As mentioned before, we drop the subscript $C$ from the notations when they are applicable for all cascades.
 
In context of the cascade network $G^{\tau}$,  we denote $G^{\tau'}$ = $(V^{\tau'}, E^{\tau'})$ as the subgraph of $G^{\tau}$, where $V^{\tau'}$ denotes the individuals who reshared a cascade $C$ in the time subsequence $\tau'$ and  $E^{\tau'}$ denotes the set of edges $e=(u, v)$ where the resharing from $u$ to $v$ happened in the time subsequence $\tau'$ or there was an interaction in the historical diffusion period indicated by the presence of $e$ in $E_D$. In our work, we create the sequence of subsequences $\tau$ = $\langle \tau'_1, \ldots, \tau'_{\mathcal{Q}} \rangle$, ordered by the starting time of each subsequence, where we denote $ \mathcal{Q}$ to be the number of subsequences for $C$, which would vary for different cascades. For the subsequences, the following conditions hold: (1) $|V^{\tau'_i}| = |V^{\tau'_j}|$ and (2) $E^{\tau'_i} \cap E^{\tau'_{j}}$ = $\emptyset$, $\forall i \neq j, \in [0, \mathcal{Q}]$. We note that $|\tau'_i|$ $\neq |\tau'_j|$ for any $i \neq j$, that is to say the time range spanned by the subsequences in itself may differ depending on the time taken by $G^{\tau'}$ to form the network.  
Similar to previous conventions, without loss of generality we will drop the subscripts and use $\tau'$ to refer to a subsequence but the operations on which are applied to all subsequences for all cascades. In our work, we keep $|V^{\tau'}|$ fixed for every cascade in our corpus. The advantage of selecting this subsequence node set size a-priori is that we can avoid retrospective analysis and be agnostic about the final cascade size $|V^{\tau}|$. Since we do not fix $\mathcal{Q}$ and let it vary based on $|V^{\tau}|$ and $|V^{\tau'}|$ for each cascade, we do not fix a specific interval for the subsequences containing the event time points for each cascade. 

A \textit{temporal representation} of a cascade is denoted by a sequence of overlapping subsequences $\mathcal{N}$ = $\langle N_1, \ldots, N_{\mathcal{S}} \rangle$ such that the following conditions hold: $V^{N_i}$ = $V^{\tau'_{i-1}}$ $\cup$ $V^{\tau'_i}$, and $E^{N_i}$ = $E^{\tau'_{i-1}}$ $\cup$ $E^{\tau'_i}$ $\forall$  $i \in [1, \mathcal{Q}]$. We perform social network analysis on the subsequences $N $ where we drop the index subscript when we generalize the analysis for all subsequences for all cascades. Such a temporal representation $\mathcal{N}$ helps us in avoiding disjoint subsequences for network analysis and replicates a sliding window approach.

We denote the subsequences containing $t_{steep}$ and $t_{inhib}$ as $\tau'_{steep}$, $\tau'_{inhib}$ respectively and the first network in $\mathcal{N}$ containing these subsequences as $N_{steep}$ and $N_{inhib}$ respectively. We note that $\tau'_{inhib}$ is not necessarily the last subsequence in the cascade, as there may be few more reshares before the cascade finally dies down but since we are interested in the subsequences before $t_{inhib}$, we discard the rest of the subsequences after $\tau'_{inhib}$ from our analysis.

Figure~\ref{fig:net_evolve} gives a visual depiction of the method of analysis performed on the cascades using the subsequences. We perform our network analysis on each $N$ in sequence of formation, under the representation described above until we reach the \textit{event subsequences} as shown in the figure. We note that the presence of historical diffusion interactions denoted by the edges of $E_D$ in $N$ introduces cycles in the structure of $N$ which otherwise would exhibit a tree structure inherent to the property of cascades.

\subsection{Network features}
\label{sec:feat_def}
For each temporal network $N$ in a cascade $C$, we compute several network features that act as indicators of the event subsequences in the cascade lifecycle and observe how these values change for the networks in $\mathcal{N} $ as the cascade progresses.

Formally for a given $G^N$ = $(V^N, E^N)$, a network feature $f$:$v \rightarrow$ $\mathbb{R}^+$ assigns a non-negative value to every node $v \in$ $ V^N$ such that the values are indicative of the role of nodes in the spread of information during the interval spanned by $N$. This node value assignment depends on the underlying structure of $N$ in terms of the edge connections between the nodes. We describe these features in detail in Section~\ref{sec:net_feat}.

\subsection{Statistical testing for feature significance}\label{sec:stat_test}
Like the one used in Bass model \cite{bass}, traditional approaches to quantify the importance
of the networked structure of social networks involve: 1) using network features (such
as the user friendship network characteristics); 2) forming regression models; 3)
estimating parameters of the regression model to reject or accept hypothesis using
statistical significance measures (as done in \cite{youtube_susarla}).
The aim of such models is to infer the linear monotonicity relation between the response and the predictor variables without explicitly incorporating the temporal variation of the measures as predictors. Such models do not implicitly characterize whether such network features would temporally be good early indicators of some phenomena like virality or inhibition in the cascading process. To resolve this issue, causality in time series data has been recently put to practice to quantify the cause and effect over time \cite{kleinberg}. However such parametric methods of logic based causality are computationally expensive. Granger causality \cite{granger_causality} has been widely used as a parametric model to measure cause and effect in time series social network data \cite{granger_inf}. We first introduce the concept of Granger causality as a tool for quantifying cause and effect and demonstrate its use for our network features. We use the causality framework to quantify the impact of node-centric features on the reshare time responses in a cascade. We use this causality framework to assess how central nodes affect the response time in the cascade, especially as it approaches inhibition.

\subsubsection{Granger Causality}
Assuming two jointly distributed vector valued stochastic variables \textbf{\textit{X}} and \textbf{\textit{Y}}, we say that \textbf{\textit{Y}} does not Granger-cause \textbf{\textit{X}} if and only if given its own past, \textbf{\textit{X}} is independent of the past of \textbf{\textit{Y}}. Formally a $p^{th}$ order vector autoregressive model (VAR) is represented by the following equation:

\begin{equation}
\mathbf{U}_t = \sum_{k=1}^{p} \mathbf{A}_k \mathbf{U}_{t-k} + \mathbf{\epsilon}_t
\end{equation} 

where $\mathbf{U}=\{\mathbf{u_1, u_2, \ldots, u_m}\}$ represents a multi-variate time series and for time $t$, $\mathbf{u}_t$ is a real valued $n$-dimensional (column) vector with elements $u_{1t}, u_{2t}, \ldots u_{nt}$. The $n$ $\times$ $n$ real-valued matrices $A_k$ are the regression coefficients and the $n$-dimensional stochastic process $\epsilon_t$ denote the residuals, which are independently and identically distributed (i.i.d.) and serially uncorrelated.
Using this notation, in the time-domain, Granger causality is motivated by the following: suppose that $\mathbf{U}_t$ is split into two interdependent processes:

\begin{equation}
\mathbf{U}_t = 
\begin{bmatrix}
\mathbf{X}_t        \\
\mathbf{Y}_t
\end{bmatrix}
\end{equation}

Under a predictive interpretation, Granger Causality from $\mathbf{Y}$ to $\mathbf{X}$ quantifies the ``degree'' to which the past of $\mathbf{Y}$ helps predict $\mathbf{X}$ in addition to the degree by which $\mathbf{X}$ is already predicted by its own past.
These comprise two regression models to test for significance of causal effect of $\mathbf{Y}$ on $\mathbf{X}$. 

\begin{equation}
\mathbf{X}_t = \sum_{k=1}^p \mathbf{A}_{xx,k} \mathbf{X}_{t-k} + \sum_{k=1}^p \mathbf{A}_{xy,k} \mathbf{Y}_{t-k} + \mathbf{\epsilon}_{x,t}
\label{eq:full_reg}
\end{equation}

and 

\begin{equation}
\mathbf{X}_t = \sum_{k=1}^p \mathbf{A^{'}}_{xx,k} \mathbf{X}_{t-k} + \mathbf{\epsilon^{'}}_{x,t}
\label{eq:half_reg}
\end{equation}

Then we apply Wald F-test to obtain a $p$-value for identifying whether or not Equation~(\ref{eq:full_reg}) results in a better regression model than Equation~(\ref{eq:half_reg}) with statistically significant better results.

 \begin{figure}[t!]
	\centering
	\includegraphics[width=9.5cm, height=4cm]{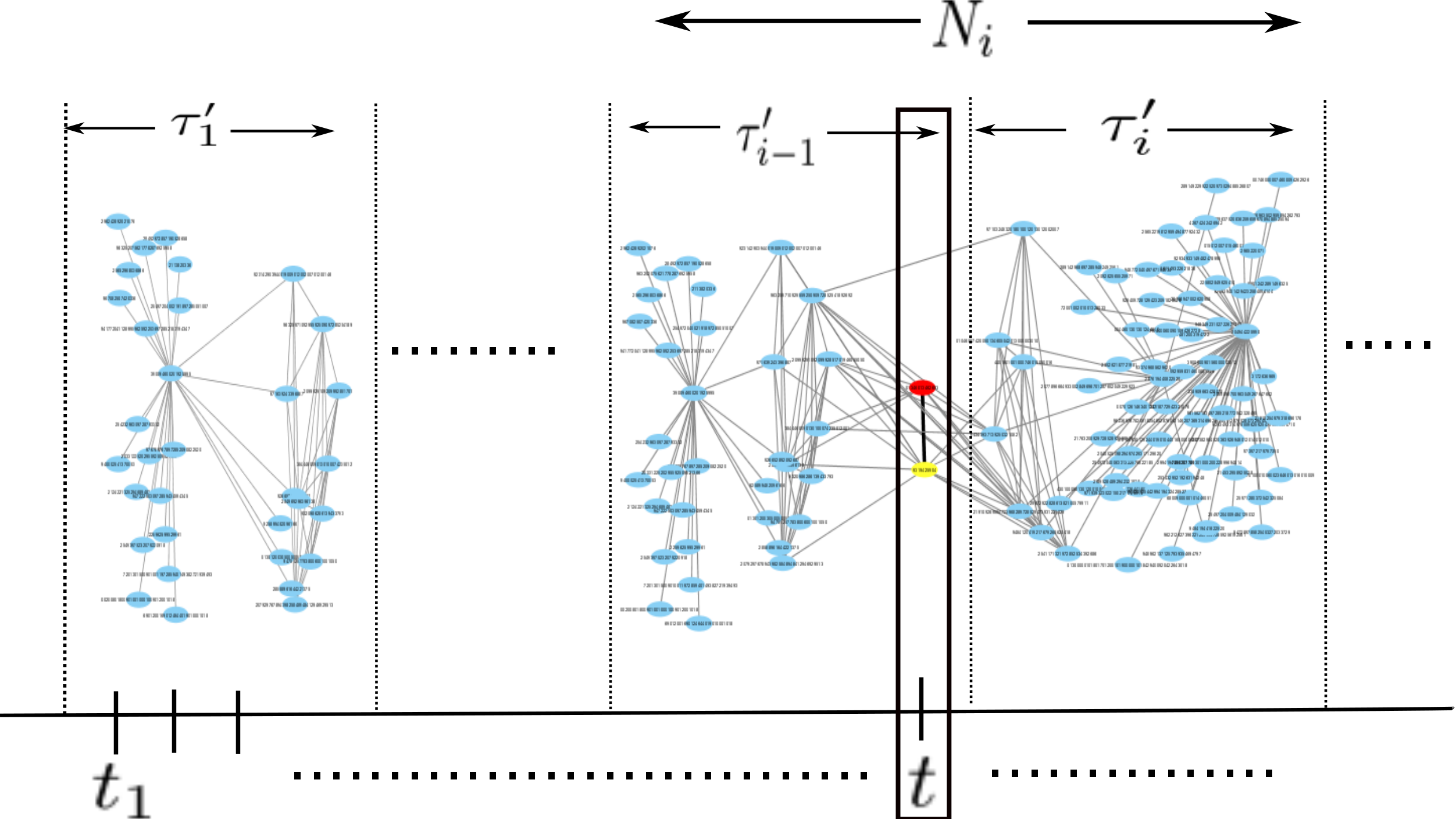}
	\caption{Example showing how the time series $\mathcal{T}_f$ for a feature $f$ is obtained. For one of the iterations considering network $N_i$ $\in \mathcal{N}$, $i$ $\in [0, \mathcal{Q}]$, for a cascade $C$,  we consider the reshares in $\tau'_{i-1}$. Let one of the reshare times be $t$ $\in \tau'_{i-1}$ and let the yellow node denote the resharer  and the red node denote the rehshared node for the reshare at $t$. We set $\mathcal{T}_f$[t] to the feature value of the red node computed using $G^{N_i}$. We repeat this operation for this iteration for all the reshares spanned in $\tau'_{i-1}$ only. }
	\label{fig:ts_demo}
\end{figure}

\subsubsection{Obtaining the feature time series}
In this paper, we consider each node centrality feature separately to assess the cause and effect significance. In this section, we describe how we model each network feature as a series in the time domain to be able to use the causality framework. A univariate time series is modeled as a function $ \mathcal{T}$:$t \rightarrow$ $x $ where $x$ denotes the value of the series at time $t$ and in our case $t$ is represented in the discrete time domain. 

We model a social network feature time series for each cascade $C$ as follows: for each reshare that occurs between two nodes \textit{source} and the \textit{target}, at time $t$, we create a mapping $\mathcal{T}_f$ of the feature value of $source$ denoted by $x$ to time point $t$, where the features are computed using the cascade network till time $t$. Precisely for each network feature $f$ described in Section~\ref{sec:net_feat}, we have a function $\textit{T}_f$:$t$ $\rightarrow$ $x$. But since there can be a lot of reshares in a cascade and it would be infeasible to construct the cumulative cascade network for every new reshare and then recompute the centralities to obtain $x$ for every $t$, we resort to using a different approach taking advantage of the temporal representation $\mathcal{N}$. 

Instead of computing the node centralities of the $source$ node after every reshare, we compute the node features after each temporal network $N$ is formed. Similar to the network analysis performed on $N$ described in Section~\ref{sec:net_analysis}, we compute the node centrality values using the network induced by $N$. Since the temporal networks $N$ are overlapping in the temporal representation $\mathcal{N}$, to avoid considering reshares twice in the time series model, we only consider the reshares within the time spanned by the first subsequence in each $N$, since each $N$ consists of two consecutive subsequences. Note that although we consider reshares within the first subsequence for each $N$ when considering the iterations over the sequence of networks in $\mathcal{N}$, the feature values are computed in the context of each network $G^N$. For each reshare at $t$, we set $\mathcal{T}_f$[t] as the node centrality value for the \textit{source} node for that reshare. For reshares which occur at the same time $t$ we take the mean of the feature values of the source nodes for the reshares at $t$. Figure~\ref{fig:ts_demo} gives an example of this mapping between the node feature values for $f$ computed using networks in $\mathcal{N}$ to the time series $\mathcal{T}_f$. 

Using the above procedure, we form the time series $\mathcal{T}_f$ for feature $f$ for each cascade among our corpus which form the set of causal variables we would be using for testing causality. We measure the effect of the causal features on the response variable $\mathcal{R}$:$t$ $\rightarrow$ $\Delta t$ where $\Delta t$ is the time difference between the current and its previous reshare. For each of the events and features, we form separate time series $\mathcal{T}_{f, e}$ and $\mathcal{R}_e$ where $e$ denotes the event, by considering the time points from start of the cascade till $t_e$. We drop the $e$ from the subscripts when we explain the operations for a generic time series.

\subsubsection{Measuring causality}
To measure Granger causality between the series $\mathcal{T}_f$ and $\mathcal{R}$, we first fit the series data of each cascade separately to a Vector Autoregression model(VAR) with lag order $p$ between $[0, P]$ where $P$ denotes the maximum lag for a model. We then choose the order $p$ of the VAR model based on the $AIC$ (Akaike Information criterion) measure which helps us keep the lag order dynamic for different time series. The following equation gives the two hypothesis for measuring the causality between feature $\mathcal{T}_f$ and $\mathcal{R}$ over a cascade $C$:

\begin{equation}
\mathbf{\mathcal{R}}[t] = \sum_{k=1}^{p} \mathbf{a}_{k} \mathbf{\mathcal{R}}[t-k] + \mathbf{\epsilon}_{t}
\label{eq:granger_null}
\end{equation}
and 
\begin{equation}
\mathbf{\mathcal{R}}[t] = \sum_{k=1}^{p} \mathbf{a}_{k} \mathbf{\mathcal{R}}[t-k] + \sum_{k=1}^{p} \mathbf{b}_{k} \mathbf{\mathcal{T}}_{f} [t-k] + \mathbf{\epsilon}_{t}
\label{eq:granger_full}
\end{equation}

where Equation~(\ref{eq:granger_null}) represents the null hypothesis and Equation~(\ref{eq:granger_full}) represents the alternate full hypothesis.

Then we use the Wald F-test to  test the hypothesis where the coefficients of the first $p$ lagged values of $\mathcal{R}$ are zero in Equation~(\ref{eq:granger_full}) that is we follow Equation~(\ref{eq:granger_null}). We note that we do not perform the causality test of $\mathcal{R}$ on $\mathcal{T}_f$ as we are not concerned about that direction of causality. The rejection of the null hypothesis implies a rejection of Granger \textit{non-causality} that is to say, it supports the presence of Granger causality.

\subsubsection{Forecasting Events}
\label{sec:forecast}
In this paper, in addition to assessing the impact of the node features on the reshares using causality, we use them to predict the time when the cascade would reach the event times $t_e$. Specifically, given time difference series $\mathcal{R}_e$, we try to forecast the last value in $\mathcal{R}_e$ for the respective event $e$ using a VAR model. This is equivalent to forecasting $t_e$ since the last point in $\mathcal{R}_e$ denotes the time difference between $t_{e}$ and its previous point.

Implicitly, the significance values in Granger causality testing over a time series are time-invariant, that is to say it does not detect whether the level of significance of the causality of a particular time point or a time subsequence is higher or lower compared to other time points or interval in the same time series. Although a particular feature may be more pronounced as a causal variable in terms of its significance, we cannot infer whether it is a good indicator of $t_e$ for a particular event $e$. Also Granger causality tests from causal variable $X$ to the effect $Y$ provides incremental benefits on forecasting $Y$ combined with the additional history of $X$ instead of just using the history of $Y$. So it does not test the benefit the feature $X$ alone in forecasting the response time of the users. To achieve that, we use the following two models to forecast $t_e$ or specifically the time difference between $t_e$ and its previous reshare time, either of which can be calculated from the other:

\begin{enumerate}
	\item \textit{Model 1}: Here we use an autoregression model with the node measures $\mathcal{T}_f$ as the input features given by:
	\begin{equation}
	\mathbf{\mathcal{R}}[t]= \sum_{k=1}^{p} \mathbf{a}_{k} \mathbf{\mathcal{T}}_{f}[t-k] + \mathbf{\epsilon}_{t}
	\label{eq:model_1}
	\end{equation}
	where the symbols hold the same meaning as defined in Equation~\ref{eq:granger_full}.
	\item \textit{Model 2}: For this we simply use the full Granger model defined in Equation~\ref{eq:granger_full}, that is we use the combined effect of the node measures and the past history of the reshare response time series itself.
	
\end{enumerate}
We explain the split of the training and testing part for each time series in details in the Results section in ~\ref{sec:for_exp}.

\begin{table}[!h]
	\centering
	\renewcommand{\arraystretch}{1}
	\caption{Properties of Reposting Network and Cascades}
	\begin{tabular}{|p{5cm}|p{4cm}|}
		\hline 
		{\bf Properties} & {\bf Reposting Network}\\ 
		\hline\hline
		Vertices           & 6,470,135 \\
		\hline
		Edges & 58,308,645 \\
		\hline 
		Average Degree & 18.02    \\       
		\hline \hline 
		Number of cascades & 7,479,088 \\
		\hline
		Number of cascades over 300 & 7407\\
		\hline
	\end{tabular}
	\label{tab:table2}
\end{table}

\section{Data description and Experiment method}
\label{sec:dataset}
\begin{figure}[!t]
	\centering
	\hfill
	\begin{minipage}{0.48\textwidth}%
		\includegraphics[width=6cm, height=4cm]{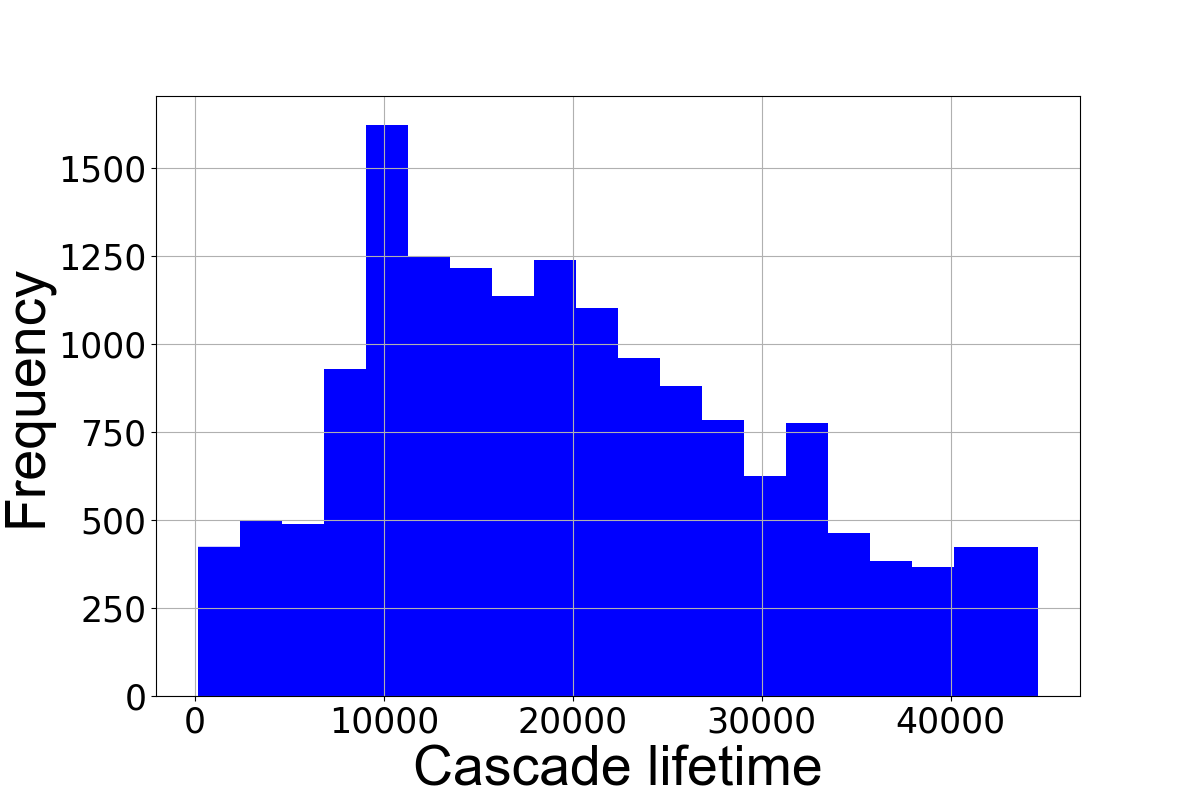}
		\subcaption{}
	\end{minipage}
	\hfill
	\begin{minipage}{0.5\textwidth}
		\includegraphics[width=6cm, height=4cm]{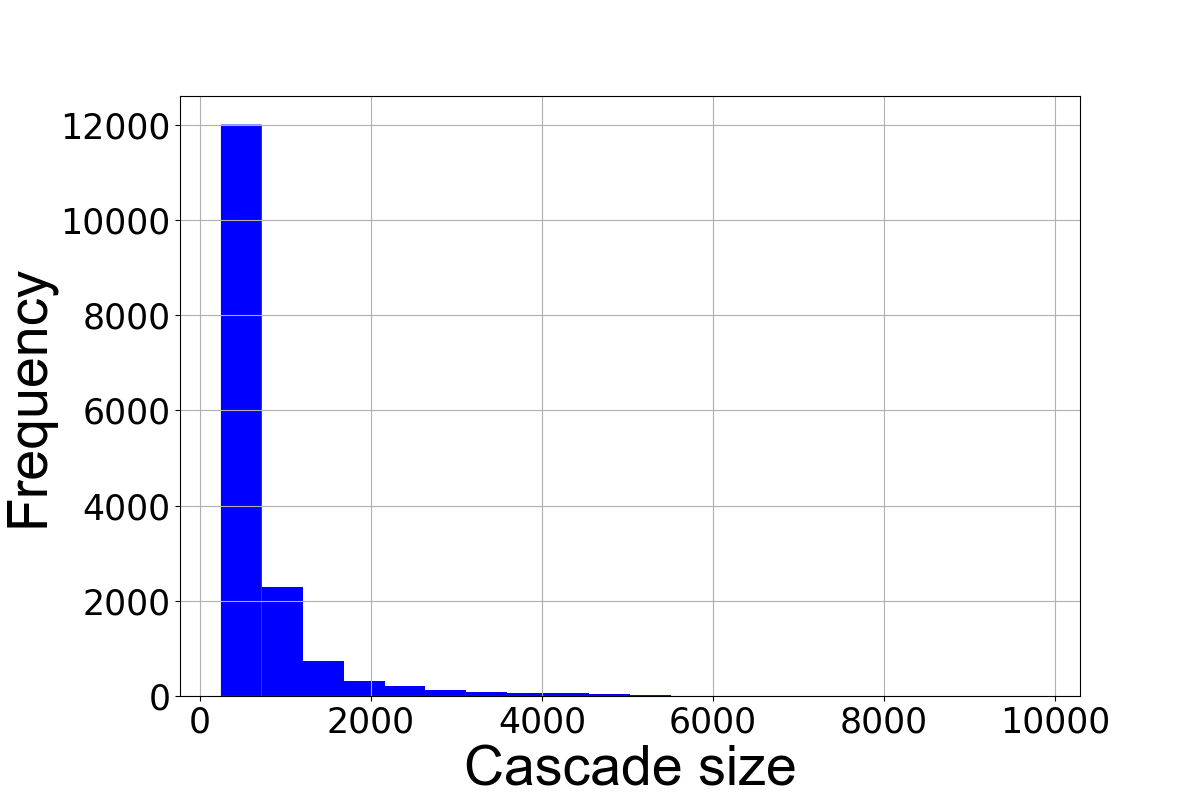}
		\subcaption{}
	\end{minipage}
	\hfill
	\caption{Histogram of (a) Cascade lifetimes in minutes (b) Cascade size.}
	\label{fig:hist_net}
\end{figure}

For building the diffusion network, we use the dataset provided by WISE 2012 Challenge\footnote{http://www.wise2012.cs.ucy.ac.cy/challenge.html} as has been previously used  in \cite{guo_cascade}. The dataset provides us with user data and the reposting information of each microblog along with the reposting times which enables us to form the cascades for each microblog separately. The diffusion network mentioned in Section~\ref{sec:net_analysis} $G_D$=$(V_D$ , $E_D)$ is created by linking any two users who are involved in a microblog reposting action within the period May 1, 2011 and August 31, 2011.  Similar to most social networks, this network also exhibits a power law distribution of degree \cite{guo_cascade}. Table~\ref{tab:table2} shows the statistics of the diffusion network and the corpus of cascades used in our experimental study. From the corpus of cascades which spanned between June 1, 2011 and August 31, 2011, we only work with cascades with more than 300 nodes. Since we are considering subsequences preceding $\tau'_{steep}  $ and $\tau'_{inhib}$ for our analysis, we discard cascades of smaller sizes in our experiments. Figures~\ref{fig:hist_net}(a) and (b) show the histograms for the cascade lifetimes measured by $T_C$ and the cascade sizes measured by $S_C$. As seen in Figure~\ref{fig:hist_net}, although the lifetimes follow a Gaussian distribution, most cascades survive for less than 600 reshares having a skewed distribution. 

Amongst the corpus of cascades, the number of Type I cascades is 5924 while the total number of Type II and Type III cascades is 1483. The total number of cascades of Type I are roughly around 80 \% of the total number of cascades that are more than size 300.
For dividing the cascade curve $C$ into intervals of size $K_C = \alpha \log(T_C)$ as described in Section~\ref{sec:steep_inhib}, we set the scaling factor $\alpha $ to 5 which we found suitable through experimental evaluations, in order to obtain window sizes which are optimal. For finding the suitable $\alpha$ and the parameters from the maximum likelihood method, we carefully choose 1000 cascades from among the entire corpus which is roughly around 13\% of the corpus, on which the evaluation is done. The sensitivity of $\alpha$ as a parameter for the optimal window sizes has been studied in detail in Appendix Section A2. We separate Type I cascades from Type II and Type III cascades by setting a threshold of time ($t_{th})$. We mark those cascades with $t_{steep} \leq t_{th}$ as Type I. We set $t_{th} $ = 5000 minutes, as we found that majority of cascades following the Type I pattern in Figure~\ref{fig:types_cascades}(a) experience the steep phase before our selected threshold time. The reason for using this threshold to label Type I cascades instead of using more complex curve fitting methods is two-fold: firstly, since the shape of curves even within Type I cascades vary based on when $t_{steep} $ occurs, it is difficult to manually select a set of Type I cascades to estimate parameters by MLE for a logistic model that are representative for all Type I cascades and secondly we observed that the Type I cascades are mainly characterized by situations where $t_{steep} $ occurs within a very short time after the cascade starts and cascades where $t_{steep}$ occurs after a certain amount of time do not exhibit the Type I pattern. As shown in Figure~\ref{fig:hist_net}(a), the median for the lifetimes occur at around 10000 minutes and hence $t_{th}$ = 5000 means that the steep growth happens within the first half for majority of the cascades.

For our social network analysis method described in Section~\ref{sec:net_analysis}, we fix $|V^{\tau'}|$ to 40 for all the cascades. Following this, $|V^{N}|$ $\leq$ 80 although $|E^{N}|$ would vary for each $N$.  For evaluation of the intervals leading to $\tau'_{steep}$ and $\tau'_{inhib}$, we consider the last 20 networks $N$ preceding $N_{steep}$ and also for $N_{inhib}$, since as mentioned before $\tau'_{steep}$ and $\tau'_{inhib}$ would vary for each cascade and therefore it is not possible to select any particular subsequence for analysis. This is also to ensure that we do not miss out on any time subsequences that may be early signs of an approaching $steep $ or an $inhibition$ interval. We obtain two sets of plots for the regions preceding $\tau'_{steep}$ and $\tau'_{inhib}$ - this comparative analysis of the two phases helps us contrast the structural properties in the network during those phases and is instrumental in making some concrete conclusions about the inhibition time phase.

\section{Network measures for events}
\label{sec:net_feat}
For each temporal network $N$ in a cascade $C$, we compute the features described in this section and observe how these values change for the networks in $\mathcal{N} $ as the cascade progresses.
Centrality of nodes or identification of `\textit{key}' nodes in spread of information has been an important area of research in social network analysis \cite{cent_nodes}. But in majority of these analyses, measurement of importance through some network statistics are performed on static networks evolving on a cumulative basis. 
We consider individual node features over time and observe how they change over the lifecycle of the cascade that is to say, whether emergence of crucial nodes with high or low feature values in the middle of the cascade maximizes the reshares rate or the absence of such nodes augments the rate of cascade decay. The setup for the analysis of evolution of the cascade networks in this paper mentioned in Section~\ref{sec:net_analysis} allows us to use the already existing centrality measures for understanding the diffusion mechanism from the perspective of node significance at different instances of time. We refrain from using temporal centralities since our level of granularity for the static networks is equal to the span of $N$. We also avoid using diffusion centralities as each $G^N$ does not include any node or edge attributes. Briefly the node measures we use in this paper can be categorized into the following:

\begin{enumerate}
	\item Degree analysis: Nodal degree and Degree entropy 
	\item Connectivity analysis: Clustering, Pagerank and Alpha centrality
	\item Path analysis: Betweenness
\end{enumerate}

\subsection{Nodal Degree}
We observe the nodal degree $k_i$ of a node $i$ as a measure of how connectivity to immediate neighbors can affect the extent of diffusion spread.

\begin{figure}[h!]
	\centering
	\hfill
	\begin{minipage}{0.50\textwidth}%
		\includegraphics[width=6cm, height=4cm]{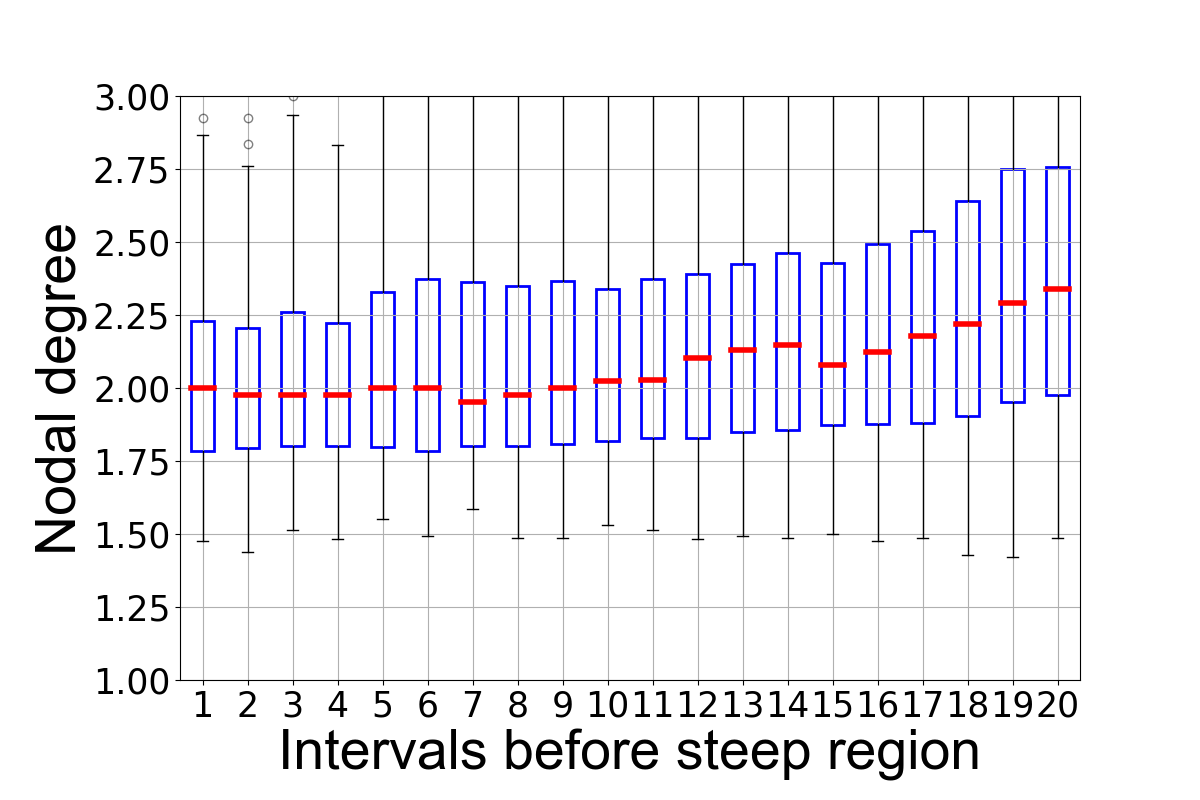}
		\hspace*{2cm}\subcaption{}
	\end{minipage}
	\hfill
	\begin{minipage}{0.40\textwidth}
		\includegraphics[width=6cm, height=4cm]{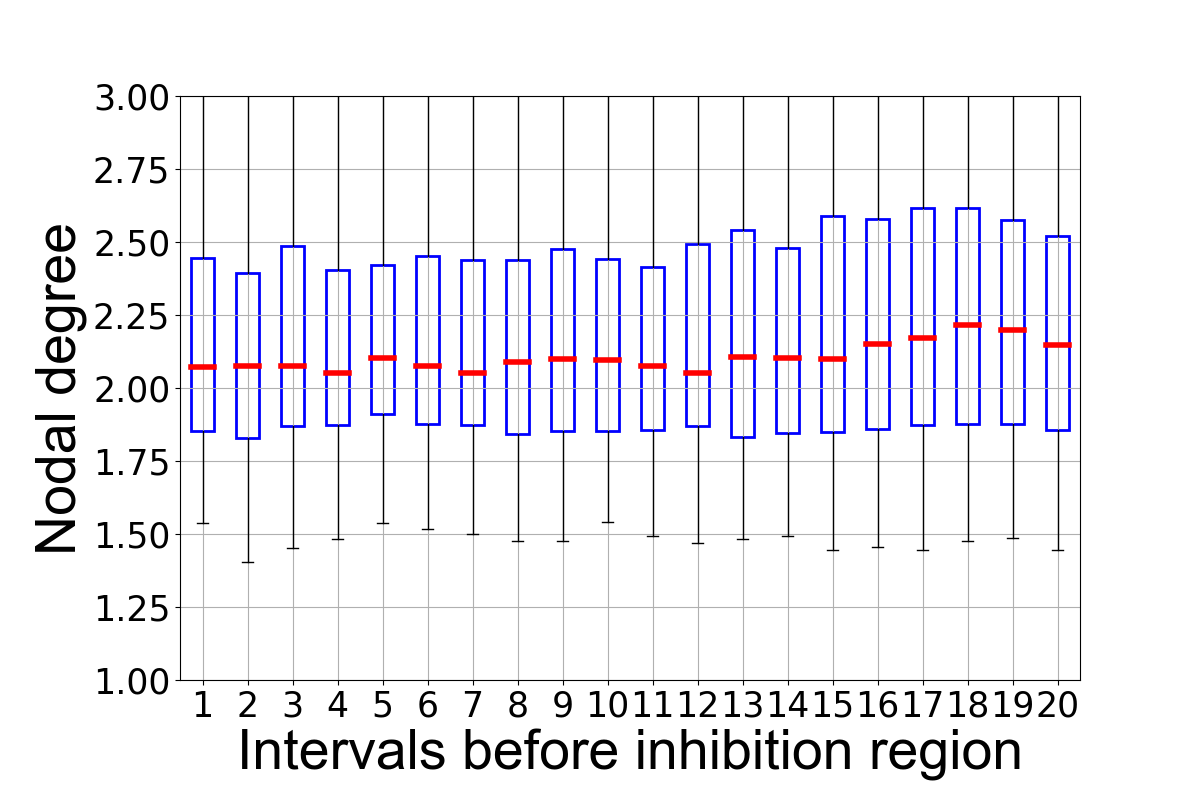}
		\hspace*{2cm}\subcaption{}
	\end{minipage}
	\hfill
	\caption{Nodal degree values}. 
	\label{fig:degree}
\end{figure}

We find that \textit{nodal degrees} in itself do not show much variations within the values owing to the sparsity of the cascade network although there are signs of slight increasing trends in the intervals preceding $\tau'_{styeep}$ observed from the plots in Figure~\ref{fig:degree}(a). The lack of evidence of any variation pattern in the evolution of \textit{nodal degree} over the temporal networks shows that the connectivity of the nodes with their immediate neighbors is not very informative per se for observing the dynamics of the networks.

\subsubsection{Degree entropy}
We take motivation from the idea proposed in \cite{deg_entropy} to see how the degree of the neighbors of a node $i$ affects its influence power when $k_i$ is in the lower percentile of the degree distribution of the network, where $k_i$ denotes the number of neighbors denoted by $n(i)$ in $N$. Although the approach in that paper used clusters to define the degree entropy of a node, we avoid using clusters and instead use the neighbors of nodes in $n(i)$ as a measure of the influence of $i$.
Traditionally, as proposed in \cite{deg_entropy}, when $k_i$ is low, influence is a function of the degree of neighbors but as $k_i$ increases, its own influence power dominates that of its neighbors in that more users reshare from $i$ itself.

We define the degree entropy as follows:
\begin{equation}
H_i = -\sum_{j=0}^{k_i} \frac{k_j}{k_i} \ log\big( {\frac{k_j}{k_i}} \big)
\label{eq:deg_ent}
\end{equation}
Equation~\ref{eq:deg_ent} states that when the node $i$'s degree $k_i$ is significantly higher than the degree of its out-neighbors, $H_i$ turns out to be on the higher percentile of the distribution of degree entropies and should be representative of the higher growth in the cascade lifecycle. On the contrary, when $k_i$ is lower compared to its out-neighbors' degree, $H_i$ is low and should be representative of the declining phase of the lifecycle.

\begin{figure}[h!]
	\centering
	\hfill
	\begin{minipage}{0.50\textwidth}%
		\includegraphics[width=6cm, height=4cm]{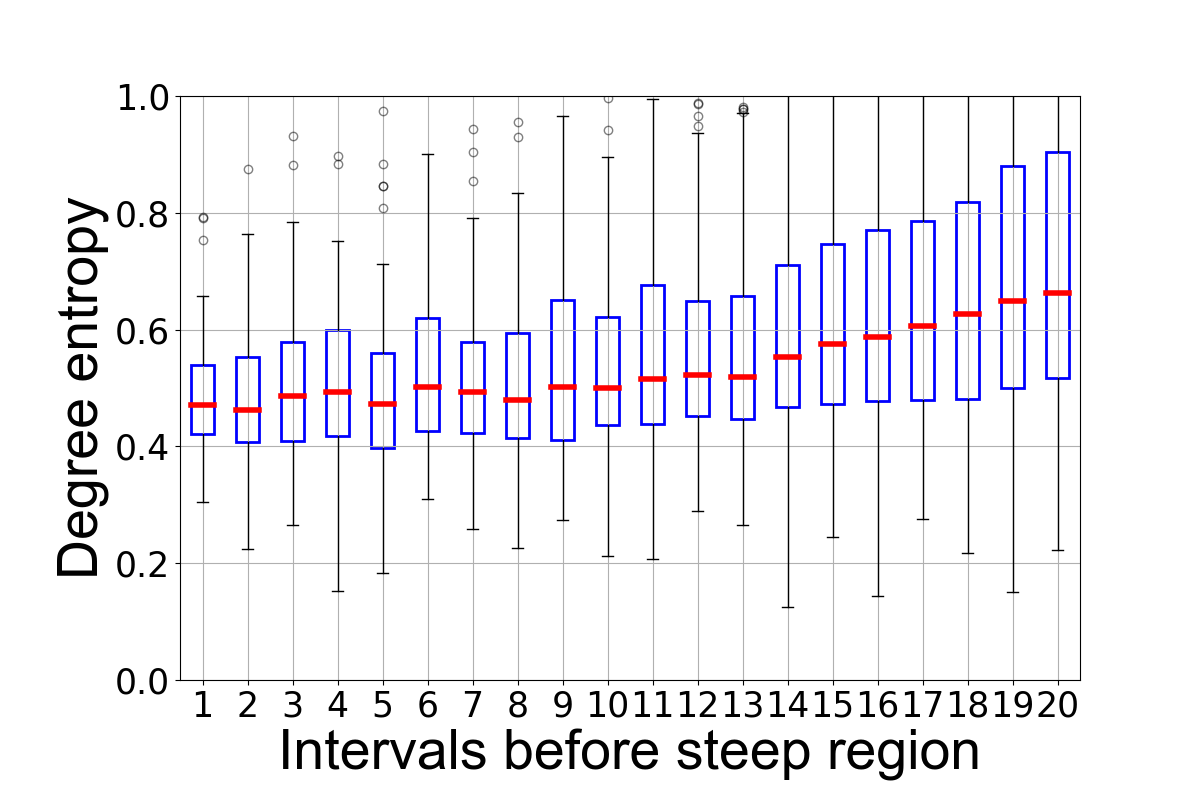}
		\hspace*{2cm}\subcaption{}
	\end{minipage}
	\hfill
	\begin{minipage}{0.40\textwidth}
		\includegraphics[width=6cm, height=4cm]{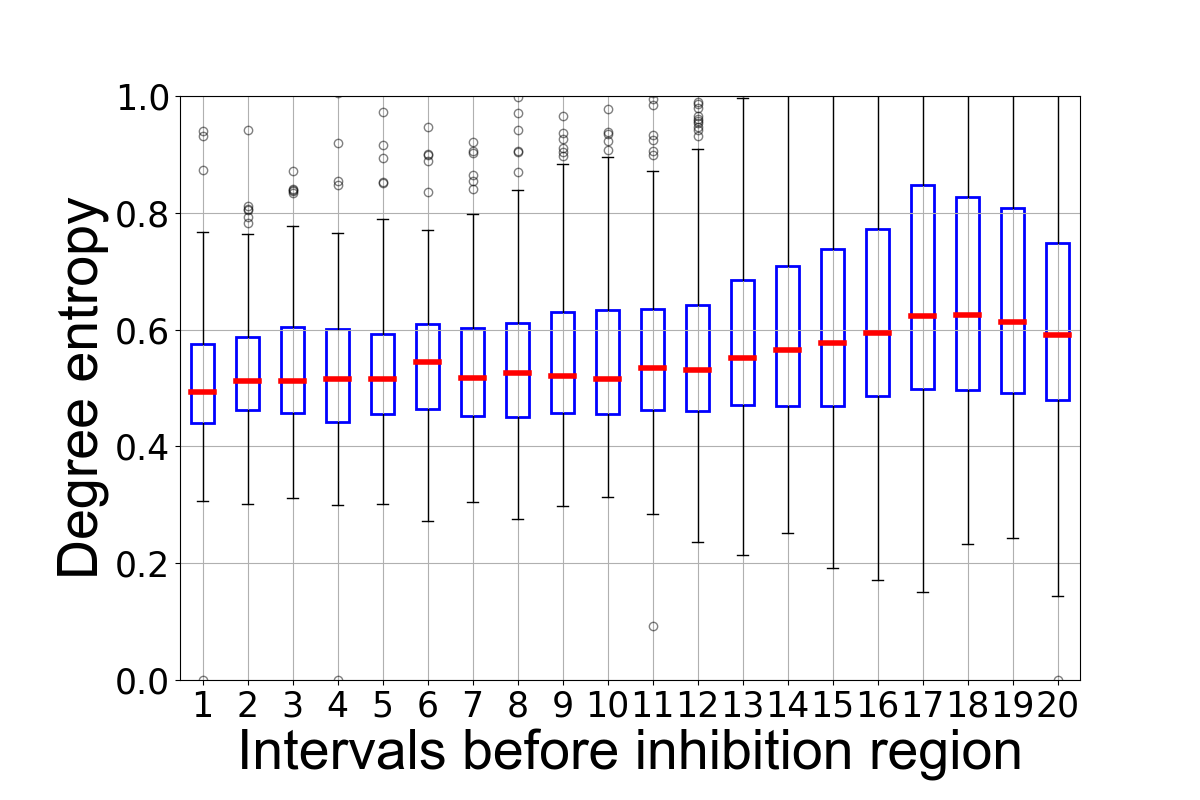}
		\hspace*{2cm}\subcaption{}
	\end{minipage}
	
	\hfill
	\caption{Degree entropy values}. 
	\label{fig:deg_entropy}
\end{figure}

The box plots in Figures~\ref{fig:deg_entropy} reveal that while the intervals preceding $\tau'_{steep}$ show a steady increase in the values, the intervals preceding $\tau'_{inhib} $ show a gradual increase indicated by the increasing medians of the boxes before starting to decay to finally reach the inhibition zone. This is more representative of the general adoption dynamics that we expect from Type I cascades. However in degree entropy, the presence of high degree neighbors undermine the node power itself as far as the degree is concerned and therefore in a way it distinguishes itself from the nodal degree itself. This evaluation will be crucial when we try to forecast the event times for a cascade in Section~\ref{sec:for_exp}.

\subsection{Clustering Coefficient}
The clustering around a node $u$ is quantified by the clustering coefficient $C_u$, defined as the number of triangles in which node $u$ participates
normalized by the maximum possible number of such triangles, formally

\begin{equation*}
C_i = \frac{2t_i}{k_i(k_i-1)}
\end{equation*}
where $t_i$ denotes the number of triangles around node $i$ and $k_i$ denotes its degree in the network. Traditionally, clustering has been believed to be an interference in the cascade progress \cite{cluster_inhibit}.The main crux in connectivity through forming quick and small clusters has been studied before and pose two important explanations: first while higher clustering suggests groups forming circles to an extent the message does not circulate beyond certain nodes and secondly more small clusters suggest that users who exhibit higher clustering coefficients are more eager to form such loops easily thereby increasing the diffusion rate. To test these two theoretical observations, we use clustering coefficients of nodes as a measure of information diffusion spread.

\begin{figure}[]
	\centering
	\hfill
	\begin{minipage}{0.50\textwidth}%
		\includegraphics[width=6cm, height=4cm]{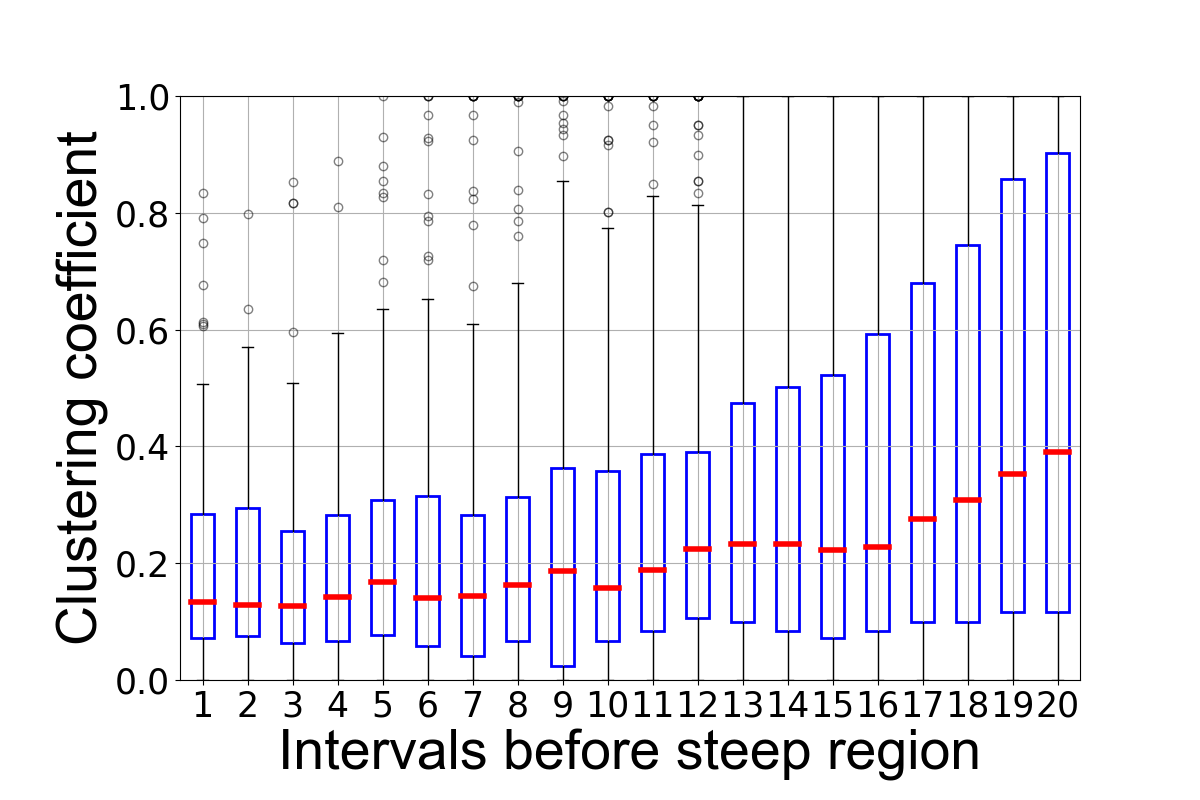}
		\hspace*{2cm}\subcaption{}
	\end{minipage}
	\hfill
	\begin{minipage}{0.40\textwidth}
		\includegraphics[width=6cm, height=4cm]{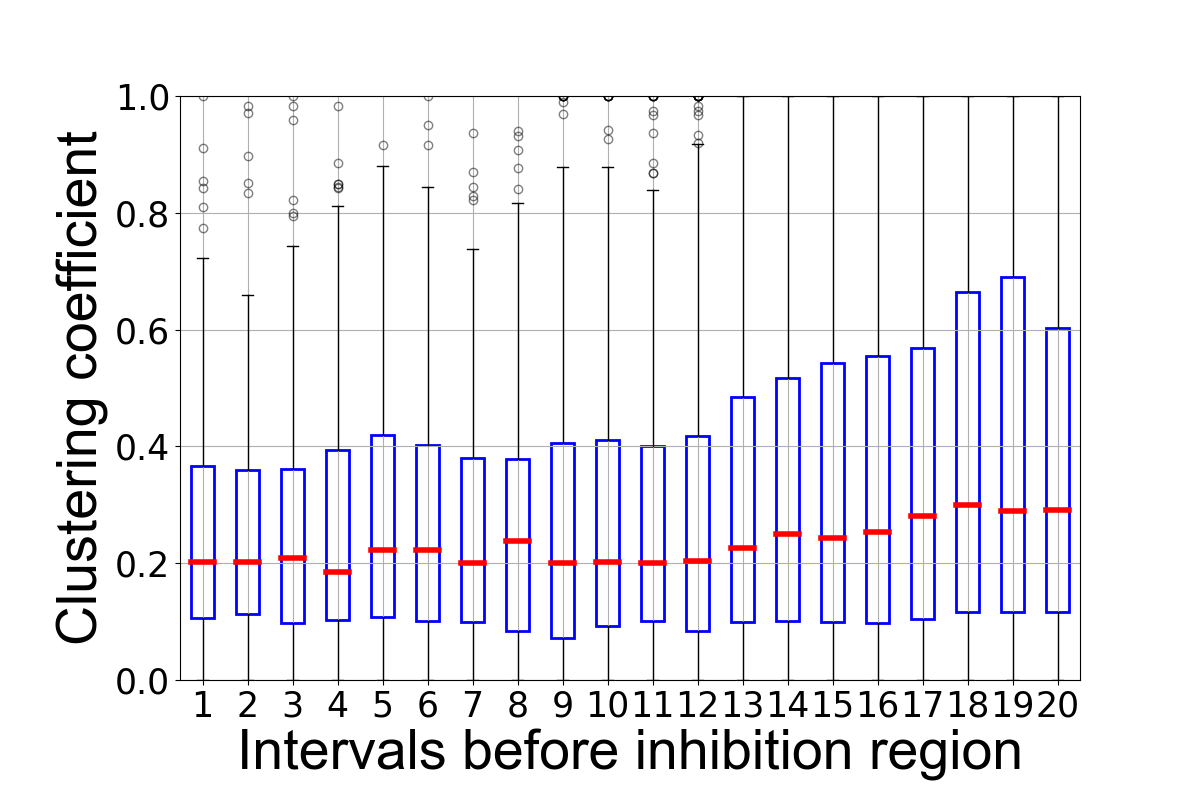}
		\hspace*{2cm}\subcaption{}
	\end{minipage}
	
	\hfill
	\caption{Clustering coefficient values}. 
	\label{fig:cluster_coeff}
\end{figure}

To test our hypothesis about nodes forming quick clusters being an amplifier of diffusion spread or being a bottleneck in the spread, we observe the clustering coefficient of nodes with the progress of subsequences in time. As shown in Figure~\ref{fig:cluster_coeff}, we observe that the clustering coefficient rapidly increases in the intervals preceding $\tau'_{steep}$, whereas although there may be a slight increase in the intervals preceding $\tau'_{inhib}$, the change is not visible. So generally we find that the hypothesis of higher clustering coefficients being a bottleneck does not hold in a cascade setting. The failure of the hypothesis to show clusters as a bottleneck towards slow growth in cascades will be enhanced when we use it for forecasting event times.

\subsection{PageRank}
PageRank centrality \cite{PageRank} has been used for ranking the spreading capability of users in diffusion networks and till date, most of the research done on PageRank has been on simulation of spreading dynamics to validate it as a strong predictor of influence.  In \cite{Ghosh_pr}, the authors study the PageRank centrality in relation to stochastic processes and conclude that in general, the PageRank measure does not perform well when it comes to predicting the information spread. Similarly, in \cite{sei_spreaders}, Pei et al. study different indicators of influential users, but find that PageRank performs poorly as compared to k-core. However there have been a lot of literature presenting such mixed views although only a very few focus on cascade settings. 

\begin{figure}[h]
	\centering
	\hfill
	\begin{minipage}{0.50\textwidth}%
		\includegraphics[width=6cm, height=4cm]{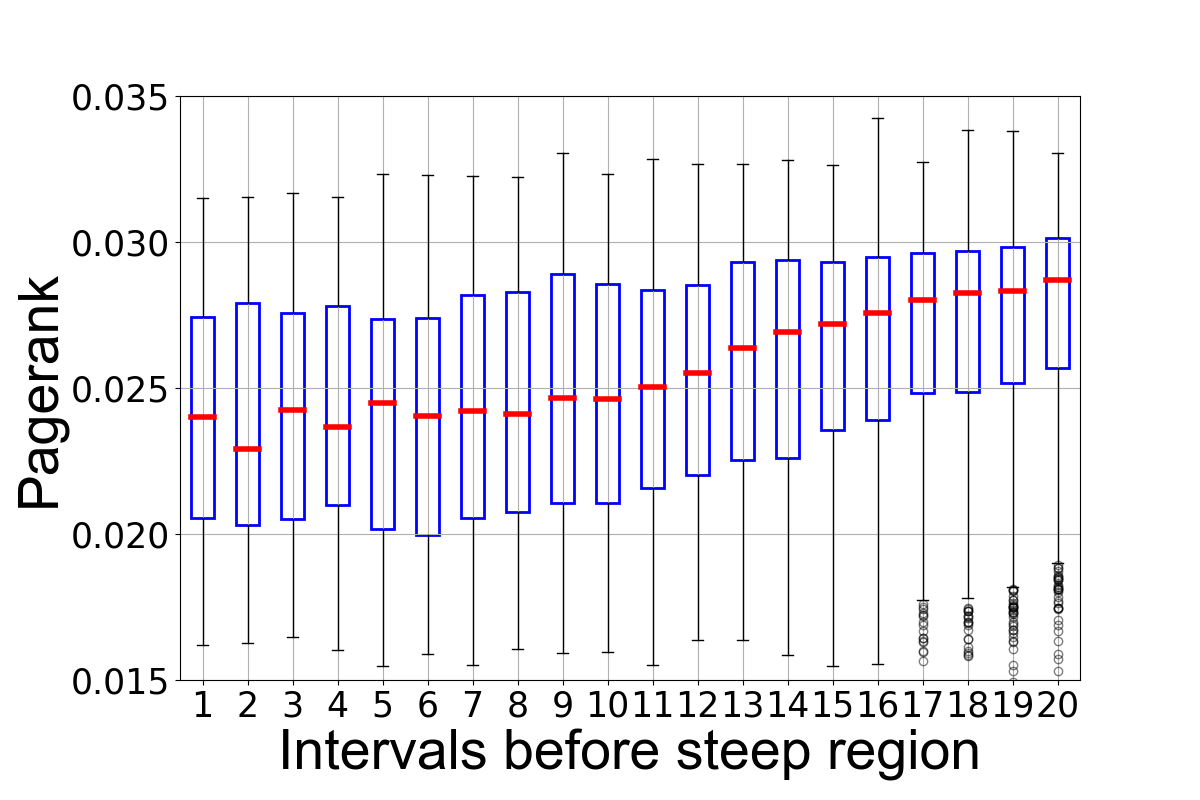}
		\hspace*{2cm}\subcaption{}
	\end{minipage}
	\hfill
	\begin{minipage}{0.40\textwidth}
		\includegraphics[width=6cm, height=4cm]{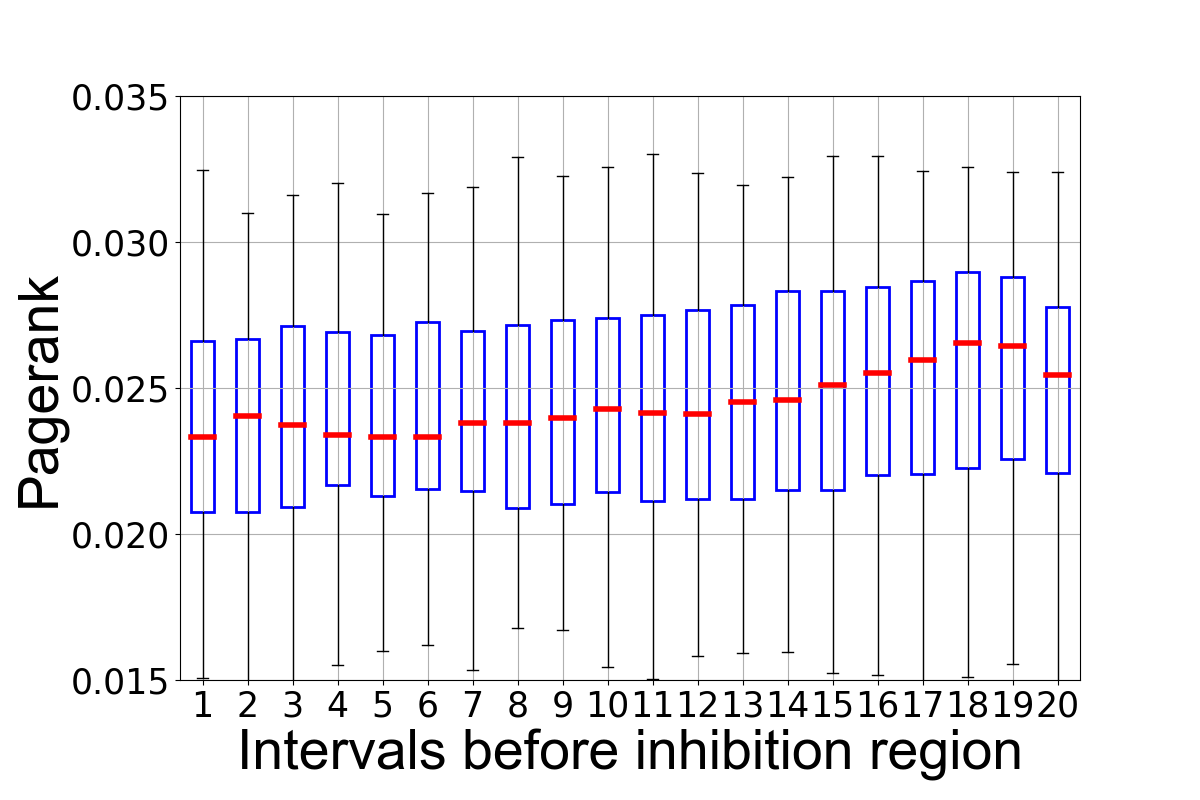}
		\hspace*{2cm}\subcaption{}
	\end{minipage}
	\hfill
	\caption{PageRank centrality values}.
	\label{fig:pr} 
\end{figure}

The box plots in Figures~\ref{fig:pr} reveal that the measures \textit{PageRank} and \textit{degree entropy} of nodes follow the same structural dynamics. Intuitively there is a link between the way Pagerank centrality is calculated in an undirected network and the degree entropy - both of them point to the fact that the strength of a node in diffusion spread is characterized by the presence a node with many out-neighbors. In Pagerank, additionally the centrality of a node is augmented by the presence of large number of out-neighbors with high degree.

\subsection{Betweenness centrality}
Betweenness centrality measures the extent to which a node lies on information diffusion paths between other nodes. Nodes with high \textit{betweenness} may have considerable influence within a network by virtue of their control on the information flow between others. They are also the ones whose removal from the network should disrupt flow of information between other nodes because they lie on the largest number of shortest paths taken by messages.

\begin{figure}[h!]
	\centering
	\hfill
	\begin{minipage}{0.50\textwidth}%
		\includegraphics[width=6cm, height=4cm]{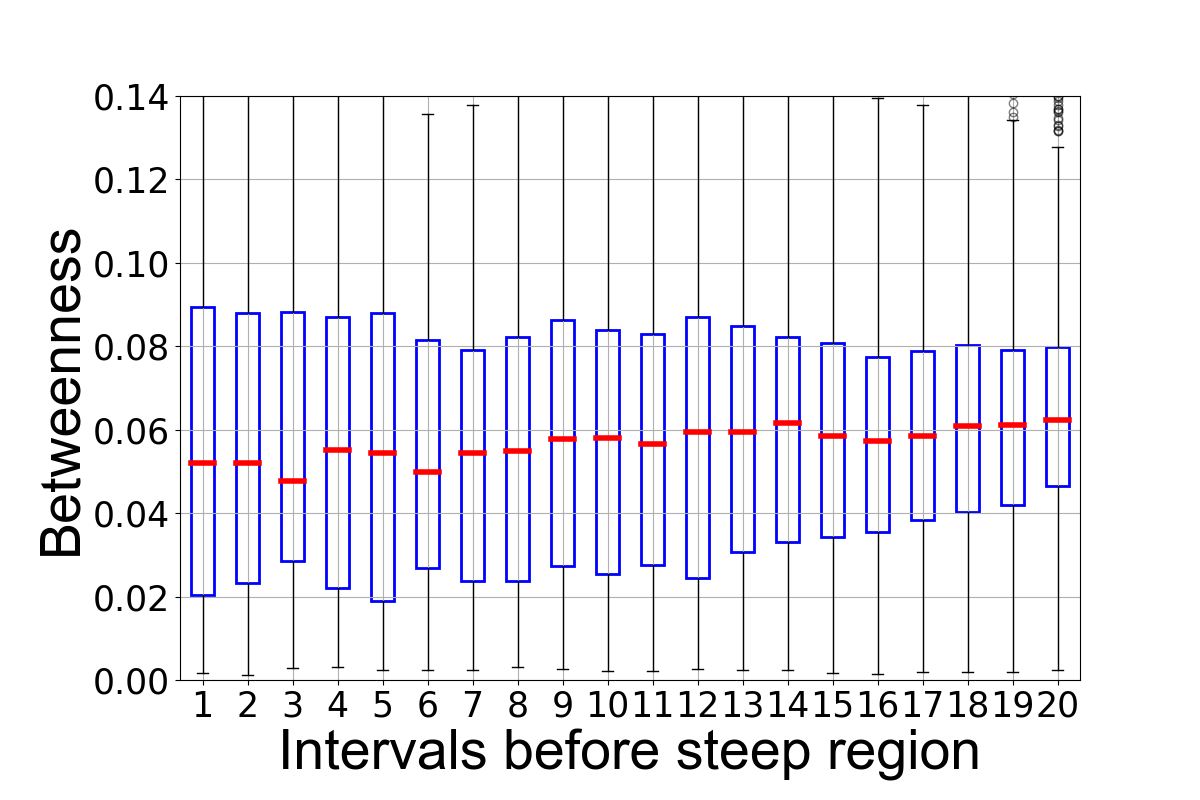}
		\hspace*{2cm}\subcaption{}
	\end{minipage}
	\hfill
	\begin{minipage}{0.40\textwidth}
		\includegraphics[width=6cm, height=4cm]{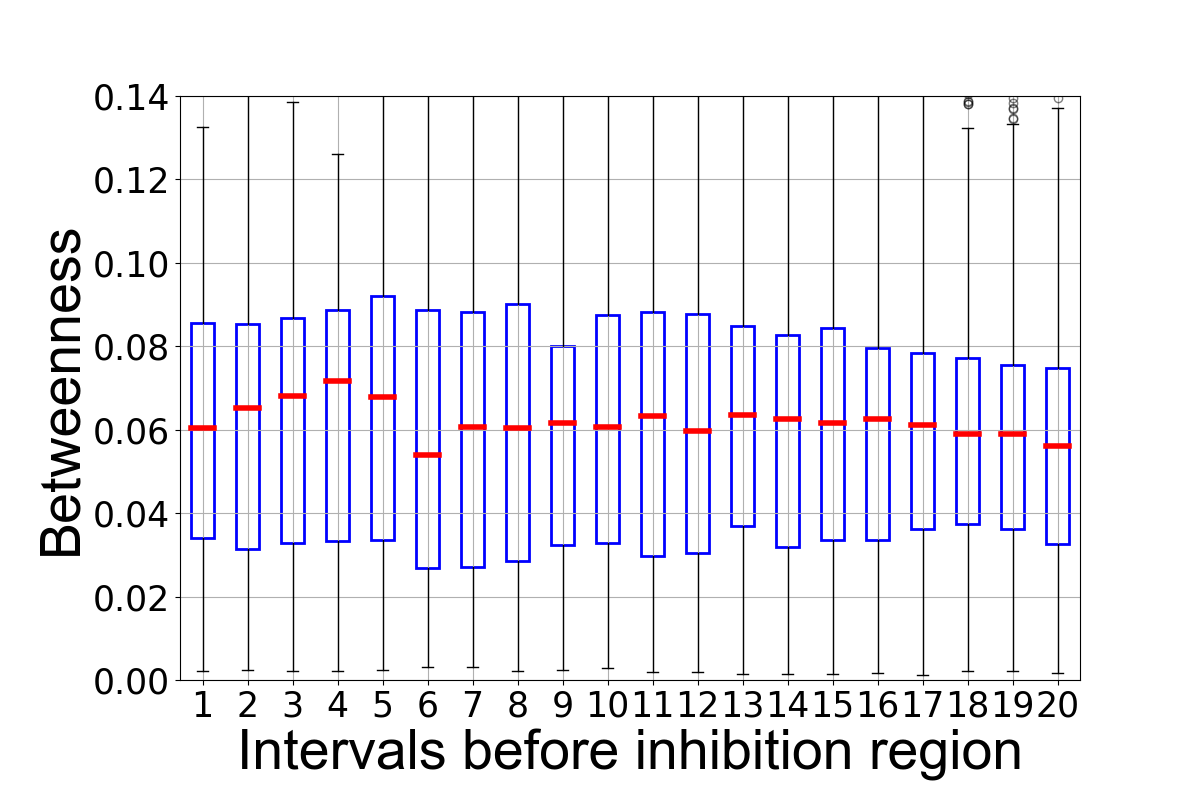}
		\hspace*{2cm}\subcaption{}
	\end{minipage}
	
	\hfill
	\caption{Betweenness values}. 
	\label{fig:betweenness}
\end{figure}

Similar to nodal degrees, we do not observe any significant change or trend pattern in the values in the intervals preceding both the intervals.

\subsection{Alpha Centrality}
\begin{figure}[]
	\centering
	\hfill
	\begin{minipage}{0.50\textwidth}%
		\includegraphics[width=6cm, height=4cm]{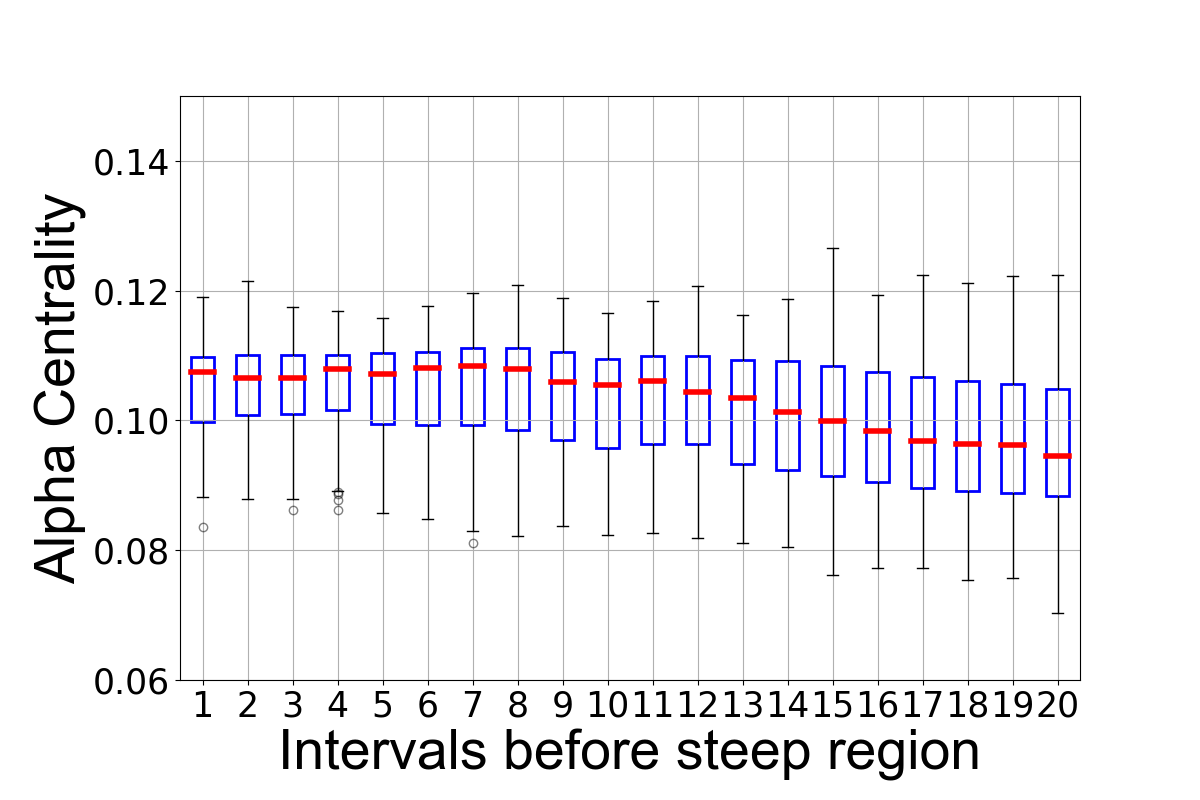}
		\hspace*{2cm}\subcaption{}
	\end{minipage}
	\hfill
	\begin{minipage}{0.40\textwidth}
		\includegraphics[width=6cm, height=4cm]{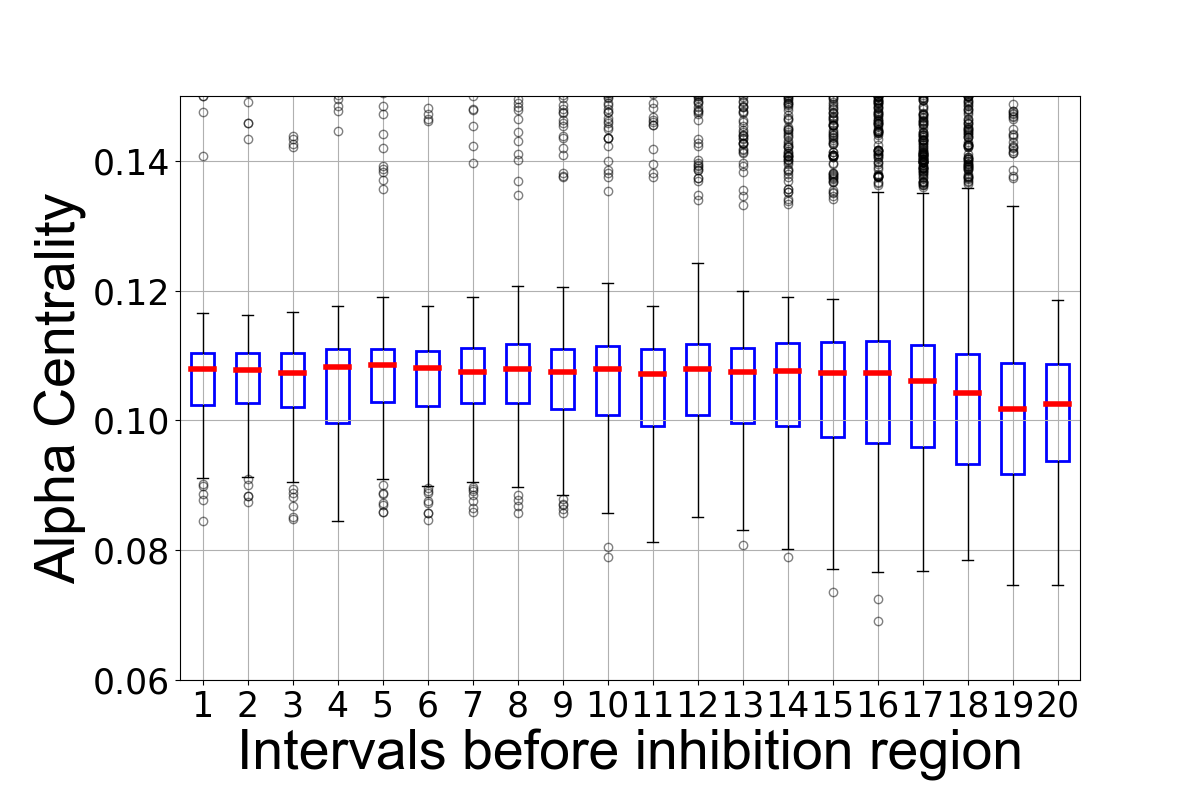}
		\hspace*{2cm}\subcaption{}
	\end{minipage}
	
	\hfill
	\caption{Alpha centrality values}. 
	\label{fig:alpha_cent}
\end{figure}

Alpha Centrality measures the number of paths originating from a node, exponentially attenuated by their length \cite{AlphaCent, AlphaLerman}.  
Formally, Alpha centrality is defined as
\begin{equation*}
\mathbf{C_{\alpha}} = (\mathbf{I} - \alpha \mathbf{A})^{-1} \mathbf{e}
\end{equation*}
where $\mathbf{C_{\alpha}}$ denotes the vector of alpha centralities for each node, $\mathbf{A}$ denotes the adjacency matrix of the cascade graph, $\mathbf{e}$ denotes a vector of ones, and $\alpha$ is a parameter that controls the influence from the neighboring nodes. Intuitively, the parameter $\alpha$ determines, how far on average the nodes' effect would be propagated. As mentioned in \cite{DiffusionCent}, when $\alpha$ $\cong$ 0, nodes with higher degree centralities correspond to nodes with high Alpha centralities and when $\alpha$ reaches the inverse of the maximum eigenvalue of $\mathbf{A}$, it is similar to eigenvector centrality. In this paper, we set $\alpha$ to half the value of the inverse of $\lambda_{max}$ which is the largest eigenvalue of $\mathbf{A}$. From Figure~\ref{fig:alpha_cent}, we find that the dip in the alpha centralities in intervals preceding the steep region is slightly higher than that before the inhibition region, suggesting that it is not highly correlated with the nodal degree as would otherwise be expected in the critical regime \cite{DiffusionCent}.

\begin{figure}[t!]
	\centering
	
	\begin{minipage}{0.50\textwidth}%
		\includegraphics[width=6cm, height=4cm]{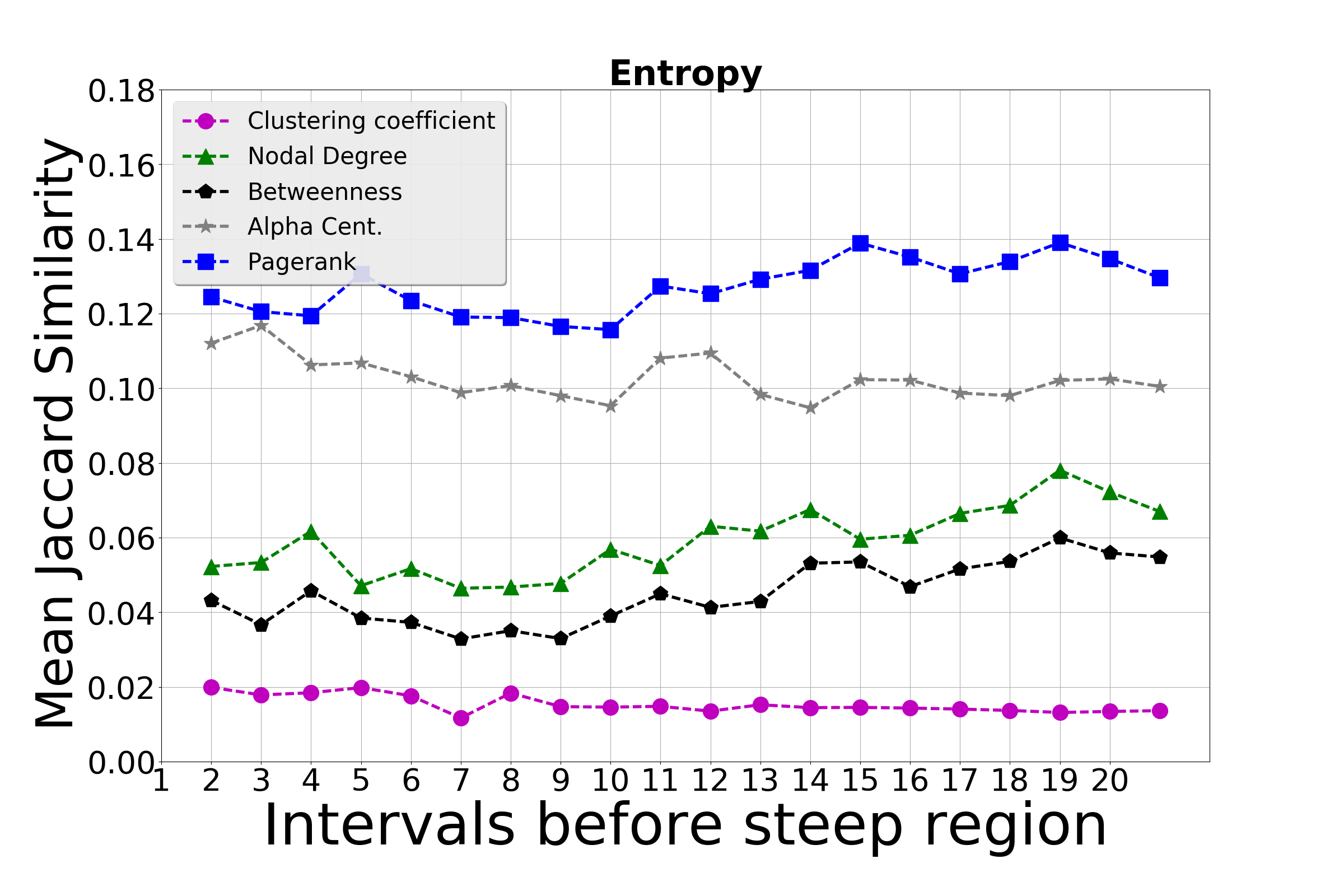}
		\hspace*{2cm}\subcaption{}
	\end{minipage}
	\hfill
	\begin{minipage}{0.40\textwidth}
		\includegraphics[width=6cm, height=4cm]{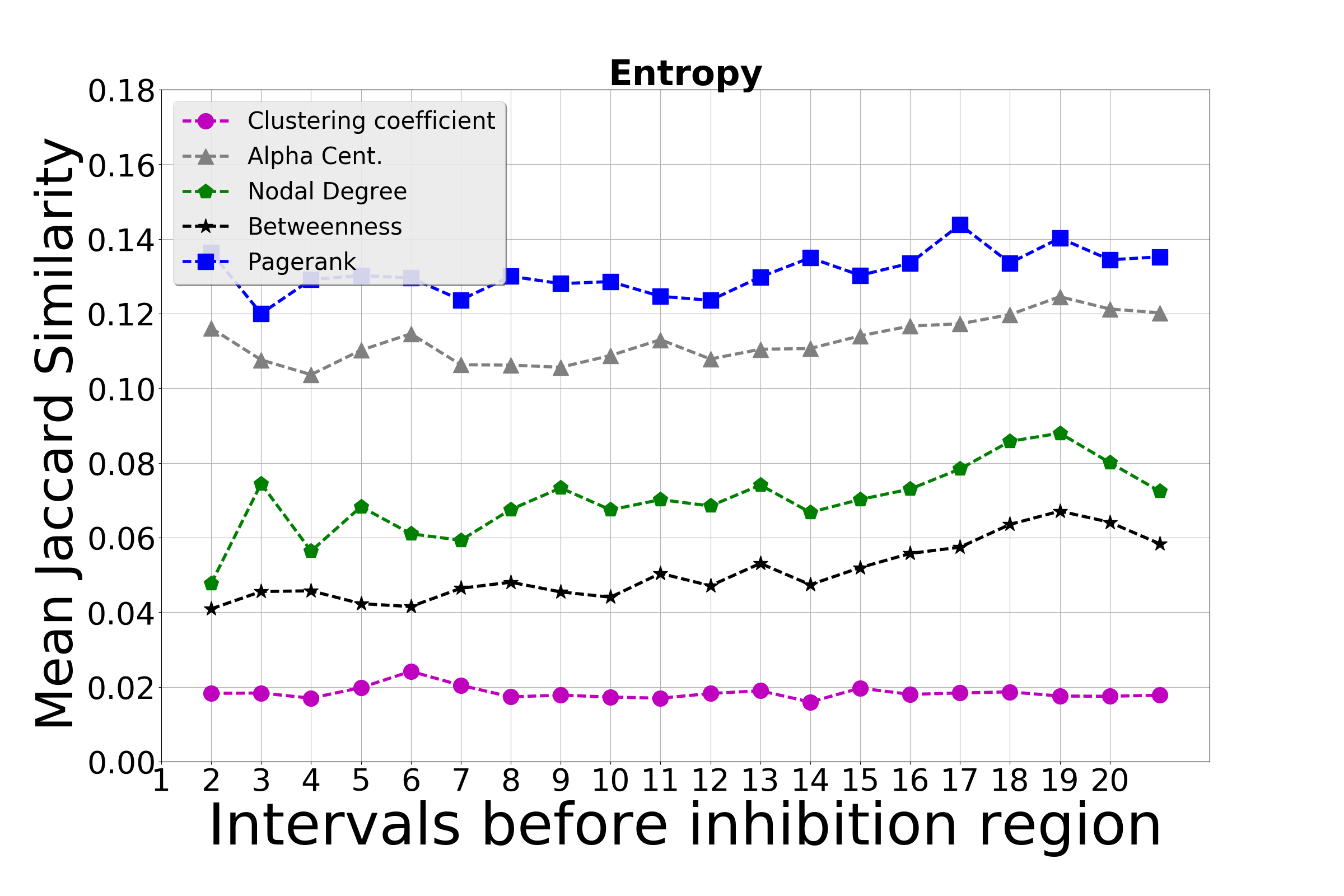}
		\hspace*{2cm}\subcaption{}
	\end{minipage}
	\hfill
	\begin{minipage}{0.50\textwidth}%
		\includegraphics[width=6cm, height=4cm]{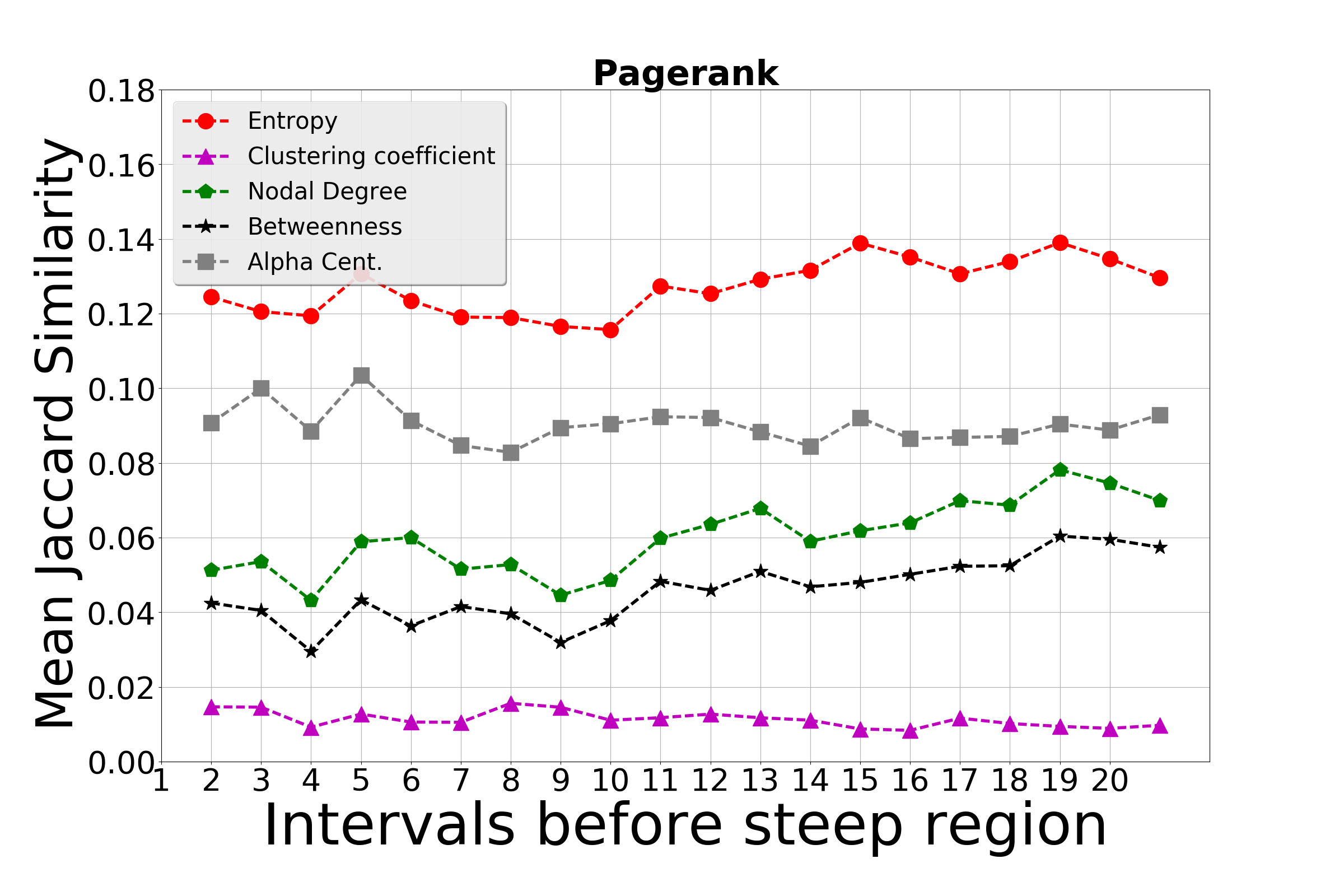}
		\hspace*{2cm}\subcaption{}
	\end{minipage}
	\hfill
	\begin{minipage}{0.40\textwidth}
		\includegraphics[width=6cm, height=4cm]{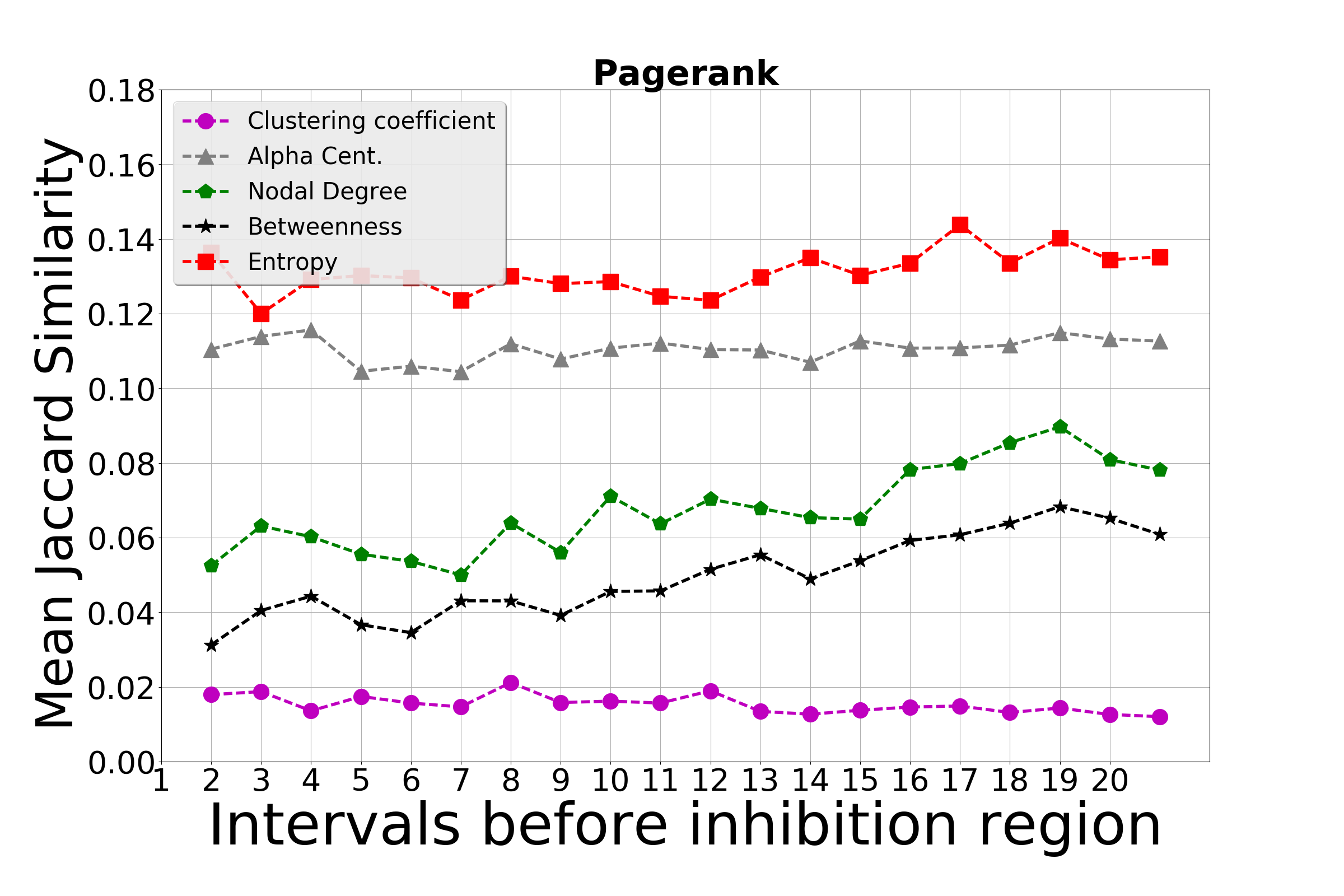}
		\hspace*{2cm}\subcaption{}
	\end{minipage}
	\hfill
	\begin{minipage}{0.50\textwidth}%
		\includegraphics[width=6cm, height=4cm]{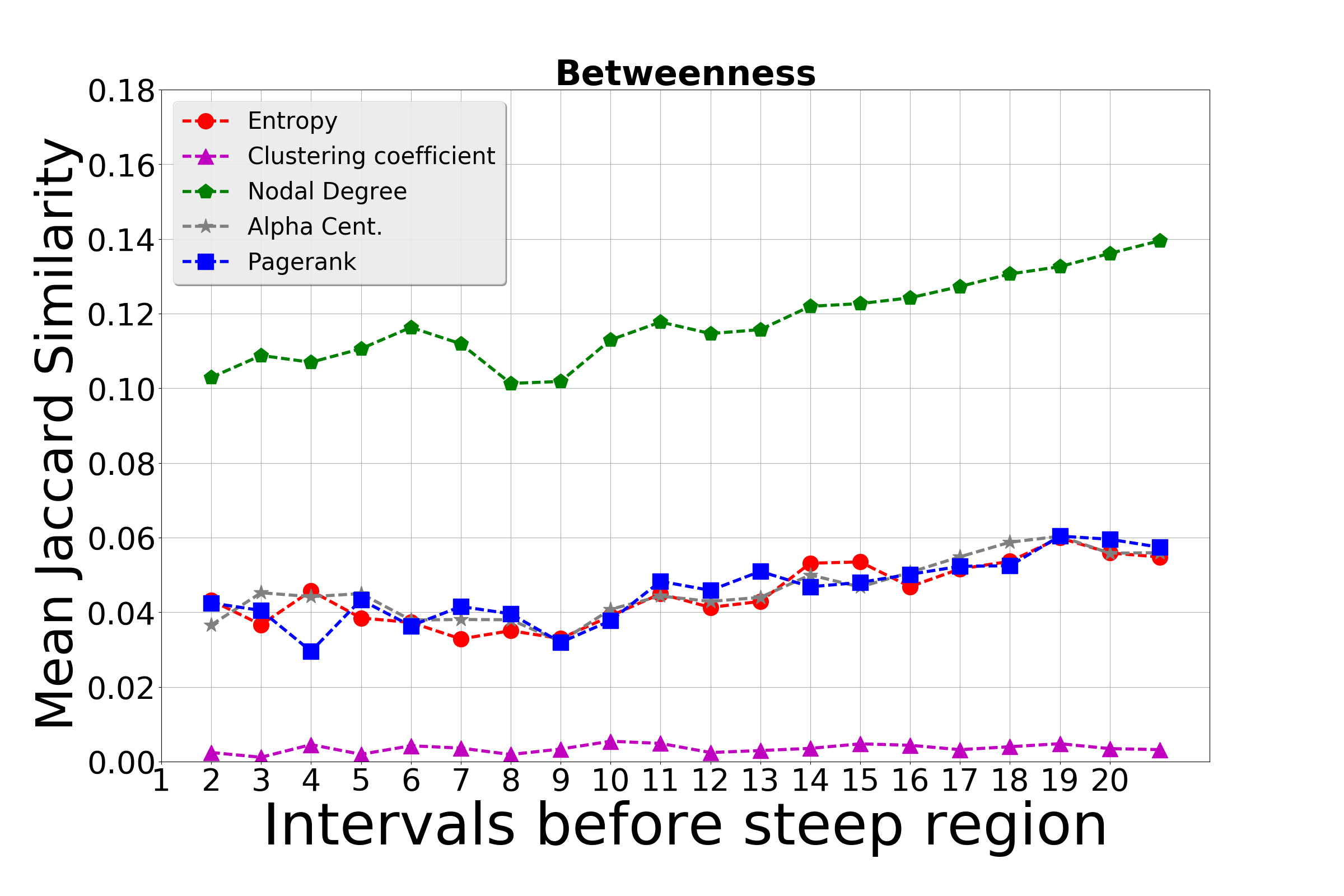}
		\hspace*{2cm}\subcaption{}
	\end{minipage}
	\hfill
	\begin{minipage}{0.40\textwidth}
		\includegraphics[width=6cm, height=4cm]{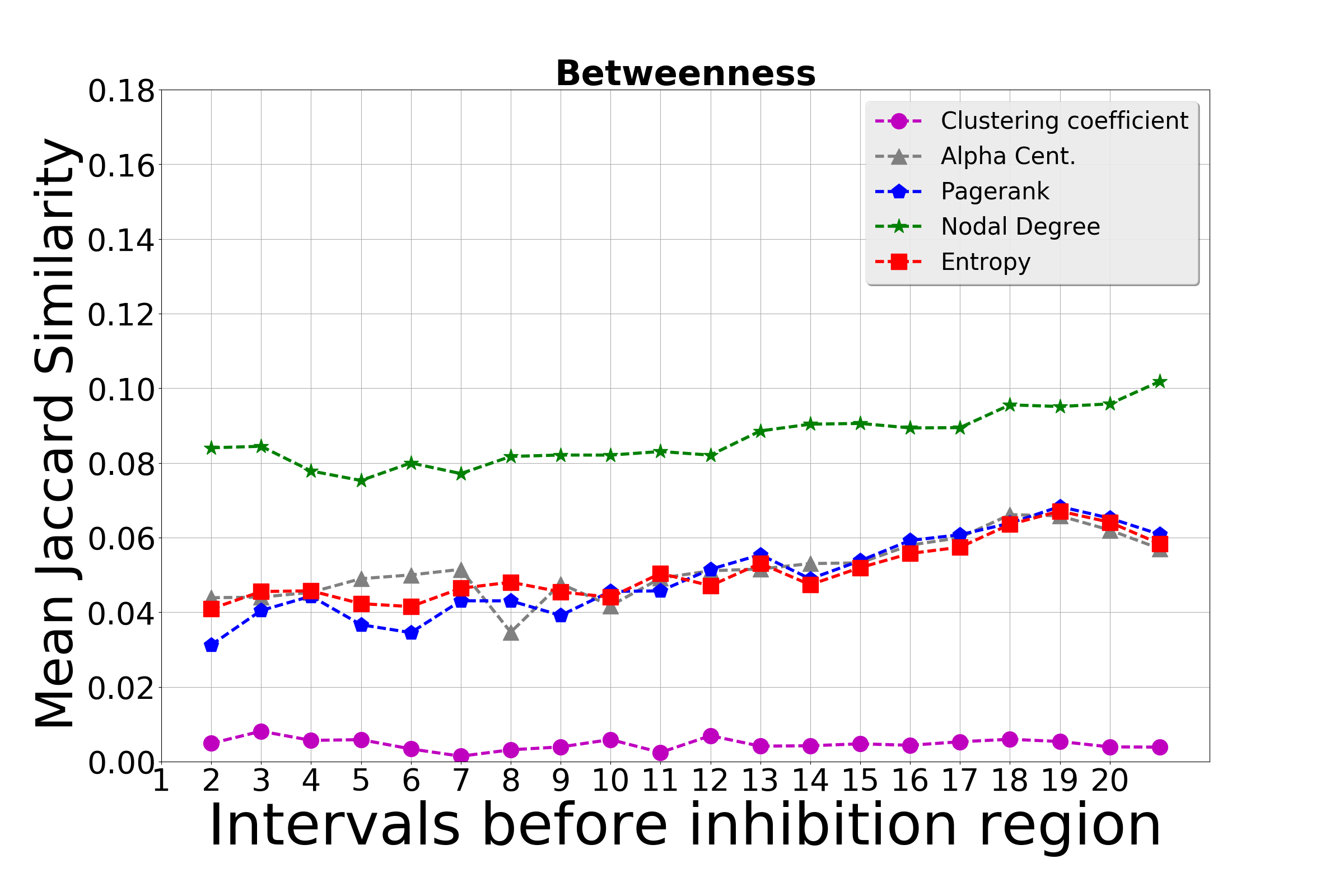}
		\hspace*{2cm}\subcaption{}
	\end{minipage}
	\hfill
	\caption{Mean Jaccard similarity for top 20 nodes in each subsequence. Each plot shows the Jaccard similarity of that measure with other measures considering the top 20 ranked nodes. Here the mean value considering all cascades have been plotted.}
	\label{fig:jaccard_1}
\end{figure}

\begin{figure}[t!]
	\centering
	
	\begin{minipage}{0.50\textwidth}%
		\includegraphics[width=6cm, height=4cm]{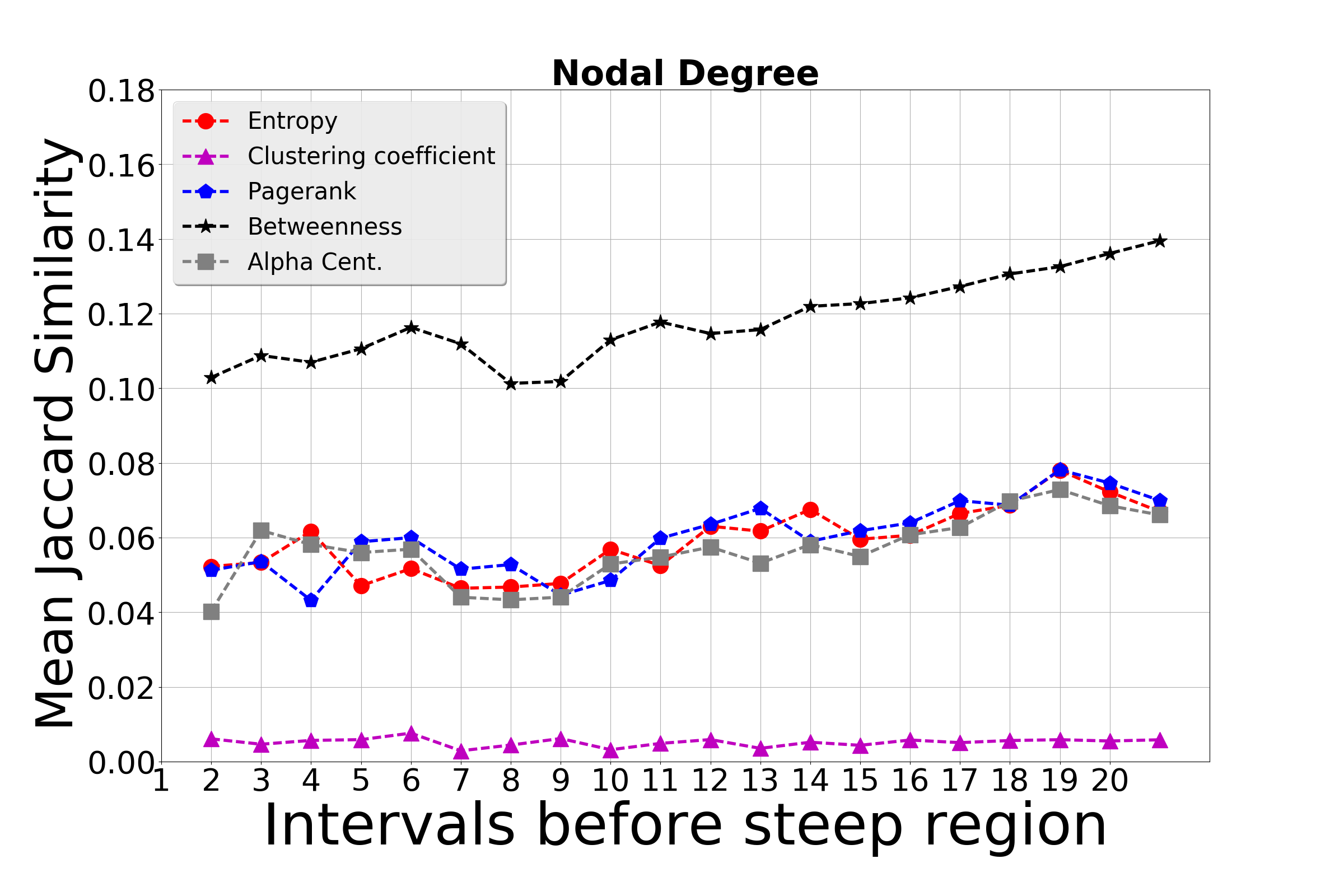}
		\hspace*{2cm}\subcaption{}
	\end{minipage}
	\hfill
	\begin{minipage}{0.40\textwidth}
		\includegraphics[width=6cm, height=4cm]{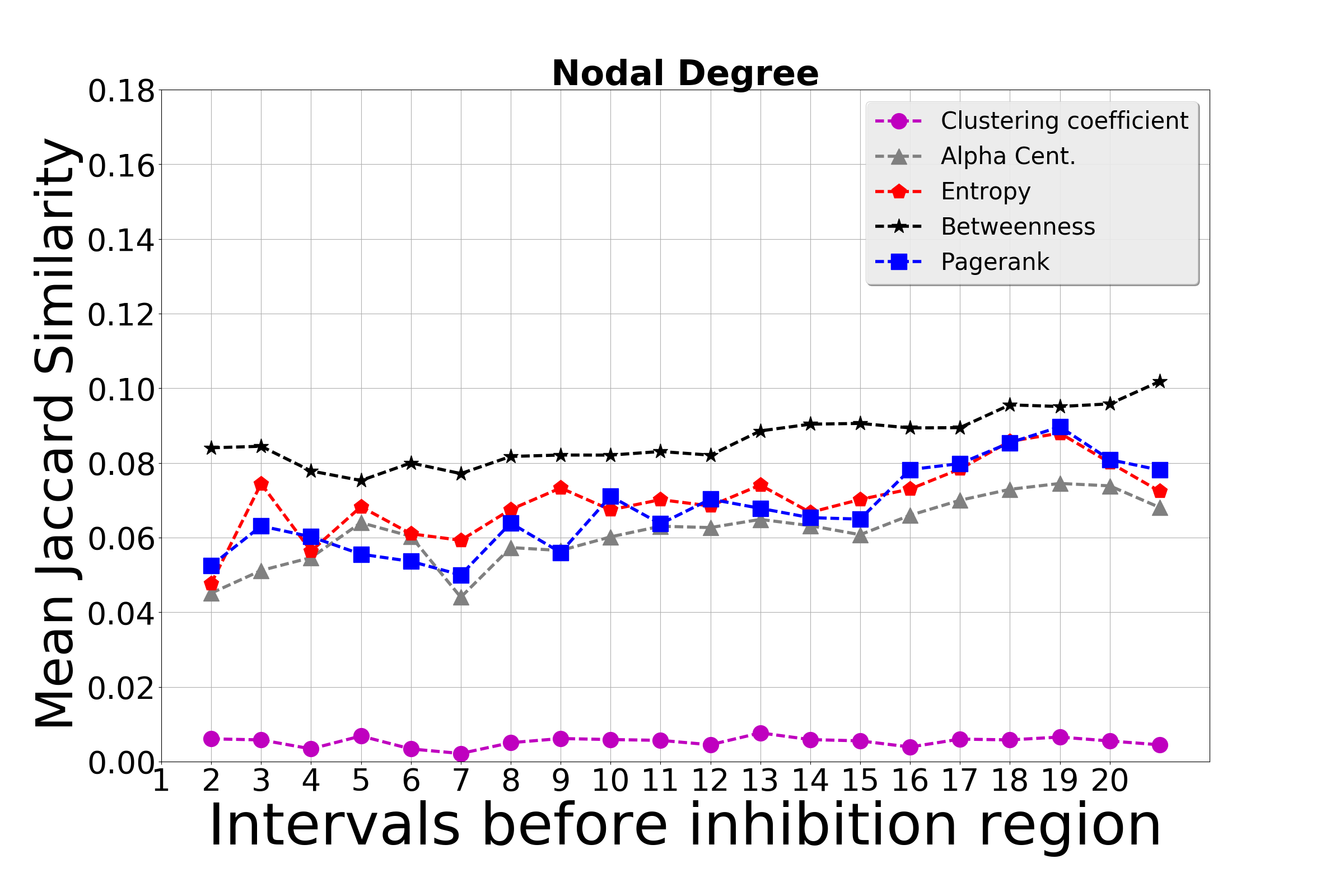}
		\hspace*{2cm}\subcaption{}
	\end{minipage}
	\hfill
	\begin{minipage}{0.50\textwidth}%
		\includegraphics[width=6cm, height=4cm]{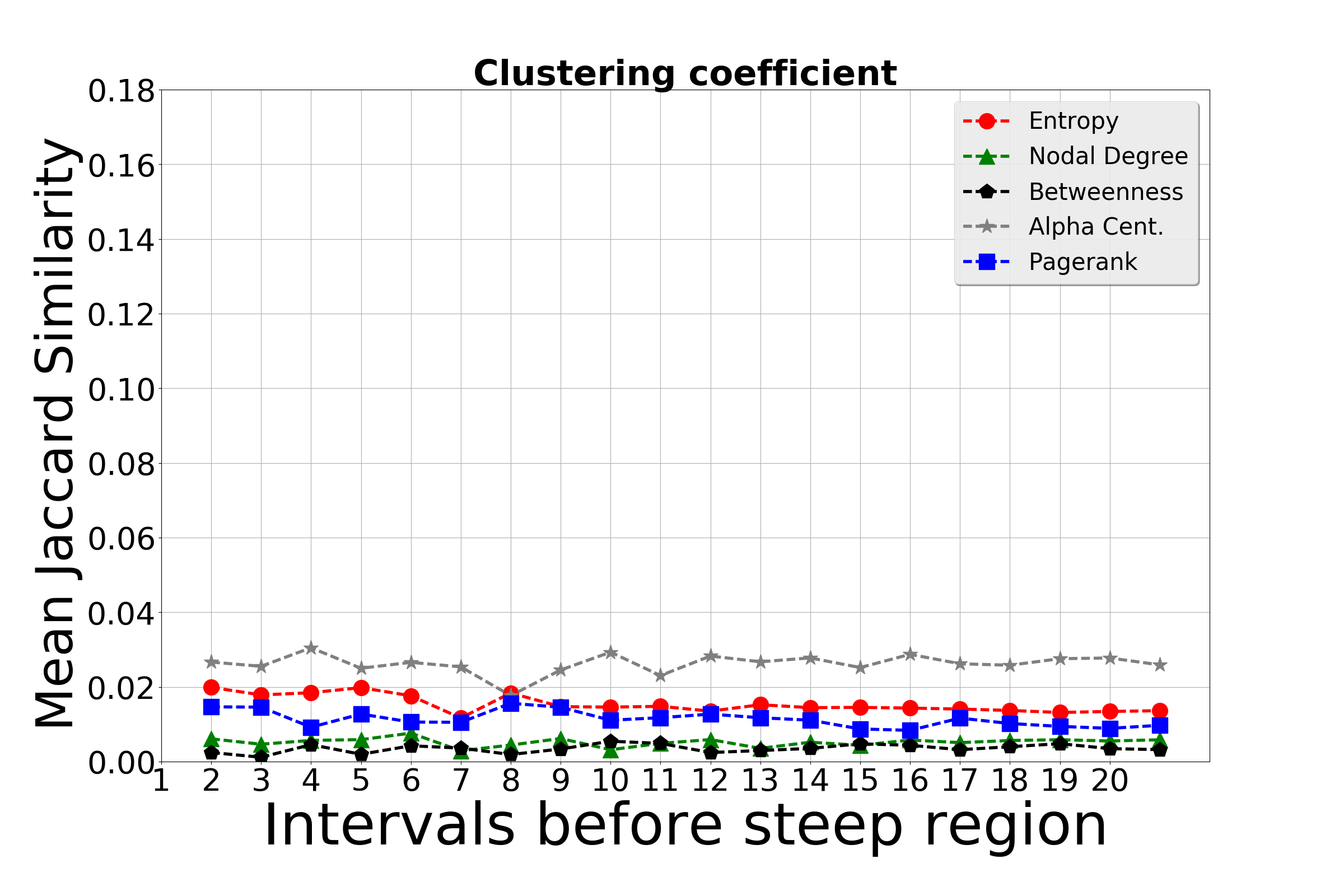}
		\hspace*{2cm}\subcaption{}
	\end{minipage}
	\hfill
	\begin{minipage}{0.40\textwidth}
		\includegraphics[width=6cm, height=4cm]{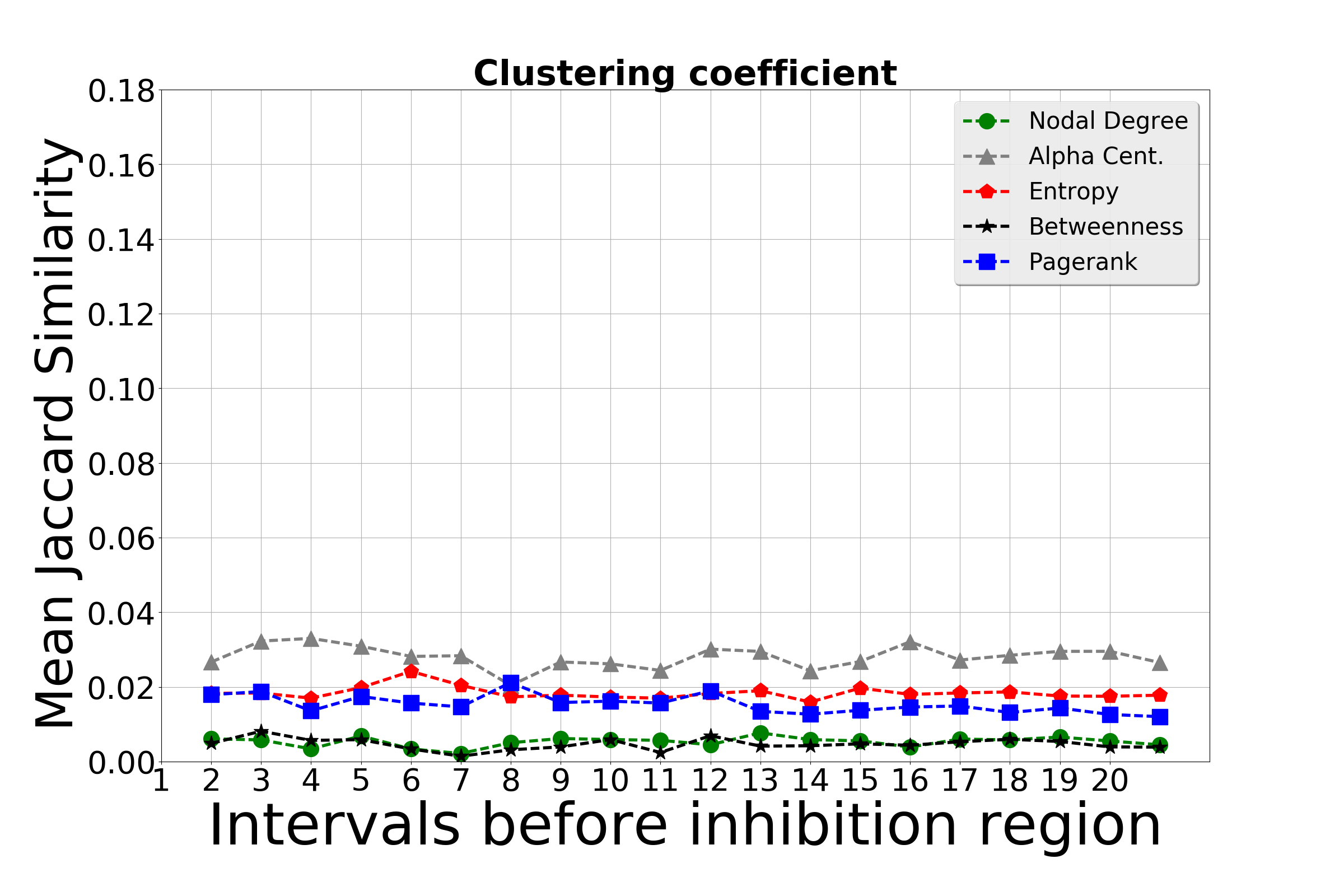}
		\hspace*{2cm}\subcaption{}
	\end{minipage}
	\hfill
	\begin{minipage}{0.50\textwidth}%
		\includegraphics[width=6cm, height=4cm]{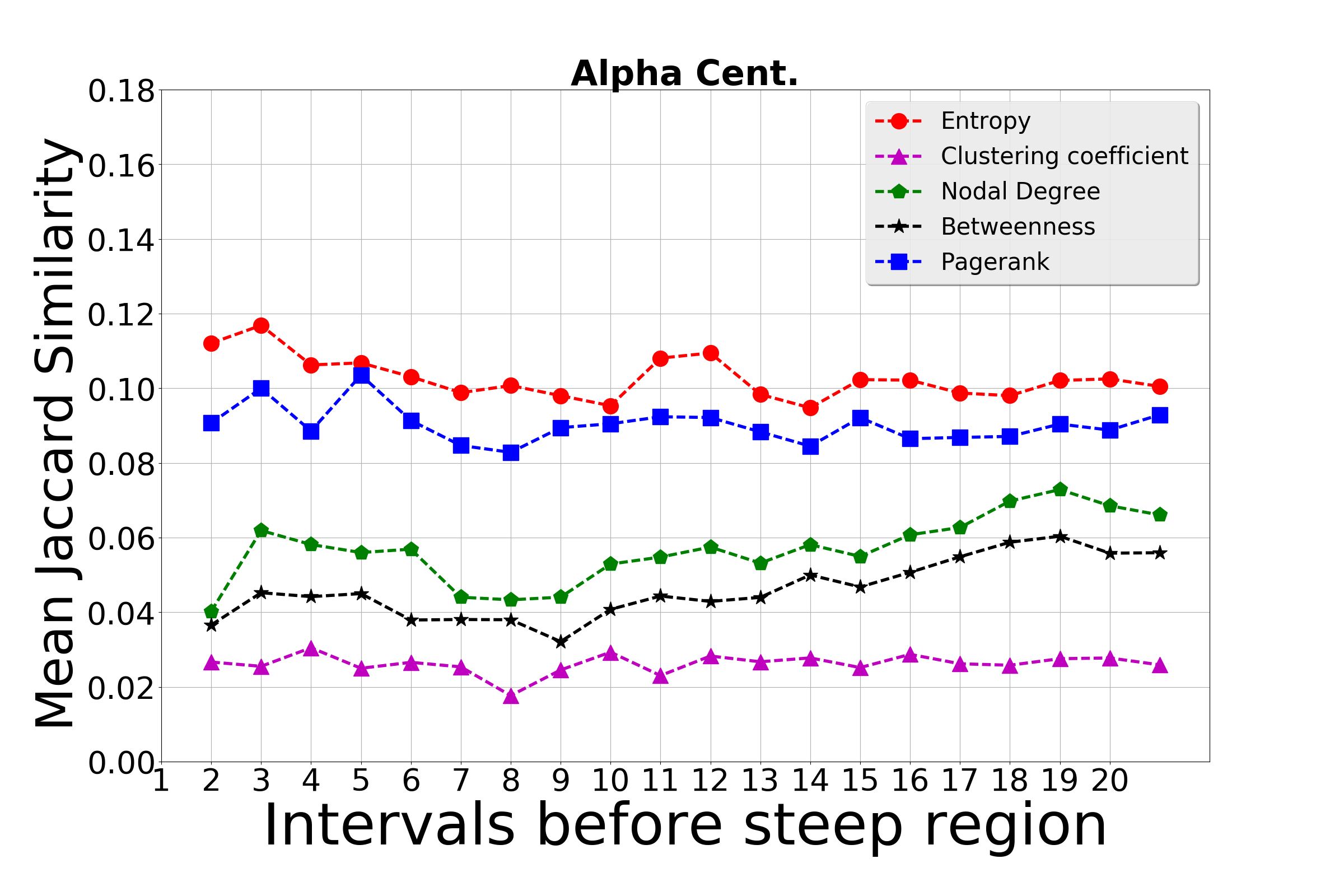}
		\hspace*{2cm}\subcaption{}
	\end{minipage}
	\hfill
	\begin{minipage}{0.40\textwidth}
		\includegraphics[width=6cm, height=4cm]{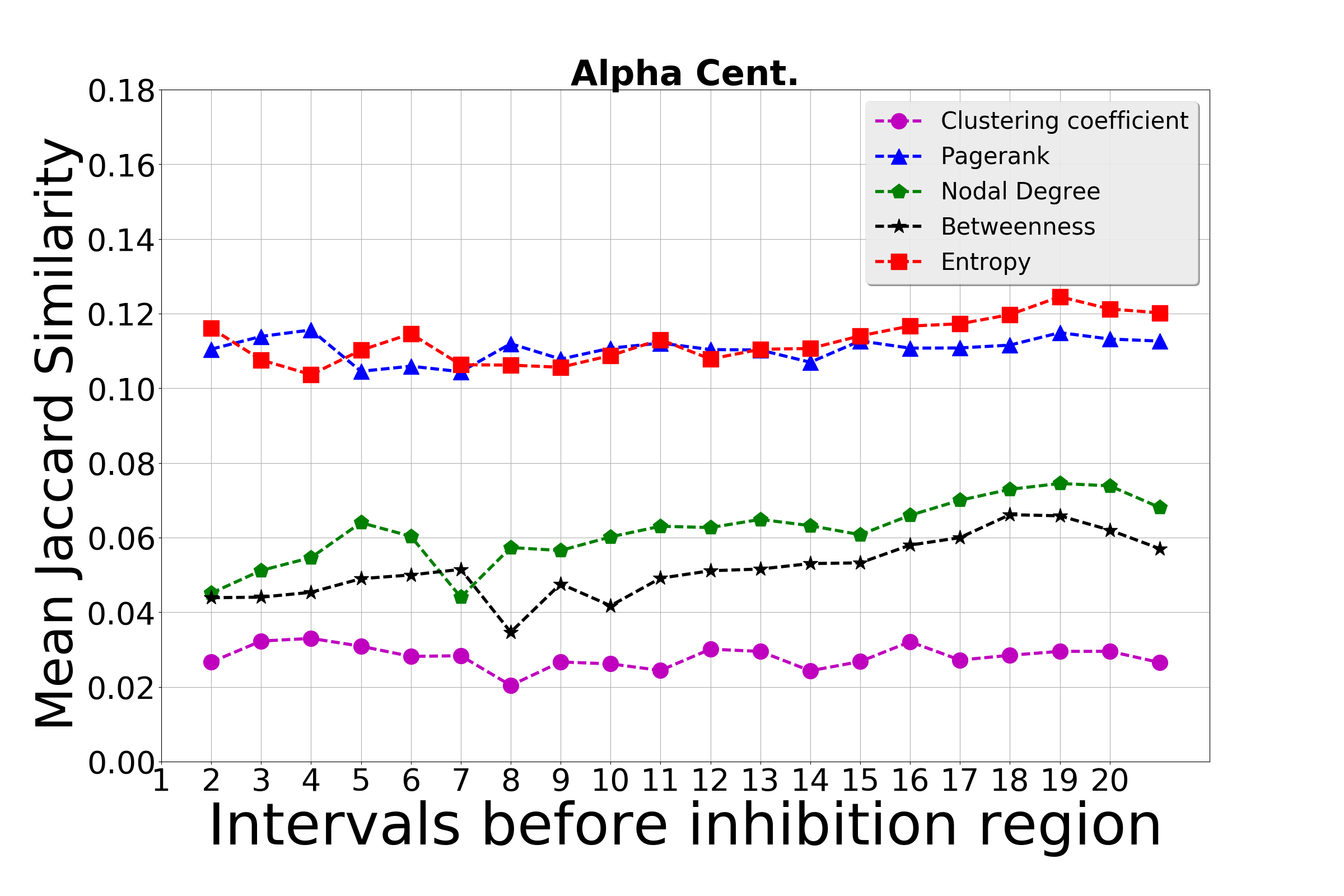}
		\hspace*{2cm}\subcaption{}
	\end{minipage}
	\hfill
	\caption{Mean Jaccard similarity for top 20 nodes in each subsequence. Each plot shows the Jaccard similarity of that measure with other measures considering the top 20 ranked nodes. Here the mean value considering all cascades have been plotted.}. 
	\label{fig:jaccard_2}
\end{figure}

\subsection{Feature Correlation}
To examine how correlated these centralities are during the two phases of the diffusion mechanism studied in this paper, we examined how similar each of these measures are with respect to the most central nodes ranked according to the measures described above. To this end, we consider the top 20 nodes in terms of rank for each measure considering the temporal networks $G^N$ for each temporal interval $N$ and find the pairwise Jaccard Similarity \cite{Jaccard} subsequence-wise. This would allow us to observe whether a node with high nodal degree also exhibits higher Pagerank as the network progresses over time and similar hypotheses for other pairs of measures considered in the paper. From Figures~\ref{fig:jaccard_1}(a), we find that \textit{Degree entropy} has higher similarity in terms of the Jaccard measure with \textit{Pagerank} and \textit{Alpha centrality}, which hints at the fact that the these measures are similar in that they consider the extent of neighborhood connectivity as a measure of diffusion and not just the immediate neighbors for the measurement. On the other hand, from Figures~\ref{fig:jaccard_1}(e) and \ref{fig:jaccard_2}(a) we find that \textit{Nodal degree} shares more similarity with \textit{Betweenness} and the similarity increases rapidly towards the intervals preceding the steep region. We find that \textit{Clustering coefficient} shares the least similarity in terms of the top ranked nodes shown in Figure\ref{fig:jaccard_2}(e) and (f) and one of the major reasons behind this is that since the clustering coefficients are low in a cascade setting, many top ranked nodes share similar values and hence taking the top ranked nodes would mean randomly picking one from among two similar ranked nodes.

\section{Quantitative analysis}
\label{sec:quant_res}
\subsection{Granger Causality results}
\begin{figure}[h!]
	\centering
	\includegraphics[width=7cm, height=4cm]{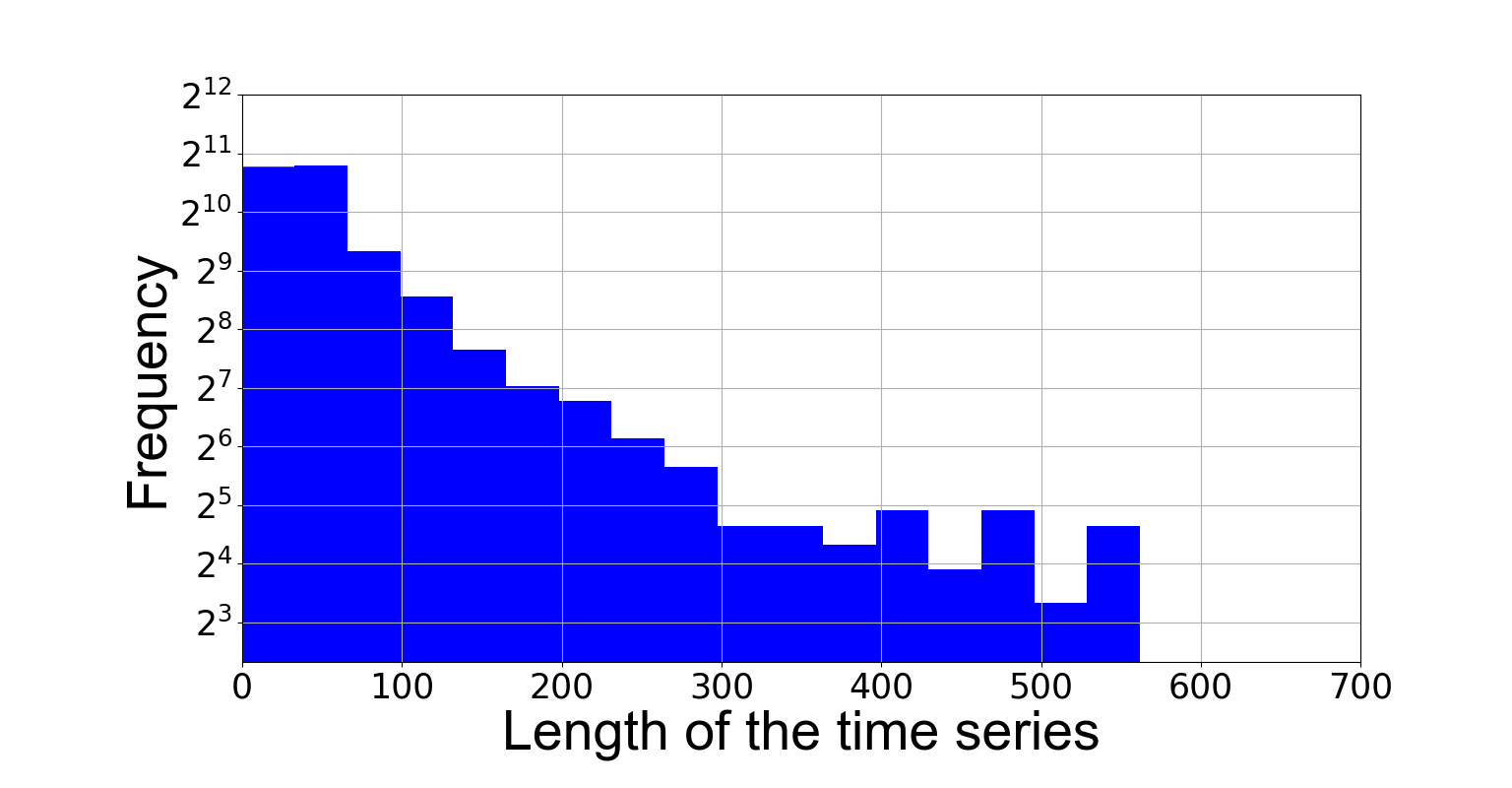}
	\caption{Histogram of time series length.}
	\label{fig:hist_series_len}
\end{figure}

Lag order selection in Vector Autoregressive models (VAR) has been comprehensively studied in \cite{lag_order}. We set the maximum lag order parameter $P$ of the autoregression model, explained in Section~\ref{sec:stat_test} to 5 although as mentioned, for each time series we select the best order $p$ in the range $[1, P]$ based on $AIC$ criterion which we finally use in Equations~\ref{eq:granger_null} and \ref{eq:granger_full}. This helps us to keep the lag order dynamic for different time series representing different cascades and is helpful as the length of the time series varies for the cascades.  We refrain from using orders higher than 5 for the model since as shown in Figure~\ref{fig:hist_series_len}, most of the time series have lengths in the range $[0, 100]$, so having higher orders would not be suitable as training data.

For checking the stationarity of the data, we performed the Dickey-Fuller test \cite{dickey} for each of the time series and in cases where there was an evidence of non-stationarity, we performed a first-order difference of the time-series and used that as input. However from our empirical observations we found that for $\mathcal{R}$, since we are already using reshare time differences as the time series which implicitly exhibited a first-difference of $\tau$ for the cascade, for most of the cascades they are stationary.

\begin{figure}[!h]
	\centering
	\includegraphics[width=9cm, height=5cm]{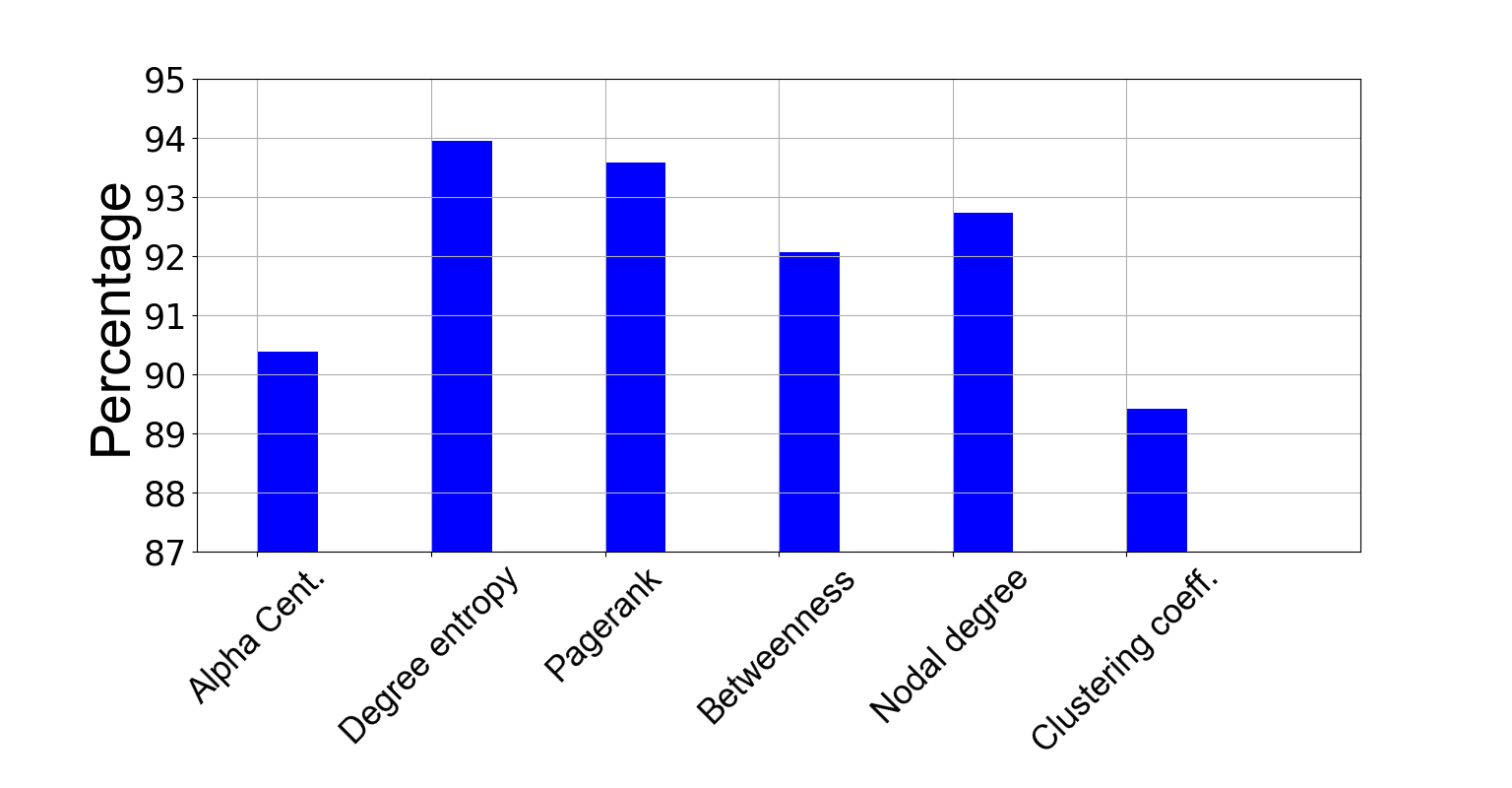}
	\caption{Granger causality results. The bar plot shows the percentage of cascades where the network feature $\mathcal{T}_f$ Granger causes $\mathcal{R}$.}
	\label{fig:causality_plot}
\end{figure}

\begin{figure}[h!]
	\centering
	\includegraphics[width=9cm, height=5cm]{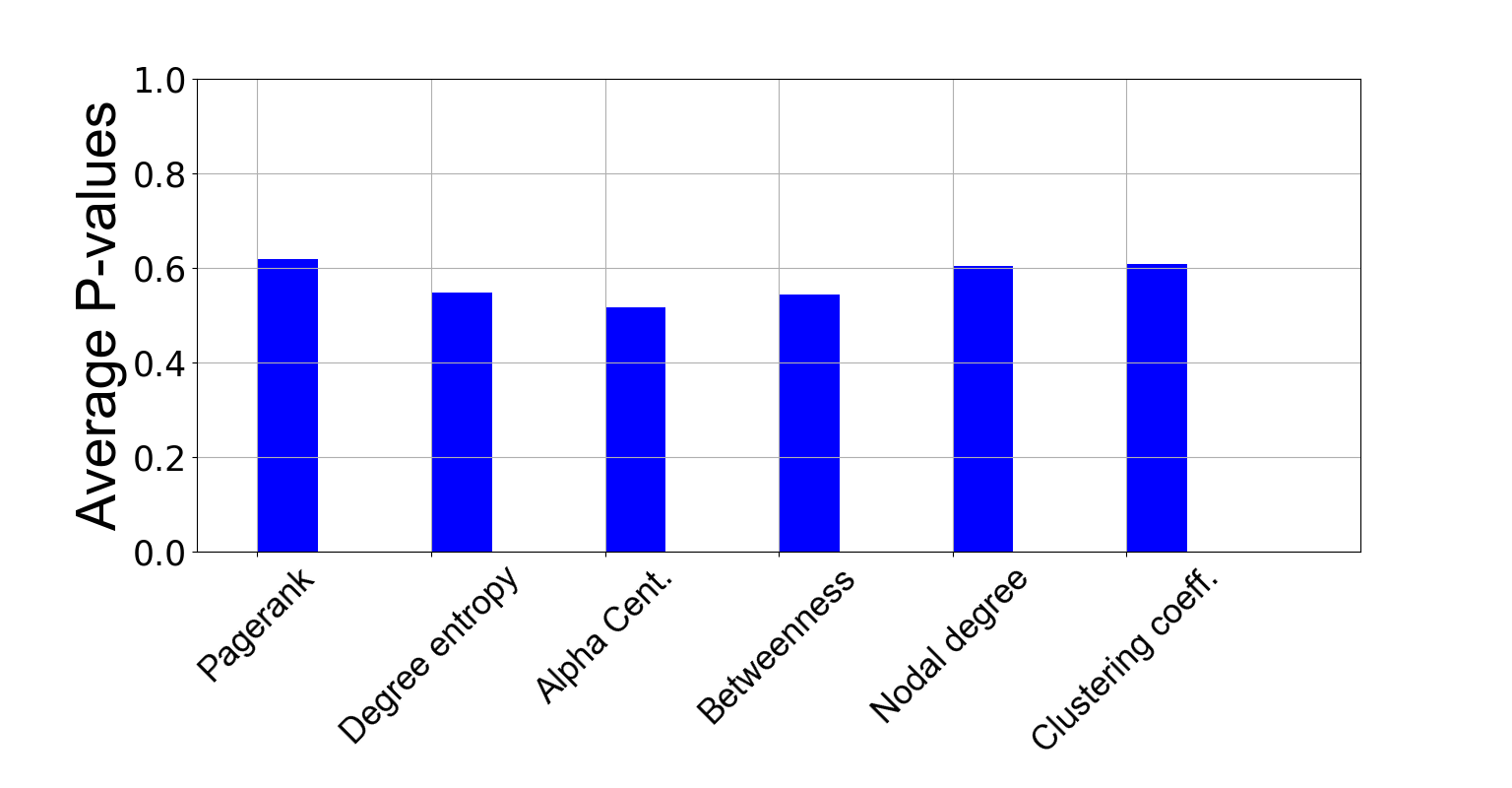}
	\caption{Average $p$-values corresponding to the test-statistic used for the VAR model in Equations~\ref{eq:granger_null} and \ref{eq:granger_full}.}
	\label{fig:p_values}
\end{figure}

For our Granger causality framework, for each of the measures we count the number of cascades where the feature time series $\mathcal{T}_f$ granger-caused the reshares response time $\mathcal{R}$ indicated by the rejection of the null model and acceptance of the full model in Equation~\ref{eq:granger_full}. The Granger causality results in Figure~\ref{fig:causality_plot} and the corresponding average $p$-values for the time series' in Figure~\ref{fig:p_values} show that the while degree entropy and PageRank as individual measures both prove to be the better indicators of the cascade dynamics leading to inhibition, clustering coefficient proves to be weakly causal among those examined. One of the many reasons for the failure of clustering coefficient in incrementally adding to the prediction of future is that since clustering measures the number of triangles around a node, it usually helps to have a large network for evaluation and with temporal networks which are constrained by the size of the cascades hinder formation of large clusters. A second reason for the poor performance of clustering coefficient as an indicator of inhibition is that unlike in social networks containing all cascades where the traces of individuals are recorded over a period long enough to measure individual tendencies towards large group formation, in a cascade setting however, the ``influencers'' keep changing very rapidly so that it becomes very difficult for an individual to form large groups within such a short span. 

\subsection{Forecasting results}
\label{sec:for_exp}
In this section, we first describe the method we used to split the data to form training and test data sets. Since the autoregression model in Equation~\ref{eq:granger_full} has a lag order of $p$, we partition the data into two groups: the training set having all the data points save the last $p+1$ points prior to $t_e$ and we reserve the last $p+1$ points to form our test set with the first $p$ among them as the input data for the regression model and the last point corresponding $t_e$ as the test output for validation. Let $\mathcal{R}_{e, c}[t_e]$ be the actual value at $t_e$ for a cascade $c$ for the event $e$ and $\hat{\mathcal{R}}_{e, c}[t_e]$ be the estimated value by the VAR model. For the forecasting $t_{inhib}$, we perform an additional control experiment on the length of the time series. We only use the points in $\mathcal{T}_{f, inhib}$ ranging between $t_{steep}$ and $t_{inhib}$ and similarly for $\mathcal{R}_e$ which we call \textit{clipped series}. We use \textit{clipped series} to see whether points just in the near past of $t_{inhib}$ improve the forecast results compared to the case when we include all points in the range $[0, t_{inhib}]$. Since the time series of the features may not be perfectly stationary, this control experiment helps us in analyzing the importance of memory in event detection. We compute the mean absolute error for all the combinations of models and time series lengths on our corpus as follows:
\begin{equation}
MAE = \frac{1}{|C|}\sum_{c=1}^{|C|} |\mathcal{R}_{e, c}[t_e]- \hat{\mathcal{R}}_{e, c}[t_e] |
\end{equation}

where $|C|$ denotes the number of cascades. 

\begin{figure}[h!]
	\centering
	\includegraphics[width=10cm, height=5cm]{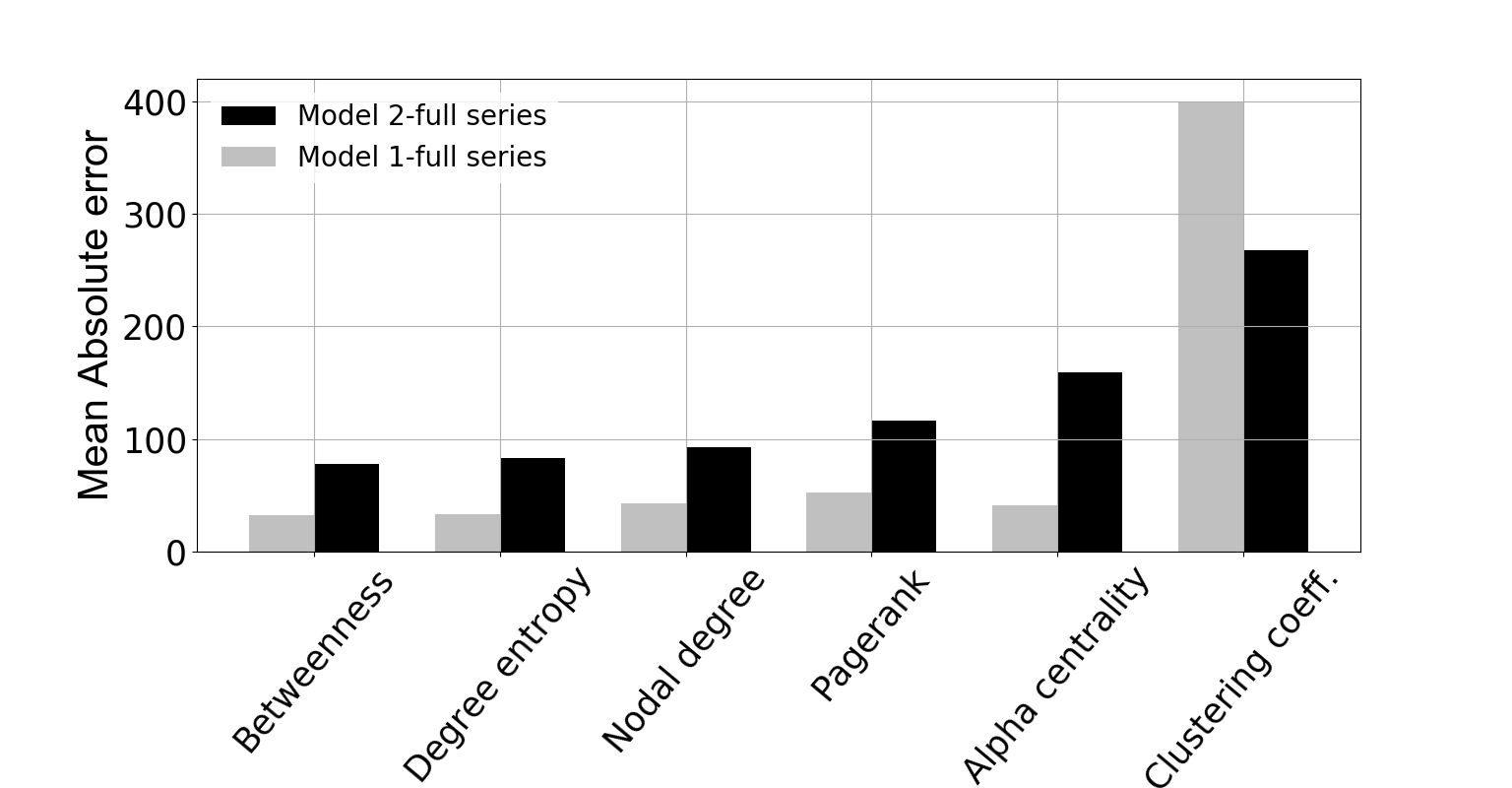}
	\caption{Bar plot of the forecast results for $t_{steep}$ for each feature.}
	\label{fig:forecast_results_steep}
\end{figure}

For forecasting $t_{steep}$, we find from Figure~\ref{fig:forecast_results_steep} that for all the network features, Model 1 performs significantly better than Model 2 in terms of the errors. The unusually high error values for clustering coefficients indicate that clustering coefficients of nodes in a cascade setting perform very less to no role in predicting the future course of the \textit{growth} phase in the lifecycle. Although other features perform equally, we find that the individual effects of the features are more useful as predictors than the combination with the response variable itself in the VAR model in Equation~\ref{eq:full_reg}.

\begin{figure}[h!]
	\centering
	\includegraphics[width=10cm, height=5cm]{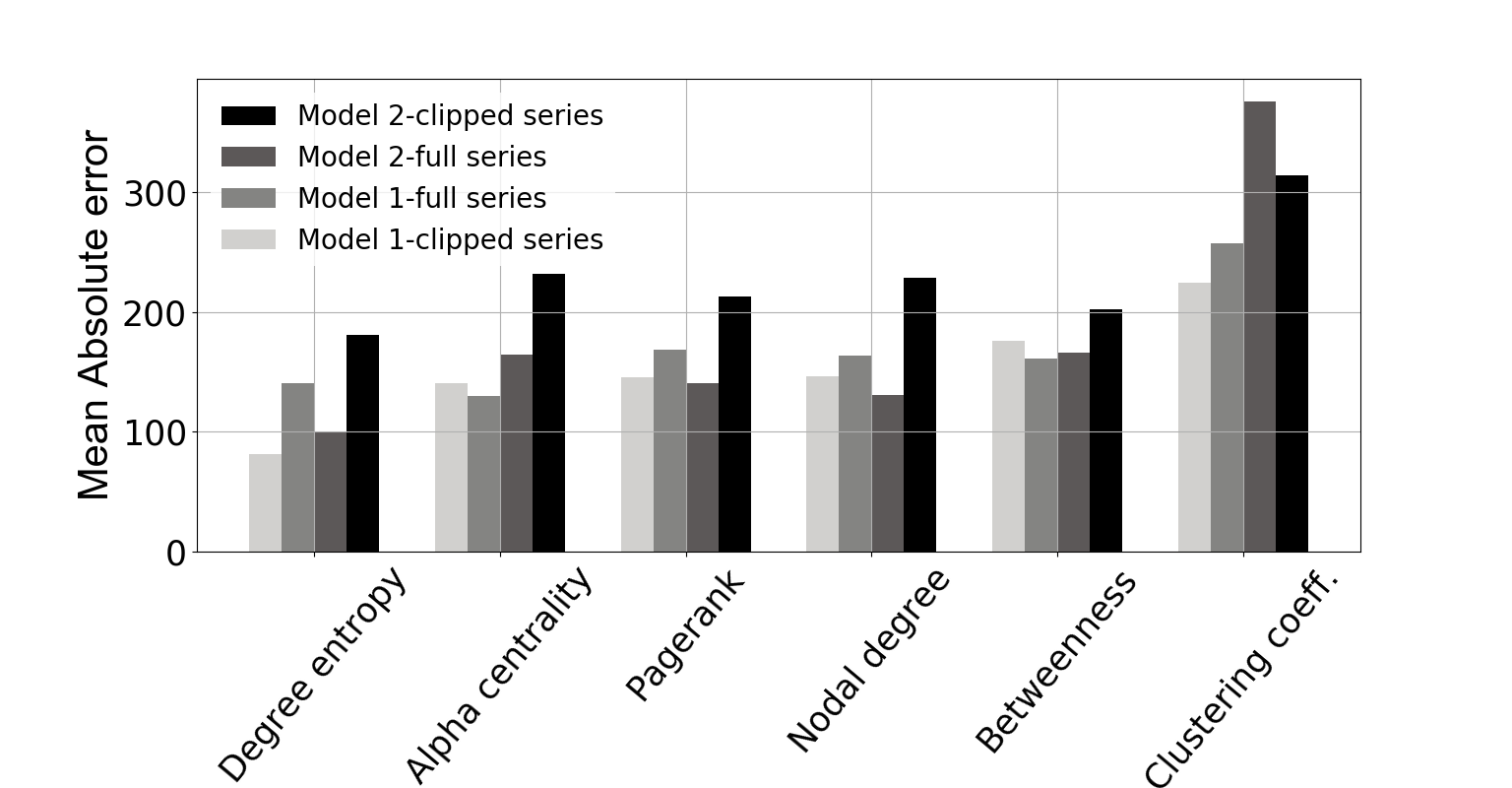}
	\caption{Bar plot of the forecast results for $t_{inhib}$ for each feature.}
	\label{fig:forecast_results}
\end{figure}

The results for the forecast of $t_{inhib}$ in Figure~\ref{fig:forecast_results} bring out the following observations: (1) firstly, the results of the Granger causality are synchronous to the forecasting results for $t_{inhib}$ from the context of the node measures as we find that for all the combinations, degree entropy turns out to be better in forecasting $t_{inhib}$ denoted by the least MAE in all the cases while clustering coefficient turns out to be a bad feature for measurement having the highest MAE among all, (2) in general, when we include only the points between the steep and the inhibition interval for measuring $t_{inhib}$, \textit{Model 1} works better than \textit{Model 2} which means that when we use the node measures in silos, the points closer to inhibition are more informative and it is not necessary to look into the far past. Secondly, when we include all the points in $\mathcal{T}_f$ for the models, we do not find any concrete evidence to suggest the better among the two, (3) In fact when we compare the two models, we find that for \textit{Model 2} which replicates the full Granger model, taking the whole time series for the regression model yields better results in terms of lower MAE while for \textit{Model 1} which tests the performance of individual measures as predictors, it is better to take only the points in the vicinity of the inhibition region.

\section{Conclusion and Future Work}
Identifying and validating the event intervals is a complicated task due to lack of ground truth for evaluation. In this paper we try to formalize the event interval identification mechanism using point processes and then define certain structural features which qualitatively as well as quantitatively measure causes behind the dynamics of growth during the two event intervals in the cascade lifecycle. In this paper, we consider the cascade network structure as the main component for analysis instead of the contact or friendship network that is traditionally used for analysis of node importance. The temporal evaluation of the cascade based on structural properties is a first in this kind of study although there have been close approaches for measuring temporal features in networks but avoiding cascades individually. The saliency of this paper lies in the structural analysis of the temporally ordered cascade networks to explain how the properties change and whether they could be indicators of the approaching steep or inhibition subsequences. We analyze the network using node-centric features which help us understand the resharing process from the user's perspective. We conclude that a user's influence in terms of higher node degree compared to its neighbors can be instrumental in driving the cascade augmented by the fact that the degree entropy has a high causal relationship with the resharing response time among all the features examined. It also exhibits the lowest MAE when we use it for forecasting the future inhibition and steep event times. On the other hand, we find that clustering coefficient of a node is unable to explain the diffusion process of the cascades leading to steep and inhibition subsequences explained by the fact that it was able to causally affect the reshare responses, the least among the features examined. Additionally it also exhibited highest MAE when forecasting the occurrence times of the events. Although the entire paper has been focused on certain type of cascades, this work can be extended to all cascades by relaxing the definition of ``inhibition'' to include all points in  a cascade where there is slackness in growth. One of the other areas where this work can be extended is to introduce measures and algorithms for minimizing the inhibition and to study what settings could boost information spread in cascades \cite{LinkInj}.

\noindent \textbf{Acknowledgements} Some of the authors are supported through the AFOSR Young Investigator Program (YIP) grant FA9550-15-1-0159, ARO grant W911NF-15-1-0282, and the DoD Minerva program grant N00014-16-1-2015.




\section*{Appendix}

\textbf{A1. Defining steep and inhibition intervals:}
\label{app:A1}
In this section we describe the 3-step algorithm used for identifying the \textit{steep} and \textit{inhibition} intervals and that serve as the two intervals for social network analysis for Type I cascades. The lack of ground truth in validating the two regions makes it difficult to identify the regions with accuracy, however we resort to an approach which is unsupervised in a way that we use the information from correctly identified intervals in majority cascades to rectify and remove the incorrect ones.

Now, we briefly introduce the two main concepts in point processes which have been used to model the reposting events of \textit{C}.
\\

\textbf{Definition 1: Point Processes} Let $t_i$ be a set of random variables 
$\forall i \in [N] $. Each of these values map a certain time interval in which an event occurs (the event being reposting in our case). The stochastic process quantifying these events defined by these variables $t_i $ gives rise to a point process.  The counting process $N(t) = \sum_{i, t_i < t} 1 $ is an alternative description of the point process. A point process can be defined in terms of N(t) as below:

\begin{equation}\label{eq:hawkes_prob}
P[N(t+\Delta t) - N(t)] = \xi(t)\Delta t + o(\Delta t)
\end{equation}

Equation~\ref{eq:hawkes_prob} says that the probability of an event occurring in a small time interval $\Delta t $ is proportional to the time-varying intensity function $\xi(t) $ (which is a probability density function) added to the time-invariant function $o(\Delta t)$. So if $\xi(t) $ is constant like for example, a Poisson process with $\xi(t) = \mu $, the process has no memory or the intra-event duration  $\Delta t $ does not depend on previous events and thus is i.i.d. To overcome this shortcoming we use the improvement introduced by Hawkes.\\

\begin{figure}[H]
	\centering
	\includegraphics[width = 7cm, height = 3cm]{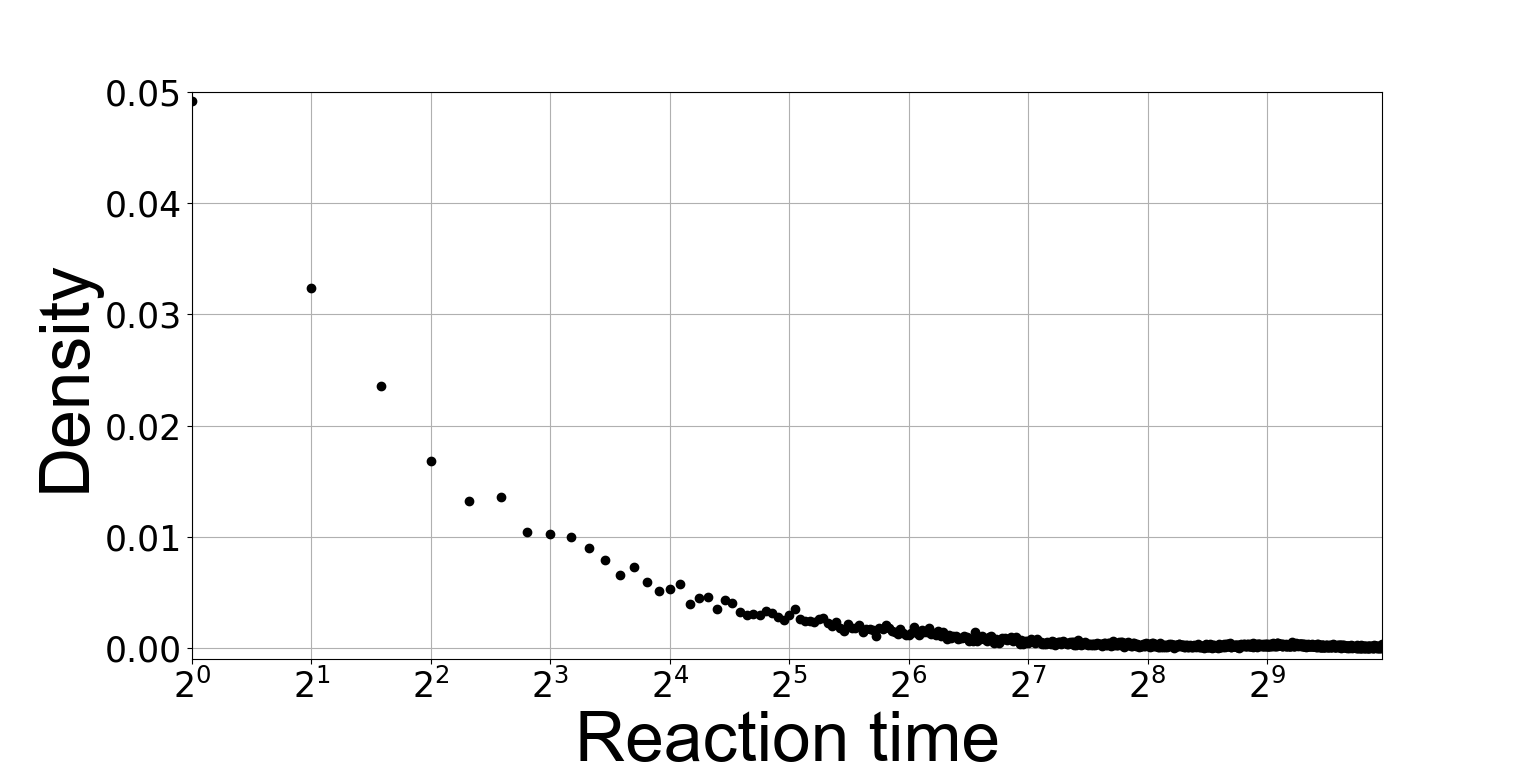}
	\caption{Probability density function $h $ or memory kernel - Reaction time distribution. The reaction time is plotted in logarithmic axes.}
	\label{fig:pdf_hawkes}
\end{figure}

\textbf{Definition 2: Single variable Hawkes process:} Hawkes process is a self-exciting point process that incorporates a response function (or kernel, in fact which is a probability density function) $h(t-t_i)$ which uses the influence of history of events on the current event to define $\xi (t) $.

\begin{equation}\label{eq:hawkes_eq}
\xi(t) = \mu(t) + \sum \limits_{t_i<t, i>0}h(t-t_i)
\end{equation}

The first term in Equation~\ref{eq:hawkes_eq} ($ \mu (t) $) is the base intensity of the model that determines the rate of arrival of first order events per unit of time, which in our case is assumed to be 0 as we consider the final intensity to be independent of this  base intensity (or which, without loss of generality could be assumed constant). The response function $h(t) $ is the probability density function of the human reaction times shown in Figure~\ref{fig:pdf_hawkes}. It follows a long-tailed power law with high density only at the beginning. In our work we define the reaction time of a user as the time gap between the time at which the user reposts the blog and the time at which its parent node reposted or posted the blog. This is the kernel or the response function $h(t) $ described in Hawkes equation. 

The concept of Hawkes process as a self exciting process can be observed from the fact that each of the previous $t_i $ observations in Equation~\ref{eq:hawkes_eq}  contribute to the intensity at time $t$. Our definition of Hawkes intensity at time $t $ is based on the measure used in \cite{seismic} which defines the rate of obtaining a re-share as below:

\begin{equation}
H[t] = p_t \sum \limits_{(t-\Delta t) \leq t_i \leq t} n_i h(t-t_i), \ t \geq t_0
\label{eq:hawkes_seismic}
\end{equation}

where $p_t$ is the infectiousness parameter of the cascade and $n_i$ refers to the out-degree of the node that contributed to the reposting at time $t_i $. The $\Delta t $ in the limits of the sum is where we deviate from the normal Hawkes equation in that we do not consider the contribution of all the past events in the intensity value at time $t $, but only events occurring in the time interval $\Delta t$ before event at $t $ occurs. 

The parameter $p_t $ measures the infectiousness of the cascade at time $t $ or in other words it is a parameter that defines the influence of infectivity of events at time $t $ on the intensity $\xi_t $. Learning parametric models to define $p_t $ by maximizing a likelihood function has been studied in \cite{hawkes_zha}. However, in our work we assume it to be constant over time and that this parameter does not affect the resposting event over time. So essentially we assume that it is an interplay of the reaction time of users as well as the user out-degree that contributes to the intensity. The node reaction time PDF $h $ is a distribution of the time taken by the users to adopt that cascade as shown in Figure~\ref{fig:pdf_hawkes}.

The advantage of using Hawkes intensity measure over the normal peak detection method of slope curve for detecting the \textit{steep} and \textit{inhibition} intervals is that Hawkes process is driven by spike dynamics and hence any sharp increase or decrease in the growth of the cascade is captured by the Hawkes intensity equation defined in Equation~\ref{eq:hawkes_seismic}.
\\

\noindent \textbf{Algorithm for identifying Steep and Inhibition regions}
For estimating the steep and the inhibition regions, we follow a three step algorithm: \\

\textbf{Step 1 - Obtaining the Hawkes curve:} In the first step, we obtain the Hawkes curve $H $ using the cascade curve $C $ and which would be used for subsequent estimations of the steep and inhibition regions. 

In our work, the interval size $\Delta t$, the number of time intervals before time $t_i $ for calculating the Hawkes intensity at time $t_i $ is defined as $\Delta t = \alpha * e^{\frac{t_i}{T_C}} $, 
where $\alpha $ is a scaling parameter which is constant in our case.

\textbf{Step 2: First estimates of Steep and the inhibition time points: } 
For estimating the peak points or the points denoting maximum growth or a sharp downfall in the Hawkes curve, we first divide the entire cascade curve into $K $ intervals of fixed window size given by $K_C = \alpha * \log(T_C) $, where $\alpha $ is the same scaling parameter used in Step 1 of the algorithm. We then sum the Hawkes intensity of the time points within each interval so as to get a single Hawkes value for each interval. We denote this new curve obtained after the split as $HI $. Here on, every time point in the Hawkes curve $ HI $ we talk about, would refer to the start time of the corresponding interval in $H$. 

The idea of identifying the \textit{steep} interval and the \textit{inhibition} interval can be derived from the fact that the points of local maxima in $H$ map to regions in $C $ where there was a surge in reposting compared to previous time points and similarly points of local minimum in $HI $ point to regions where there is a decrease in activity relative to its previous time intervals in $ C $. So $HI $ closely resembles the slope of the logistic curve of $C$, where each point $HI[t] $ would resemble the derivative at time point $t $.

After obtaining $HI $, we perform a 2-step procedure for obtaining the estimates of the steep and the inhibition intervals. For detecting the points of local minima and local maxima, we filter out points in $HI $ which are larger than their immediate neighborhood or, $ t_i > t_i-\Delta d$ and $t_i > t_i + \Delta d $ for local maximum and  $ t_i < t_i-\Delta d$ and $t_i < t_i + \Delta d $ for local minimum $\forall t_i \in HI $ $\forall$ $ \Delta d \in {1, 2, 3} $. However this step identifies a lot of outliers as peak points. To remove the outliers, we calculate the moving mean curve $M $ of the intensities for each time point $t_i $ in $HI $ based on the previous $\Delta k$  = 300 minutes interval of $t_i $. Based on this simple technique of finding the maximum among neighborhood, we first find all the local maxima and local minima points. However, since the $steep $ interval shows the maximum rate of growth in the cascade lifecycle through our empirical observations (for Type I cascades), we set the global maximum of $HI $to be $t_{steep}$.

However the same cannot be concluded about the inhibition time points. Therefore using $M $ we filter out all the inhibition time points which are greater than the Hawkes value at those points, that is to say, we keep only those inhibition points $t $ in $HI$ where $HI [t] < M[t] $. The reason behind this step of the algorithm is that the inhibition point $t$ exhibits a very sharp change (and hence lower $HI[t]$) in the reposting rate as compared to its previous time points $t_i$ where the intensities are not lower than $HI [t] $ with $t_i - \Delta k \leq t_i$ $<$ $t$. Therefore any point $t$ which has lower values $HI[t_i] $ will have $ HI[t] > M[t] $ and hence not an inhibition point. Let us denote all the inhibition points obtained after this step as $I^C $ where $|I^C| \geq 1$.

\begin{figure}[H]
	\centering
	\begin{minipage}{0.3\textwidth}
		\includegraphics[width=5cm, height = 3cm]{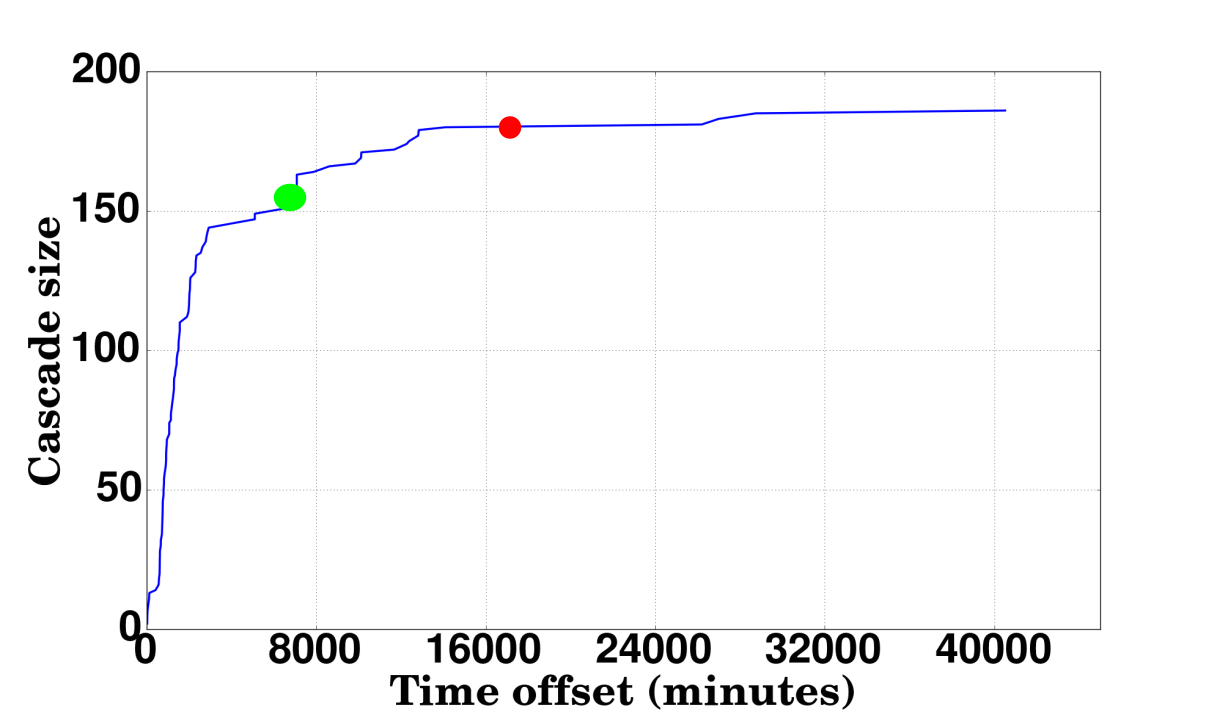}
		\hspace*{2cm}\subcaption{}
	\end{minipage}
	\hspace{3cm}
	\begin{minipage}{0.3\textwidth}
		\includegraphics[width=5cm, height=3cm]{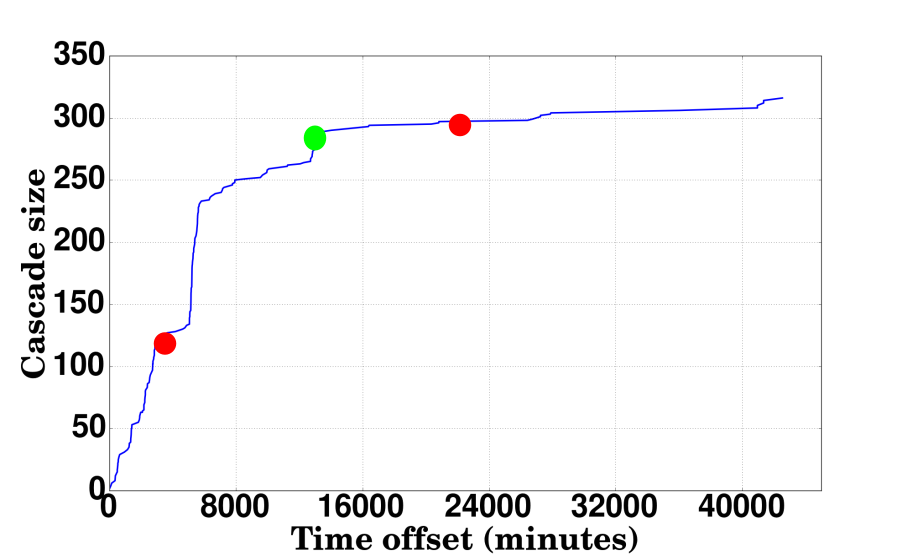}
		\hspace*{2cm}\subcaption{}
	\end{minipage}
	\caption{Two examples of incorrect identification (a) one where the inhibition point gets identified before the actual start. (b) inhibition point gets identified after the actual start. Green points denote the correct $t_{inhib}$. Red points show the incorrect region which the algorithm identifies as $t_{inhib}$.}
	\label{fig:wrong_points}
\end{figure}

\textbf{Step 3- Final Estimates of Steep and Inhibition region:}
As can be seen from Figure~\ref{fig:wrong_points}, even after using the moving mean filter, the output would include multiple inhibition points, some of which are clearly not the inhibition regions that we desire to obtain, two examples of which are shown in Figures~\ref{fig:wrong_points}(a) and (b). Unlike in a prediction problem, due to lack of ground truth values to verify our inhibition points, we resort to a likelihood estimation of the probable inhibition region. 

\begin{figure}[H]
	\centering
	\begin{minipage}{0.3\textwidth}
		\includegraphics[width=5cm, height = 3cm]{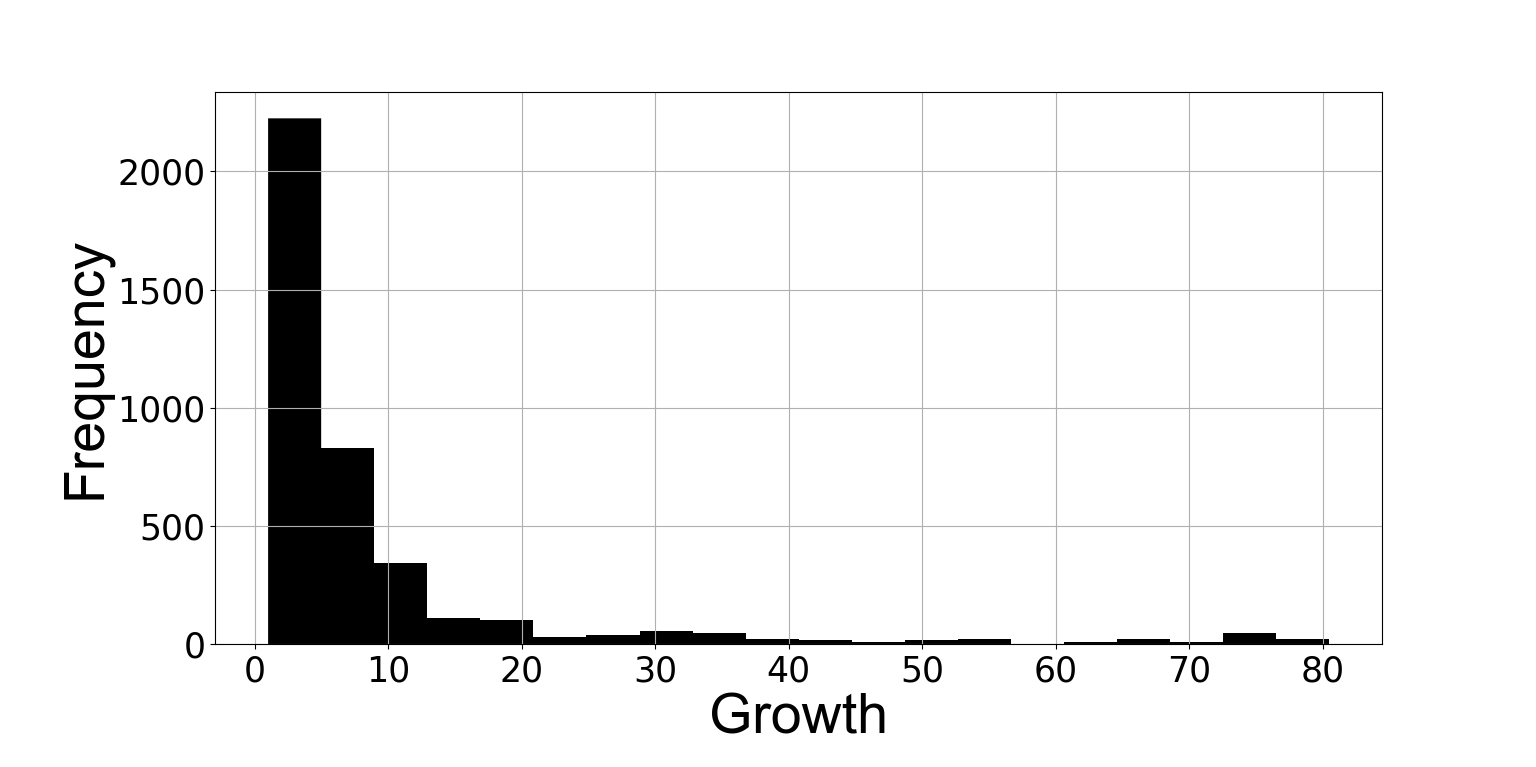}
		\hspace*{2cm}\subcaption{}
	\end{minipage}
	\hspace{3cm}
	\begin{minipage}{0.3\textwidth}
		\includegraphics[width=5cm, height=3cm]{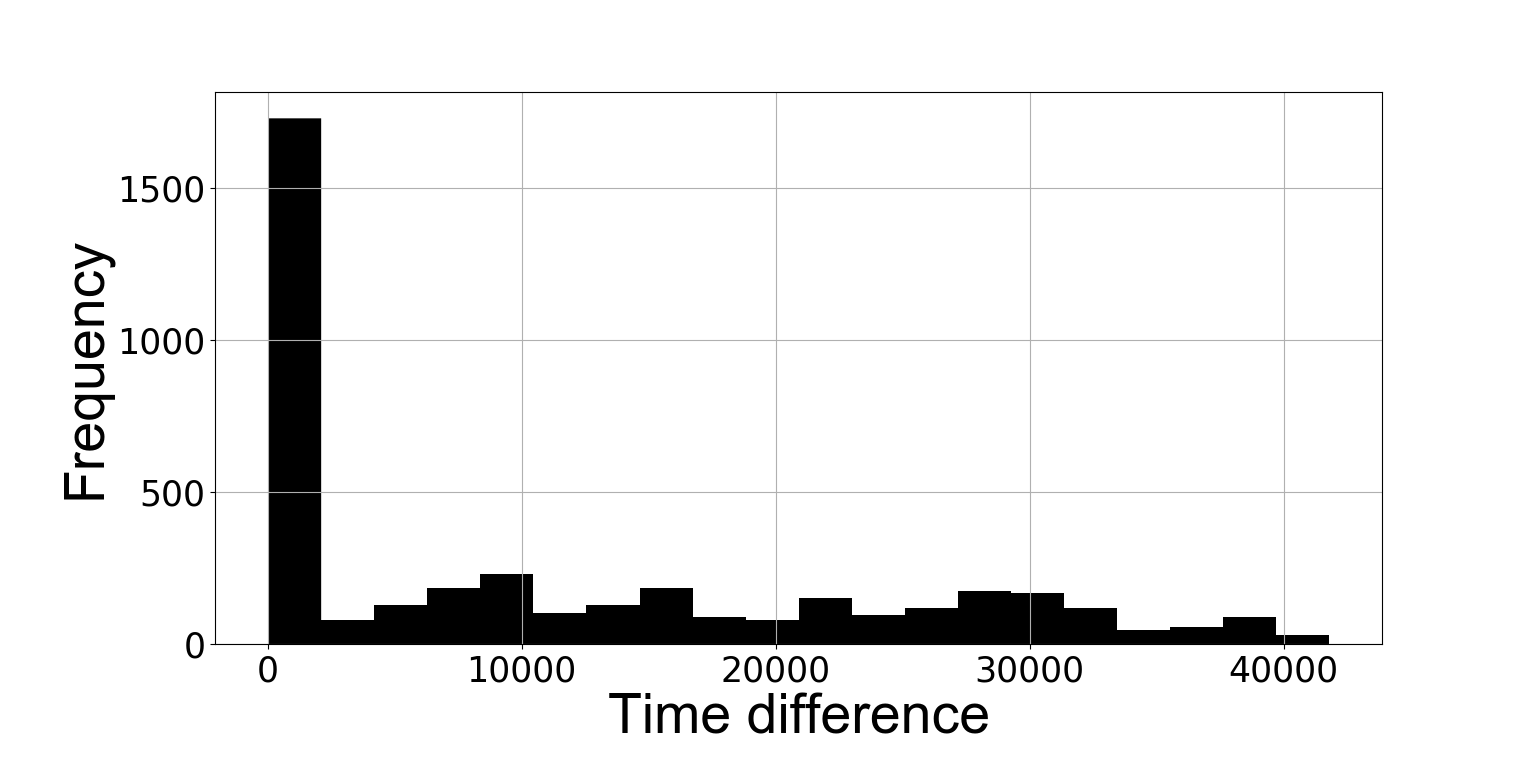}
		\hspace*{2cm}\subcaption{}
	\end{minipage}
	\caption{(a) Growth ratio histogram (b) Time gap histogram}
	\label{fig:growth_hist}
\end{figure}

For this, we run Step 2 on all the cascades, and $\forall ic \in I_C$  $\forall C \in C_{all}$, $C_{all} $ denoting the set of all cascades in our data, we obtain the following two attributes:

1) Time gap $\Delta TG[ic] $ = $ic $  - $t_{steep}^C $

2) Growth $ g[ic] $ = $\frac{S[ic]}{S[t_{steep}]} $

So essentially, $\Delta TG $ and $g $ would be a list of values for all the inhibition points for all cascades- the correctly identified as well the incorrect ones.

We minimize the negative log likelihoods of $\Delta TG $ and $g $ by fitting a function to the probability distributions of $\Delta TG $ and $g$ separately and using an optimization algorithm to obtain the parameters of the function. We use a Poisson density $ f(x|\beta) = \frac{\beta^x}{x!} e^{-\beta}$ with parameters $\beta $ and the corresponding log likelihood function  $L(\beta, x) = ln \ \prod_x f(x| \beta) $, where $x$ refers to the values in $\Delta TG $ and $g $. Here we run the optimization procedure twice separately for $\Delta TG$ and $g $ to obtain two $\beta$ values - one each for $\Delta TG$ and $g $. We use the Nelder-Mead optimization algorithm \cite{nelder} for obtaining the parameter $\beta$. Once we compute the parameters, we consider the most probable values for the two attributes $TG_p$ and $g_p$ as the mean of the Poisson density functions estimated. Since the mean of a Poisson density is $\beta$ itself, we take the parameter $\beta$s of $\Delta TG$ and $g $ as $TG_p$ and $g_p$ respectively. We obtain $TG_p$ and $g_p$ as 4171.25 and 4.25 respectively where the time gap $TG$ is measured in minutes.
We select the first point $t $ in $C $ whose $\Delta TG[t] $ and $g[t] $ are both greater than $TG_p$ and $g_p$ respectively. We denote that $t$ to be $t_{inhib} $. 
\\

\textbf{A2. Parameter sensitivity evaluation for $\alpha$ for window scaling:}
\\

In our method for calculating Hawkes interval intensity at time $t$, we consider intervals of size $K_C$ where $K_C = \alpha * \log{T_C} $. As mentioned before, $\alpha$ controls the window size and to evaluate the sensitivity of $\alpha$ on the final inference of $t_{inhib}$ and $t_{steep}$, we evaluate the parameters $\beta$ in the Poisson distribution by considering $\alpha$ in the list [$1, 3, 5, 7, 10, 15]$, that is we consider $K_C$ very small as well as extremely large. 

\begin{figure}[H]
	\begin{minipage}{0.3\textwidth}
		\includegraphics[width=5cm, height = 3cm]{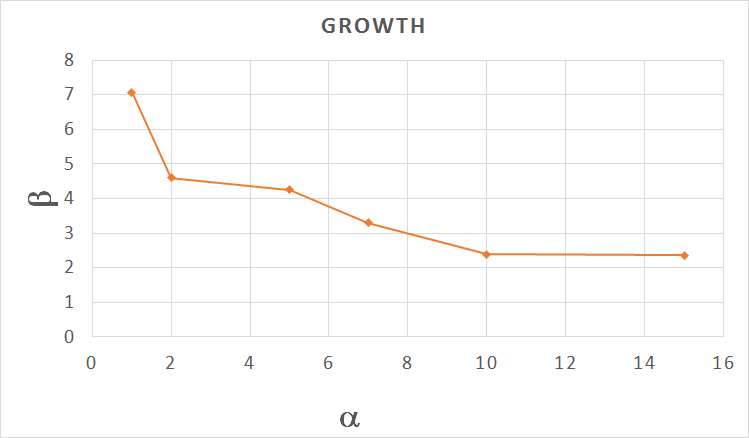}
		\hspace*{2cm}\subcaption{}
	\end{minipage}
	\hspace{3cm}
	\begin{minipage}{0.3\textwidth}
		\includegraphics[width=5cm, height=3cm]{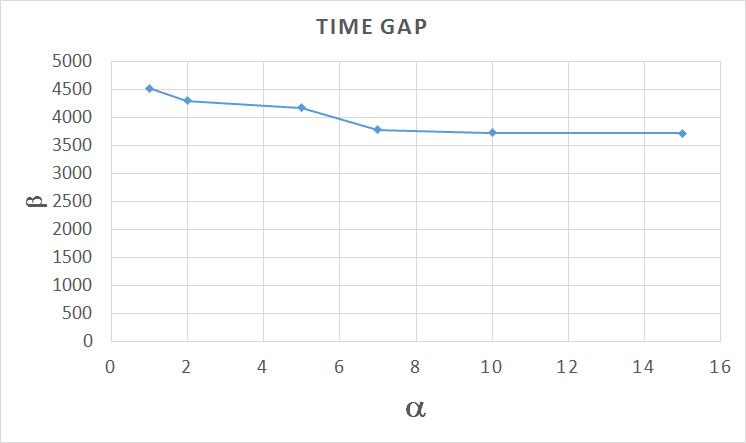}
		\hspace*{2cm}\subcaption{}
	\end{minipage}
	\caption{$\beta$ vs $\alpha$ for (a) Growth ratio  (b) Time gap}
	\label{fig:parm_sens}
\end{figure}

The procedure is briefly listed as follows: after we obtain the set of potential $t_{steep}$ and $t_{inhib}$ for all cascades for each $\alpha$ from Step 2, we fit the Growth ratio and Time Gap values each to Poisson distribution and obtain the set of parameters $\beta$ using MLE. We plot the values of $\beta$ vs $\alpha$ for Growth ratio Figure~\ref{fig:parm_sens}(a) and Time Gap Figure~\ref{fig:parm_sens}(b). We observe a monotonically non-increasing relation and as $\alpha $ increases $\beta$ decreases. Smaller $\beta$ denotes skewed values of Growth whereas very high values of $\beta $ denote a very uniform distribution both of which point to bias and randomness. To avoid that, we pick the parameter value 5 which is an optimal value in that sense. 

\end{document}